\documentclass{emulateapj}

\usepackage{natbib}
\usepackage{amsmath}
\shorttitle{Nuclear cluster}
\shortauthors{FRITZ ET AL.} 

\begin{document}

\title{The Nuclear Cluster of the Milky Way: Total Mass and Luminosity\footnote{Based on observations collected at the ESO Paranal Observatory (programs
060.A-9026, 063.N-0204, 70.A-0029, 071.B-0077, 71.B-0078, 072.B-0285, 073.B-0084, 073.B-0085, 073.B-0745, 073.B-0775, 273.B-5023, 075.B-0547
 076.B-0259, 077.B-0014, 077.B-0503, 078.B-0520, 078.B-0136, 179.B-0261, 179.B-0932,
 179.B-2002, 081.B-0568, 082.B-0952, 183.B-0100, 183.B-1004, 086.C-0049, 087.B-0117, 087.B-0182, 087.B-028, 088.B-0308)
}
}
\author{ T.K.~Fritz \altaffilmark{1,$\#$},  S.~Chatzopoulos\altaffilmark{1}, O. Gerhard\altaffilmark{1}
, S.~Gillessen\altaffilmark{1},  R.~Genzel\altaffilmark{1,2}, 
O.~Pfuhl\altaffilmark{1}, 
 S.~Tacchella\altaffilmark{3}, F.~Eisenhauer\altaffilmark{1}, T.~Ott\altaffilmark{1}
}

\altaffiltext{1}{Max Planck Institut f{\"u}r Extraterrestrische Physik, Postfach 1312, D-85741, Garching, Germany.}
\altaffiltext{2}{Department of Physics, University of California, Berkeley, 366 Le Comte Hall, Berkeley, CA 94720-7300}
\altaffiltext{3}{Institute for Astronomy,ETH Zurich, Wolfgang-Pauli-Strasse 27, CH-8093, Z\"urich, Switzerland}
\altaffiltext{$\#$}{E-mail: tkf4w@virginia.edu}

\keywords{  Galaxy: center - Galaxy: fundamental parameters}

\begin{abstract}
Like many other late-type galaxies, the Milky Way contains a nuclear star cluster. In this work we obtain 
the basic properties of its dominant old stellar population. Firstly, we derive its structural properties by constructing a stellar surface density map of the central 1000$\arcsec$ using extinction corrected                
star counts from VISTA, WFC3/IR and VLT/NACO data. We can describe the profile with a two-component models.              
The inner, slightly flattened                
(axis ratio of $q=0.80\pm0.04$) component is the nuclear cluster,               
while the outer component corresponds to the stellar component of the               
circumnuclear zone. We measure for the nuclear cluster a half-light radius               
of $178\pm 51\,\arcsec \approx 7\pm 2\,$pc and a luminosity of M$_{\mathrm{Ks}}=-16.0\pm 0.5$.             
Secondly, we enlarge the field of view over which detailed dynamics are available from $1\,$pc to $4\,$pc.               
We obtain more than 10000 individual proper motions from NACO data, and more than 2500 radial velocities from VLT/SINFONI data.               
We determine the cluster mass by means of isotropic spherical Jeans modeling.               
We fix the distance to the Galactic Center and the mass of the supermassive black hole.               
We model the cluster either with a constant mass to light ratio or with a power law mass model                
with a slope parameter $\delta_\mathrm{M}$. For the latter               
we obtain $\delta_\mathrm{M}=1.18 \pm0.06$.                
 Assuming spherical symmetry, we get a nuclear cluster mass within 100$\arcsec$ of $M_{100\arcsec}=(6.09 \pm 0.53|_{\mathrm{fix} R_0}\pm 0.97|_{R_0}) \times 10^6 $ M$_{\odot}$ for both modeling approaches.                
A model which includes the observed flattening gives a 47\% larger mass, see Chatzopoulos et al. 2015.                
Our results slightly favor a core over a cusp in the mass profile.                
By minimizing the number of unbound stars within 8$\arcsec$ in our sample we obtain a distance estimate of  $R_0=8.53^{+0.21}_{-0.15}\,$kpc, where               
an a priori relation between $R_0$ and SMBH mass from stellar orbits is used. Combining our mass and flux we obtain $M/L=0.51 \pm 0.12  M_{\odot}/L_{\odot,\mathrm{Ks}}$.                
This is roughly consistent with a Chabrier IMF.               
\end{abstract}

\maketitle

\section{Introduction}

In the centers of many late-type galaxies one finds massive stellar clusters, the nuclear star clusters 
\citep{Phillips_96,Matthews_97,Carollo_98,Boeker_02}. The nuclear clusters are central light overdensities on a scale of about 5~pc \citep{Boeker_02}. 
Also the central light concentration 
of the Milky Way \citep{Becklin_68}, is a nuclear star cluster \citep{Philipp_99,Launhardt_02}. 
Nuclear clusters are comparably dense as globular clusters, but are typically more massive \citep{Walcher_05}.
In some galaxies the clusters coexist with a supermassive black hole (SMBH), see e.g. \citet{Graham_09}.
The formation mechanism of nuclear stars cluster is debated \citep{Boeker_10}.
There are two main scenarios: on the one hand formation of stars in dense star  clusters, which are possibly globular clusters, followed by cluster 
infall \citep{Tremaine_75,Andersen_08,Capuzzo_08}. On the other hand in situ star formation from the cosmological gas inflow 
\citep{Milsavljevic_04,Emsellem_08}.

Due to the proximity of the center of the Milky Way the nuclear cluster of the Milky 
Way can
 be observed in much higher detail than any other nuclear cluster \citep{Genzel_10}. It is useful to shed light on the properties and the origin
of nuclear clusters in general.  
The access is hampered by the high foreground
 extinction of A$_{\mathrm{Ks}}=$2.42 \citep{Fritz_11}. 
 Therefore the light profile is uncertain: \citet{Becklin_68,Haller_96,Philipp_99} find a central light excess with a size of at least 400$\arcsec$
on top of the bulge. In contrast, \citet{Graham_09,Schoedel_11a} claim that the nuclear 
cluster transits at 150$\arcsec$ to the bulge. On a larger scale \citep{Launhardt_02} find
that there is another stellar component between nuclear cluster and bulge,
an edge-on disk of 3$^{\circ}$ length, the nuclear disk, which is flattened by a factor five. It corresponds roughly to the central molecular zone \citet{Launhardt_02}. 
The flattening is qualitatively confirmed in
 \citet{Catchpole_90} and \citet{Alard_01}. Further in, the flattening is less well constrained.
\citet{Vollmer_03} mention an ellipsoid of 1.4:1 in the range of about 200$\arcsec$.
Within 70$\arcsec$ the light distribution seems to be circular \citep{Schoedel_07} although a quantification is 
missing therein.    
The majority of the stars in the central R$\approx$2.5 pc are older than 5 Gyrs  
\citep{Blum_02,Pfuhl_11}. Only in the center (r$\approx\,$0.4 pc) the light
 is dominated by 6 Myrs old stars \citep{Forrest_87,Krabbe_91,Paumard_06,Bartko_09} 
with a top-heavy IMF \citep{Bartko_10,Lu_13}.

The mass of the SMBH is well-determined to be $ 4.3 \times10^6$ M$_{\odot}$ with an error of less than 10\% 
\citep{Gillessen_09,Ghez_09}. In contrast the mass of the nuclear cluster is less well constrained. The mass within 
r$\leq1$pc is $ 10^6$ M$_{\odot}$ with about 50 \%  systematic uncertainty \citep{Genzel_10}. In the central parsec the potential is dominated by the SMBH which makes measurements of the additional stellar mass  
more difficult. 
Further, possibly due to the surprising core in the profile of the stellar distribution \citep{Buchholz_09,Do_09,Bartko_10}
also recent Jeans modeling attempts \citep{Trippe_08,Schoedel_09a} to recover the right SMBH mass fail. This possibly biases also the stellar mass determination there. 
As a result also the newer works of \citet{Trippe_08} and \citet{Schoedel_09a} have still about 
50 \% stellar mass uncertainty in the central parsec, similar to \citet{Haller_96} and \citet{Genzel_96} who used fewer radial velocities.
The mass determination outside the central parsec is mainly based on relatively few radial velocities of late-type stars, either 
maser stars \citep{Lindqvist_92b,Deguchi_04}, or stars with CO band-heads \citep{Rieke_88,McGinn_89}. With the absence of proper motion information outside the center the extent of anisotropy is there also not well constrained.
Also radial velocities of gas in the circumnuclear disk  (CND) were used for mass determinations outside the central 
parsec \citep{Genzel_85,Serabyn_85,Serabyn_86}.

Thus, although the central cluster of the Milky Way is the closest
nuclear cluster, its mass and luminosity profiles are still
poorly constrained.  Here and in \citet{Chatzopoulos_14} we improve
the constraints on these parameters. In this paper, we first present
improved observational data: we extend the area for which all
three stellar velocity components are measured, to r$\approx 4$ pc. We also
construct a surface density map of the nuclear cluster out to r$_{\mathrm{box}}= 1000\arcsec$. Then we
present a first analysis of the new data, using simple isotropic
spherical Jeans models. With these assumptions, the analysis is
relatively fast and we can easily investigate several systematic effects. Because these models do not fully match the nuclear cluster, we employ
more detailed axiymmetric models in \citet{Chatzopoulos_14} which
fit the data well.

In Section~\ref{sec:dataset} we present our data, and describe the extraction of velocities in Section~\ref{sec:deriv_vel}. In Section~\ref{sec:lum_prop} we derive 
the surface
density properties of the nuclear cluster
 and fit it with empirical models. In Section~\ref{sec:an_method} we describe the kinematic properties mostly qualitatively and
use Jeans modeling to estimate the mass of the nuclear cluster.
We discuss our results in Section~\ref{sec:discussion} and conclude in Section~\ref{sec:conclusions}. Where a distance to the GC needs to be 
assumed, we
adopt $R_0=8.2\,$kpc \citep{Reid_93,Genzel_10,Gillessen_13,Reid_14}.
Throughout this paper we define the projected distance from Sgr~A* as R and
the physical distance from Sgr~A* as r \citep{Binney_08}.

\section{Data Set}
\label{sec:dataset}

In this section we describe the observations used for deriving proper 
motions, radial velocities, and the luminosity properties of the nuclear 
cluster. 

\subsection{High Resolution Imaging}
\label{sec:highres_im}

For deriving proper motions and for determining the stellar 
density profile in the center (R$_{\mathrm{box}}\approx20\arcsec$)
we use adaptive optics images with a resolution of
$\approx 0.080\arcsec$.
In the central parsec we use the same NACO/VLT images 
\citep{Lenzen_03,Rousset_03} as described in
\citet{Trippe_08} and \citet{Gillessen_09}. We add images obtained in 
further epochs since then, in the 13 mas/pixel scale matching to the
\citet{Gillessen_09} data set and in the 27 mas/pixel scale extending 
the \citet{Trippe_08} data set.
The images are listed in Appendix~\ref{sec:ap_prom_mot}.

For obtaining proper motions outside
the central parsec we use adaptive optics images covering a larger field of view. These 
are four epochs of NACO/VLT images,
one epoch of MAD/CAMCAO  at the VLT \citep{Marchetti_04,Amorim_06} and 
one epoch of
Hokupa'a+Quirc \citep{Graves_98,Hodapp_96} Gemini North images, see Appendix~\ref{sec:ap_prom_mot}.  
Most images cover the Ks-band, some are obtained with H-band or narrower 
filters within the K-band.
The VLT images are flat-fielded, bad pixel corrected and sky subtracted.
In case of the Gemini data we use the publicly available images\footnote{Based on the Data Set of the Gemini North Galactic Center Demonstration Science.}. 
These 
images are combinations of reduced images with nearly the same pointings.

\subsection{Wield Field Imaging}
\label{sec:wide_im}

To obtain the structural properties of the nuclear cluster
outside of the central R$_{\mathrm{box}}=20\arcsec$ we use two additional
data sets. Here, high resolution is less important, but area coverage and extinction correction are the keys.

\begin{enumerate}
\item Closer to the center we use HST WFC3/IR data\footnote{
Based on observations made with the NASA/ESA Hubble Space Telescope, obtained from the Data Archive at the Space Telescope Science Institute, 
which is operated by the Association of Universities for Research in Astronomy, Inc., under NASA contract NAS 5-26555. These observations are 
associated with program 11671  (P.I. A. Ghez).
}.
The central R$_{\mathrm{box}}\approx68\arcsec$ around Sgr~A* are covered in the filters M127, M139 and M153. We use the images in M127 and M153 in our analysis.
 We optimize the data reduction compared to the pipeline in order to achieve Nyquist-sampled final pixels. We use MultiDrizzle to combine the different images in the same filter, with a drop size parameter of 0.6 with boxes and a final pixel 
size of 60 mas. These choices achieve a high resolution and still samples the image homogeneously enough to avoid pixels without flux.
We do not subtract the sky background from the images, since it is 
difficult to find a source-free region from where one could estimate it.
Since we use only point source fluxes this does not affect our analysis. 
We change the cosmic removal parameters to  7.5, 7, 2.3 and 1.9 
to avoid removal of actual sources. Due to the brightness of the GC sources cosmics are only of minor importance.
The final images have an effective resolution of 0.15$\arcsec$.
\item On a larger scale we use the public VISTA Variables in the Via
 Lactea Survey (VVV)  data obtained with VIRCAM \citep{Saito_12}. We use from data release 1 the central tile 333 in H and Ks-band. 
The data contain flux calibrated, but not 
background subtracted images. The resolution is about 1$\arcsec$. These images cover more than one square degree around the GC. We only
use R$_{\mathrm{box}}\approx1000\arcsec$. 
In the center the crowding is severe and nearly all sources are saturated in the Ks-band. This is not a limitation, since 
there we can use the higher resolution images from NACO and WFC3/IR.
\end{enumerate}

\subsection{Spectroscopy}

For obtaining spectra of the stars we use data cubes obtained with the 
integral field spectrometer SINFONI \citep{Eisenhauer_etal2003, 
Bonnet_03}.
We use data with the combined H+K-band (spectral resolution 1500) and 
K-band (spectral resolution 4000) grating.
The spatial scale of the data varies between the smallest pixel scale 
($12.5\,$mas pixel$^{-1}$ $\times$ $25\,$mas pixel$^{-1}$) and the 
largest scale $125\,$mas pixel$^{-1}$ $\times$ $250\,$mas pixel$^{-1}$. 
The spatial resolution of the data correspondingly is between 70 
mas and 2$\arcsec$. It has been matched at the time of the 
observations to the stellar density that increases steeply toward
Sgr~A*. Thus, we can detect many sources in the center, and can sample 
also large areas at large radii.
We apply the standard data reduction SPRED \citep{Abuter_06, 
Schreiber_04} for SINFONI data, including detector calibrations (such as 
bad pixel
correction, flat-fielding, and distortion correction) and cube 
reconstruction. The wavelength scale is calibrated with emission line 
lamps and finetuned with atmospheric
OH lines.
Finally, we correct the data for the atmosphere by dividing with a late 
B-star spectrum and multiplying with a blackbody of the same temperature.

\section{Determination of Velocities}
\label{sec:deriv_vel}

The primary constraint for estimating the mass of the nuclear cluster 
 comes from velocity data. 
All our velocity dispersions are derived from individual velocities. Wrongly estimated errors for the velocities cause a bias in the calculated dispersions. Therefore, realistic error estimates are essential in our analysis.

\subsection{Proper Motions}
\label{sec:proper_motion}

\begin{figure}
\begin{center}
\includegraphics[width=0.99 \columnwidth,angle=-90]{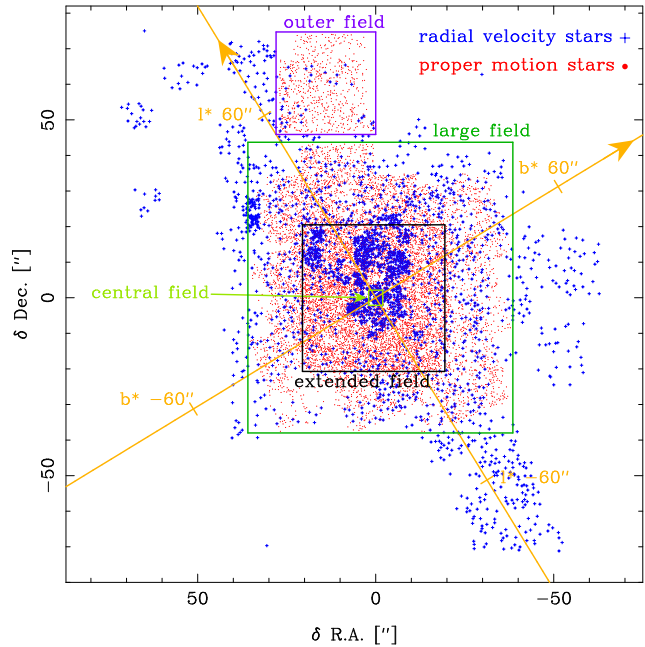} 
\caption{ 
Distribution of stars with radial velocities and proper motions. 
For the proper motions we combine four different data sets: the central field in the central 2$\arcsec$. The extended field from 2$\arcsec$ outwards 
to $\approx$20$\arcsec$, the large field from there outwards to $\approx$40$\arcsec$ and a separate, outer 
field in the north. 
The yellow lines define our 
coordinate system, shifted Galactic coordinates l$^*$/b$^*$ \citep{Deguchi_04, Reid_04}
 where the center is shifted to 
Sgr~A*. 
} 
\label{fig:_vel_map}
\end{center}
\end{figure}

We  combine in this work proper motions obtained from four different data sets: central field, 
extended field, large field and outer field, see Figure~\ref{fig:_vel_map}.  \citet{Trippe_08} used the extended field.
 In the central field the stellar crowding is high and a different analysis is needed than further out. Also, 
 the number of epochs and their similarity decrease from inside out 
which requires a more careful and separated error analysis further out.
We give here a short overview of the data and methods used.
The details are explained in Appendix~\ref{sec:ap_prom_mot}.

\begin{itemize}
\item In the central (R$\leq 2\arcsec$) we use the same method for astrometry as in \citet{Gillessen_09}, see  Appendix~\ref{sec:pm_center}. 
We now track stars out to r$_{\mathrm{box}}\approx2\arcsec$. With this increased 
field of view, we more than double the number of stars compared to \citet{Gillessen_09}.
We use 79 stars with a median magnitude of m$_\mathrm{Ks}=15.45$.
Some of the old, late-type stars have significant accelerations \citep{Gillessen_09}, but the curvature of their orbits
is not important compared to their linear motion, and thus the linear motion approximation is sufficient.
Due to the sample size the Poisson error of the dispersions dominates all other dispersion errors.
\item In the radial range between a box radius of 2$\arcsec$ and 20$\arcsec$ (extended field) we expand slightly on the data and method used by 
\citet{Trippe_08}.  
We do not change the field of view and the selection of well isolated stars compared to \citet{Trippe_08}. 
The number of stars with velocities has increased a bit due to the addition of new images.
In total we have dynamics for 5813 stars in this field. The median magnitude is  m$_\mathrm{Ks}=15.76$.
We obtain a proper motion dispersion of $\sigma_{\mathrm{1D}}=2.677 \pm 0.018$ mas/yr using all
 stars and averaging the two dimensions.  
The error includes only Poisson noise, which is likely the dominating error, because  we measure identical dispersion values and errors for the bright and faint half of 
the sample, see Appendix~\ref{sec:tri_fie}. 
\item Outside a box radius of 20$\arcsec$ (large field) nearly no proper motions were available, with the exception of a small area in \citet{Schoedel_09a}, 
which however the
authors did not use for their analysis. From the images with sufficiently good AO correction we obtain velocities for 3826 stars, see 
Appendix~\ref{sec:big_fie}. The median magnitude is  m$_\mathrm{Ks}=15.30$.
For the proper motion dispersion we obtain $\sigma_{\mathrm{1D}}=2.330 \pm 0.019$ mas/yr. 
The comparison of the dispersion for fainter and brighter stars shows that the error on the dispersion (after subtracting the velocity errors in 
quadrature) of the fainter stars is 2$\,\sigma$ larger than for the brighter stars. Since this is barely significant the errors again include only Poisson noise.  
\item For expanding our coverage of proper motions to 78$\arcsec$ we use the outer field. These data do not cover the full circle 
around Sgr~A*, but only a selected field. We have two epochs for this field: 
Gemini data from 2000 and NACO data from the 29$^{\mathrm{th}}$ May
 2011.  
We exclude in this field stars with m$_\mathrm{Ks}>15.6$, because we obtain a higher velocity dispersion for them than for brighter
 stars, see Appendix~\ref{sec:out_fie}.
In that way we select 633 proper motion stars with a median magnitude of  m$_\mathrm{Ks}=14.79$.
From these stars  we obtain  $\sigma_{\mathrm{1D}}=1.918 \pm 0.038$ mas/yr using both dimensions together.
We assume again that the Poisson error dominates over other error sources for the stars selected.
\end{itemize}

\subsection{Radial Velocities}
\label{sec:rad_vel}

Since we can resolve the stellar population in the GC, we measure velocities of single stars and obtain
 dispersions by binning stars together. 
We use two different sources for the velocities (see also Appendix~\ref{sec:app_rv}):

\begin{itemize}
\item Within R$<95\arcsec$ 
we use our SINFONI data to measure the radial velocities of the late-type giants in the GC, see Appendix~\ref{sec:app_our}
 for the details. We obtain radial velocities for 2513 stars.
The median velocity error of these velocities is about 8 km/s as 
we obtain from comparing independent measurements for multiple covered stars. 
The line-of-sight velocity dispersion of these stars is  $\sigma_z=102.2 \pm 1.4$ km/s.
The velocity errors have less than 1$\,\sigma$ influence on the measured dispersion. 
When we take into account the uneven distribution in $l^*$ (Figure~\ref{fig:_vel_map}) with binning (to avoid the influence of rotation)
we obtain as the total radial motion of the nuclear cluster 
$6.1 \pm 3.8$ km/s. By definition it should be 0.
 This indicates that also our velocity calibration is probably correct. 
In conclusion it seems likely that Poisson errors are dominating the dispersion uncertainty for our radial velocity sample, and we only include those in the final analysis.

\item
Outside of 110$\arcsec$ 
 we use radial velocities from the literature \citep{Lindqvist_92a,Deguchi_04} out to $\approx\,$3000$\arcsec$ (Appendix~\ref{sec:app_maser}).
Both of them used maser radial velocities. We match the two samples and use each star only once. 
As a side product of this matching, we confirm that the typical velocity uncertainty is less than 3 km/s as stated in \citet{Lindqvist_92a}
and \citet{Deguchi_04}.
Due to the big position errors in these radio data it is difficult to find the corresponding IR stars.
We therefore exclude from the combined list the eleven stars that overlap spatially with the areas in which we obtained spectra, with the aim of 
avoiding using stars twice. Overall we use 261 radial velocities outside the central field. 
\end{itemize}

The radial velocity stars sample is not identical with the proper motion stars sample. The majority of the faint proper motions stars have
no radial velocities and many of the outer radial velocity stars have no proper motion coverage. The radial velocity stars have a median magnitude of m$_{\mathrm{Ks}}=13.66$, while the proper motion stars are in the median 1.85 magnitudes fainter. We thus use the common (e.g. \citet{Vandeven_06,Trippe_08,Marel_10}) approach of using different stars for proper motions and radial velocities. Otherwise we would have only 1840 stars in both samples, with an inhomogeneous spatial distribution. The difference in magnitude is likely not a problem, because the stars in both samples are giants. Their different luminosities are mainly caused by different evolutionary stages, not by different ages and masses. Thus, mass dependent effects, like mass segregation, affect both samples in same way and we can safely use all 10368 proper motion and all 2774 radial velocity stars.

\subsection{Sample Cleaning}
\label{sec:out_rej}

In order to probe the gravitational potential of the GC out to more than 100$\arcsec$ we 
need to use a stellar population in dynamical equilibrium extending over a sufficiently large radial range. 
The late-type population in the GC is suited.
A small fraction of the stars in our sample does not belong to this population and is therefore excluded  as far as possible from our analysis. We have the following three exclusion criteria (see Appendix~\ref{sec:app_clean} for the details):
\begin{itemize}
\item

The young stars follow different radial profiles \citep{Bartko_10} and have different dynamics. 
We thus exclude the young stars from our sample. These are the early-type stars, the WR-, O- and B-stars \citep{Paumard_06,Bartko_09,Bartko_10}
and the red supergiant IRS7,
 that has the same age as the WR/O-stars of around 6 Myrs  \citep{Blum_02,Pfuhl_11}.
These stars are the most important contamination, especially close to Sgr A*. Due to
missing spectra we cannot clean our proper motion sample
completely from these stars. However, we choose the selection criteria such (Appendix~\ref{sec:app_early})
that in all radial ranges not more than about 4\% of the stars are young. 
\item
We also exclude stars, which due to their low extinction 
belong to the Galactic disk or bulge, similar to what is done in \citet{Buchholz_09}, see Appendix~\ref{sec:app_foreg}.
Because in some areas images in a second filter are missing we cannot clean our sample
totally from foreground stars.
Since these stars are not clustered this pollution is nowhere important. 
Integrated we include maybe about 1\% foreground stars in the GC sample.
\item
Extreme outliers in velocity are visible in some subsamples (Appendix~\ref{sec:app_fast}). About  1\% of the proper motion stars outside of the
 extended field
are outliers probably caused by measurement flukes. 
In the maser sample about 5\% are outliers (\citet{Lindqvist_92a,Deguchi_04}
 and Section~\ref{sec:fast_stars}). (See Appendix~\ref{sec:app_fast} for our procedure of outlier identification.) Their high velocities probably indicate that they belong to another population. 
The other samples are free from obvious outliers.
\end{itemize}

\section{Luminosity Properties of the Nuclear Cluster}
\label{sec:lum_prop}

For obtaining masses from Jeans modeling it is necessary to know the space density distribution of the tracer population
\citep{Binney_08}.
In principle the dynamics alone constrain the tracer distribution \citep{Trippe_08}. However, that constraint is so weak that strong priors are necessary.
Better constraints are possibly by using the surface density  \citep{Binney_08,Genzel_96,Schoedel_09a,Das_11}.
 Given the inhomogeneous and spatially incomplete nature of our dynamics sample, it would be very 
cumbersome (if not impossible) to derive the spatial distribution of tracers from that data set. 

It is easier to derive the tracer surface density of our tracers from other more complete data set.
There are two possibilities:
 \begin{itemize}
\item One can use the luminosity profile of the cluster. The usual assumption of a constant mass to light ratio will fail, however, because of the 
young stars in the center that dominate the luminosity. Hence, spectral information is needed in addition.
\item Another method for the GC cluster is to extract the surface (stellar) density profile. 
It is also necessary to correct for the early-type stars that contribute an important fraction of all stars in the center.
\end{itemize}

With these data we can obtain the density profile also in radial ranges where we have only very few velocities. Due to projection effects these radii are also important.  
In Section~\ref{sec:number_density} we obtain density maps, the radial profile of the nuclear cluster and also the profile in direction of and 
perpendicular to the Galactic plane. 
In the Jeans modeling (Section~\ref{sec:jeans_modelling}) we 
will actually fit our radial density and dynamics data at once, since also the dynamics yield a weak constraint on the 
tracer profile. However, in order to make this a problem with few parameters, we identify beforehand (Section~\ref{sec:fitting_profile}) a functional form that gives an
satisfying description of the radial distribution of the velocity tracers. 
In Section~\ref{sec:flattening} we use the density maps to separate nuclear cluster and nuclear disk.
Finally, we obtain in Section~\ref{sec:total_luminosity} the total luminosity of the nuclear cluster.

\subsection{Deriving Density Profiles}
\label{sec:number_density}

For deriving the light properties  we use three different data sources (Section~\ref{sec:dataset}).
 In case of the star density we use for all areas the dataset with the highest resolution. In case of the flux density
 we omit the WFC3/IR data because it is obtained in different filters.

We use the following steps to derive the stellar distribution:
\begin{itemize}
\item We exclude very bright foreground stars and GC clusters like the Arches. 
\item We correct the star counts for completeness if necessary, see Appendix~\ref{sec:obt_lum}.
\item We exclude stars younger than 10 Myrs from our sample.
\item We correct for extinction using two NIR filters. The resulting map is still patchy in some areas because of the optical depth being too high for
 correction.
\item We create masks to exclude the emission from these areas. 
The masks are defined such that the maps appear smooth and symmetric
in $|l^*|$ and $|b^*|$. 
Since only few pixels are masked out, the overall bias is small.

\item The areas excluded are masked out for two-dimensional fitting. They are replaced with the average of the other areas at the same $|l^*|$ and $|b^*|$ for creation of radial profiles and for visualization.
\end{itemize}
Not all steps are necessary for all data subsets, and the procedures are described in detail in Appendix~\ref{sec:obt_lum}.

In Figure~\ref{fig:_surf_bri2} we present the profiles obtained.
The profiles obtained from the integrated flux and from stellar number counts are 
similar but not identical within the errors.
On the one hand, in the case of the flux profile the assumption of screen extinction is a simplification.
On the other hand due to the high source density in the GC, magnitudes of point sources are less reliable than the extended flux in the GC.
To be conservative we use 
both profiles to fit the mass in our Jeans modeling (Section~\ref{sec:jeans_modelling}) and 
we include the scatter between the obtained masses
in the mass error budget.
Splitting the number 
counts based profile into a $|l^*|$ and $|b^*|$ component (Appendix~\ref{app:meas_flat} and Figure~\ref{fig:_surf_bri5b}) yields profiles similar to the model of \citet{Launhardt_02}.

\begin{figure}
\begin{center}
\includegraphics[width=0.70 \columnwidth,angle=-90]{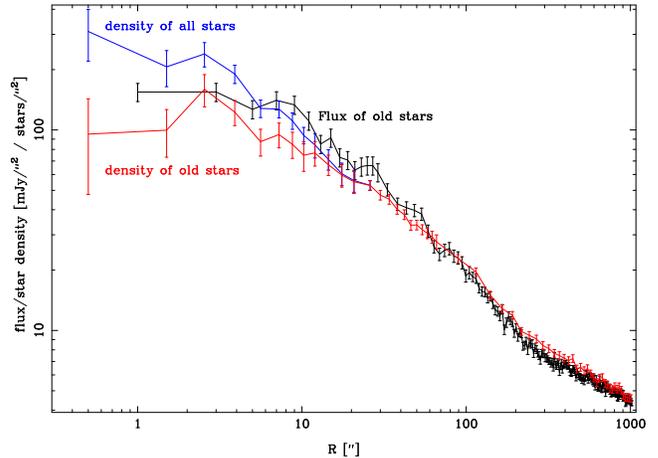}
\caption{
Radial distribution of stars. We construct the radial stellar/flux density profile from 
NACO, WFC3/IR and VISTA images (in order of increasing field of view).
} 
\label{fig:_surf_bri2}
\end{center}
\end{figure}
 
\begin{figure}
\begin{center}
\includegraphics[width=0.70 \columnwidth,angle=-90]{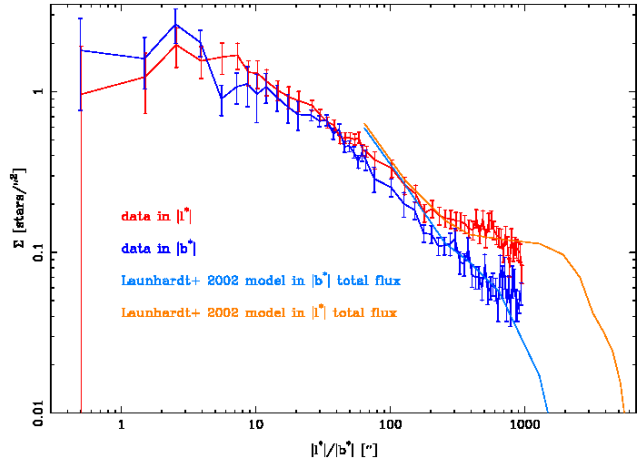}
\caption{  Stellar density in and orthogonal to the Galactic plane. 
Due to the limited number counts, the density, especially close to the center (R$<68\arcsec$), is not only measured exactly in that planes, see Appendix~\ref{app:meas_flat}.
From \citet{Launhardt_02} we show the model presented in their Figure~12, scaled to our data.
Their model does not include Galactic Disk and bulge, which are included in our data.  
} 
\label{fig:_surf_bri5b}
\end{center}
\end{figure}

\subsection{Spherical Fitting of the Stellar Density Profile}
\label{sec:fitting_profile}

We now fit the density profile assuming spherical symmetry. 
We relax this assumption in Section~\ref{sec:flattening}. 
 For the Jeans 
modeling we need a space density profile. Nevertheless, here we firstly parametrize the projected density in order to compare our data set with the 
literature before we fit the space density. Often we use power laws for first comparisons. The power law of the projected density is defined
in the following way: $\Sigma(R)=R^{-\Delta} $, and the power law of the space density as: $\rho(r)=r^{-\delta} $. $\Delta$ and $\delta$ are our definitions for power law
slopes. We use $\delta_\mathrm{L}$ for flux and star 
counts and $\delta_\mathrm{M}$ for mass.

\subsubsection{Nuker models}
\label{sec:fitting_profile_nuk}

Usually projected surface densities in the GC were fitted with (broken) power laws \citep{Genzel_03b}. A generalization to two slopes is the Nuker 
profile \citep{Lauer_95}:

\begin{equation}
\begin {split}
\Sigma(R)=\Sigma(R_b) 2^{(\beta-\Gamma)/\alpha} 
\left(\frac{R}{R_b}\right)^{-\Gamma} \left[1+\left(\frac{R}{R_b}\right)^{\alpha}\right]^{(\Gamma-\beta)/\alpha}
\end{split}
\label{eq:nuker_prof}
\end{equation}

Therein, $R_b$ and $\Sigma_b$ are the break radius and the density at the break radius. The exponent  $\Gamma$ is the inner (usually flatter) power law 
slope,  
and $\beta$ is the outer (usually steeper) power law slope. The parameter $\alpha$ is the sharpness of the transition; a large value of $\alpha$ 
yields a very sharp
transition, essentially a broken power law. Using $\alpha=100$ (fixed) our fits can be compared with the literature.

Due to the break at 220$\arcsec$ (Figure~\ref{fig:_surf_bri2}), we restrict the Nuker fits to  $r<220\arcsec$ (Table~\ref{tab:_surf_fit1}).
Rows 1 and 2 in Table~\ref{tab:_surf_fit1} 
 give our fits for stellar number counts and the flux profile, respectively. Row 1 can be directly 
 compared to the 
literature. \citet{Buchholz_09} conducted the largest area (r$<20\arcsec$) study so far. They obtained
$\Gamma=-0.17\pm0.09 $ and $\beta=0.70\pm0.09$ for $R_b=6.0\arcsec$. This fit is broadly consistent with our data, although we obtain  
for this radial range no clear sign for a power law break. The break radius (6$\arcsec$) of \citet{Buchholz_09} is smaller than ours, although these authors do not 
cite an error. 
The binned data of \citet{Buchholz_09} look similar to our data. The same is true for the data in \citet{Bartko_10} who do not attempt to fit the
 profile. 
These works find a very weak increase of the density
with radius inside of $\approx5\arcsec$ and then a somewhat stronger decrease of the density with radius further out. This is also the 
case in \citet{Do_09,Do_13}. 
\citet{Do_13} used data out to 14$\arcsec$ and find a single power law with a slope of $\Delta_\mathrm{L}=0.16\pm0.07$.

Our data seem to be consistent with a break radius for the late-type stars around 20$\arcsec$. When counting stars regardless of their age, a smaller break radius of about 8$\arcsec$ is found \citep{Genzel_03b,Schoedel_07}. 
For our large radial coverage, a single power law obviously fails to fit our data, even when we restrict our data to the 
range for which we have spectral classifications ($R<90\arcsec$). The best fit broken power law has then $\chi^2$/d.o.f$=11.43/20$ 
(Row~3 in Table~\ref{tab:_surf_fit1}), while a single power law yields $\chi^2$/d.o.f$=53.12/22$.

\begin{figure}
\begin{center}
\includegraphics[width=0.70 \columnwidth,angle=-90]{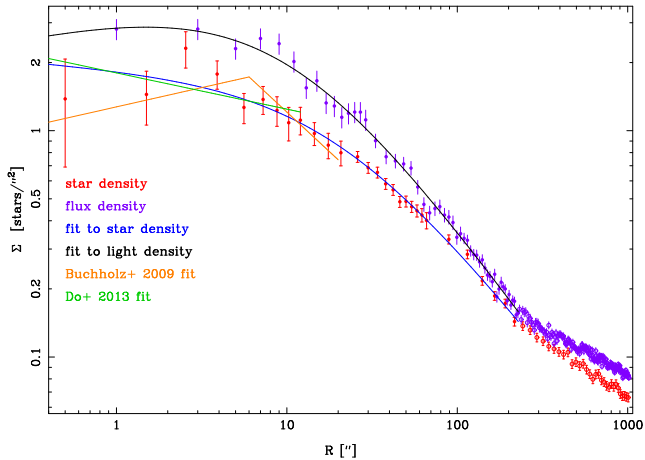} 
\caption{Nuker fits (with $\alpha=1$) to the late-type stars surface density, either their star 
density or their flux density.
We present here fits to the inner data (r$<220\arcsec$, filled dots). The outer data outside the break is presented by 
open dots.
We also plot the fits of \citet{Buchholz_09} and \citet{Do_13}.
} 
\label{fig:_surf_fit1}
\end{center}
\end{figure}

This indicates that the 
transition between the two slopes is softer than for a broken power law. 
We also test whether a very small $\alpha$ (very soft transition) can be excluded. However, there are nearly no differences in the $\chi^2$ for small $\alpha$; all $\alpha$ have nearly the same probability. Therefore, we give only upper limits for $\alpha$: they are $1.15$ and $0.65$ for the flux and star count data, respectively.
Figure~\ref{fig:_surf_fit1} shows our data together with some fits.

The older literature used single power law profiles to describe their flux density profiles.  Between 20$\arcsec$ and 220$\arcsec$ our data can be fit relatively well by a 
single power law of $\Delta_\mathrm{L}=0.765 \pm 0.018$ (stellar density) and of $\Delta_\mathrm{L}= 0.915 \pm 0.015$ (flux density). These slopes show again that the two data sets are not
 consistent.
\citet{Becklin_68,Allen_83},
 \citet{Haller_96} obtain a slope of $\Delta_\mathrm{L}=0.8$, while \citet{Philipp_99} obtain a flatter slope of
$\Delta_\mathrm{L}=0.6$.

In contrast to these data the flux profiles from \citet{Graham_09}\footnote{ \citet{Graham_09} used the profile from \citet{Schoedel_08a} constructed from public 
2MASS images.} and \citet{Schoedel_11a} 
do not follow a single power law, they flatten at about 80$\arcsec$ 
indicating that the 
 bulge dominates already there. Likely this bright bulge is an artifact caused by a missing sky subtraction.
The floor level in the profiles of \citet{Graham_09,Schoedel_11a}, much higher than the bulge floor in the COBE/DIRBE data of \citet{Launhardt_02}, fits
to typical K-backgrounds observation on earth.
Like  \citet{Vollmer_03} for 2MASS data  we subtract the light level towards dark clouds as sky. 
 None of \citet{Becklin_68,Allen_83,Haller_96,Philipp_99,Graham_09,Schoedel_11a} use two color information to correct for extinction in contrast to our work.
The single power law fit of \citet{Catchpole_90} to extinction corrected
star counts with a slope of $\Delta_\mathrm{L}=1.1$ in the radial range from 140$\arcsec$ to 5700$\arcsec$
 is not well comparable to our data set due to the different radial range. 

\begin{deluxetable*}{lllllllll} 
\tabletypesize{\scriptsize}
\tablecolumns{9}
\tablewidth{0pc}
\tablecaption{Nuker fits of the surface density profile \label{tab:_surf_fit1}}
\tablehead{ No. &data source & radial range & $\alpha$ & $\chi^2$/d.o.f & R$_b$ [$\arcsec$] &  $\Sigma$(R$_b$) & $\beta$ &  $\Gamma$ }
\startdata
1&star density & R$<220\arcsec$ & 100  	    &26.66/25	& 23 $\pm$ 3  & 0.86 $\pm$ 0.09	 &  0.771 $\pm$ 0.018   & 0.277 $\pm$ 0.054  \\
2&flux density & R$<220\arcsec$ & 100  	    &74.11/50	& 24 $\pm$ 2  & 71.2 $\pm$ 4.6   &  0.934 $\pm$ 0.016	 & 0.304 $\pm$ 0.030  \\
3&star density & R$<90\arcsec$  & 100  	    &11.43/20	& 13 $\pm$ 3  & 1.11 $\pm$ 0.15	 &  0.645 $\pm$ 0.040  & 0.186 $\pm$ 0.084  \\
4&star density & R$<220\arcsec$ & 1		    &15.28/25	& 21 $\pm$ 14 & 0.84 $\pm$ 0.28	 &  0.972 $\pm$ 0.089	 & 0.059 $\pm$ 0.146  \\
5&flux density & R$<220\arcsec$ & 1		    &50.71/50	& 10 $\pm$ 3   & 110 $\pm$ 20 &  1.048 $\pm$ 0.040	 & -0.163 $\pm$ 0.148 \\
\enddata

\end{deluxetable*}

\subsubsection{$\gamma$-models}
\label{sec:fitting_profile_gam}

For our Jeans-modeling (Section~\ref{sec:jeans_modelling})  we need a parametrization of the space density profile $\rho_{\mathrm{N}}(r)$. This is connected to the observable surface density profile $\Sigma(R)$
by the following projection integral:
\begin{equation}
\begin {split}
\Sigma(R)=2 \int_{R}^{\infty} \rho_{\mathrm{N}}(r) r dr/\sqrt{r^2-R^2} 
\end{split}
\label{eq:abel_1}
\end{equation}

We  use the spherical $\gamma$-model \citep{Dehnen_93} (This is equivalent to the $\eta$-model of \citet{Tremaine_94} under the transformation $\gamma=3-\eta$.):

 \begin{equation}
\begin {split}
\rho_{\mathrm{N}}(r)= \frac{3-\gamma}{4\,\pi}\frac{L}{r^{\gamma}}\frac{a}{(r+a)^{4-\gamma}}
\end{split}
\label{eq:eta}
\end{equation}

Therein $L$ is the total flux, respectively the star counts. 
$a$ is the scale of the core of the model. The density slope is  $-\gamma$  within the core and -4 at $\infty$.
 The $\gamma$-model has a known positive distribution function which we employ in 
\citet{Chatzopoulos_14}.
More complex profiles, like for example a three-dimensional Nuker model, can also fit the data. The projected profiles are nearly independent of the parametrization in that case, e.g. for the star counts
the $\Delta\chi^2$ between the $\gamma$-model fit and the Nuker fit is only 1.3. However, the Nuker model has more degrees of freedom. Further, complex profiles contain poorly constrained parameters (e.g. $\alpha$ in Nuker) for
our data set; they overfit the data.

The GC light profile has two breaks (Figure~\ref{fig:_surf_fit1}).
The breaks in the profiles around 200$\arcsec$ are probably a sign of a two component nature of the nuclear
light distribution, as suggested by \citet{Launhardt_02}. 
They call the inner component the nuclear (stellar) cluster and the outer one the nuclear (stellar) disk,
in analogy to other galaxies.
In contrast, \citet{Serabyn_96} assumed that the central
 active star forming zone inside the inactive bulge of the Milky Way consists of a single component,
a central stellar cluster of R$=100\,$pc.

Due to the breaks we cannot fit the full data range with one $\gamma$-model. We use instead two independent 
$\gamma$-models. The use of two models is the main reason that we cannot use the Nuker model since many parameters are then ill determined. 
Still the central slope of the outer component is difficult to determine also for $\gamma$ models
from the data and we fix it to be flat by setting $\gamma_\mathrm{outer}=0$. A smaller value would correspond to a central depression, and values $>0.5$ 
can create profiles in which the outer component dominates again at very small distances. 
 We obtain the fits presented in Row 1 and 2 in Table~\ref{tab:_surf_fit2}.

\begin{deluxetable*}{lllllllll} 
\tabletypesize{\scriptsize}
\tablecolumns{9}
\tablewidth{0pc}
\tablecaption{$\gamma$-model fits of the surface density profile \label{tab:_surf_fit2}}
\tablehead{ No. &data source & radial range &  $\chi^2/d.o.f.$ &$L_\mathrm{inner}$ & $a_\mathrm{inner}$ & $\gamma_\mathrm{inner}$ &  $K_\mathrm{outer}$ & $a_\mathrm{outer}$ }
\startdata
1&star density & all R &29.1/55 & $6.73\pm0.88 \cdot10^4$ stars &$194\pm 33$'' & $0.90\pm0.11$ &$7.05\pm1.49 \cdot10^6$ stars & $3396\pm 458$'' \\
2&flux density & all R &220.3/204 & $3.42\pm0.14 \cdot10^6$ mJy &  $117\pm 10$'' & $0.76\pm0.08$ & $5.48\pm0.38 \cdot10^8$ mJy &$3711\pm 158$''\\
3&star density & R$<220\arcsec$ & 18.8/26 & $28.8\pm3.4 \cdot10^4$ stars &$626\pm 93$'' & $1.14\pm0.06$ &- & - \\ 
4&flux density & R$<220\arcsec$ &96.87/51 & $14.52\pm0.83 \times10^6$ mJy &  $511\pm 44$'' & $1.21\pm0.04$ &- & -\\
\enddata
\end{deluxetable*}

\begin{figure}
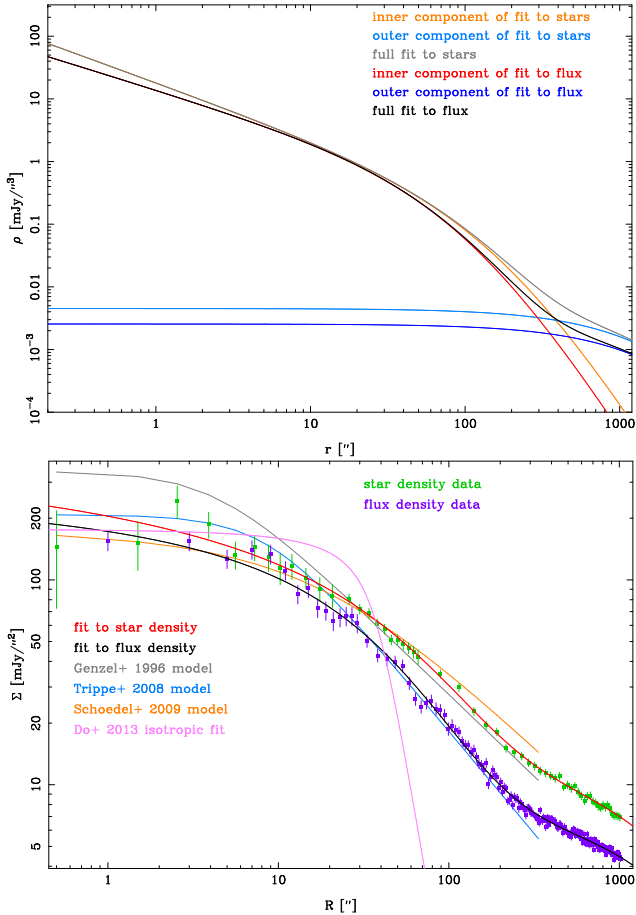

\begin{center}
\includegraphics[width=0.70 \columnwidth,angle=-90]{f5a.eps} 
\includegraphics[width=0.70 \columnwidth,angle=-90]{f5b.eps} 
\caption{Space and projected density fits in comparison with the star and flux density data. 
For illustration purposes the stellar density data and fits are shifted by a factor 105. We fit two $\gamma$-models. We also show the models of
\citet{Genzel_96,Trippe_08,Schoedel_09a},
 and \citet{Do_13b}.
} 
\label{fig:_densities_1}
\end{center}
\end{figure}

The two best fitting profiles for stars and light are similar (Figure~\ref{fig:_densities_1}) to each other, but again not consistent within their errors
as already noticed during the Nuker profile fits.
The central slope of the cluster is consistently $\gamma_\mathrm{inner}=0.83\pm0.12$.
However, the small error is a consequence of the functional form that we use. Other functions like Nuker(r) yield a range of inner slopes that
appears to be consistent with the uncertainty reported by \citet{Do_09}.

The outer $\gamma$-model is needed inside of 220$\arcsec$. Row 3 and 4 in 
Table~\ref{tab:_surf_fit2} show fits with a single component. 
The resulting 
 $\gamma_\mathrm{inner}$ are unrealistically large ($>1$) when comparing with  \citet{Do_09}.
 To quantify the mutual consistence of the two data set, we fit the two data sets with the best single $\gamma$ fit of the other allowing only the scaling to change. 
The star density data gives $\chi^2/d.o.f.$=70.0/28; the flux density data $\chi^2/d.o.f.$=194.0/53.

The space density models in the literature do not describe our data well, even when we restrict the comparison to the inner 220$\arcsec$, 
see Figure~\ref{fig:_densities_1}. The density model of \citet{Schoedel_09a} is a bad fit to both of our data sets ($\chi^2/d.o.f.=83.37/28$ and $\chi^2/d.o.f.=494.67/53$, for star and flux density, respectively).
The reason is the combination of a large break radius with a relatively small outer slope. It thus overestimates the density further out and underestimates it in the center. 
The models of \citet{Trippe_08}  ($\chi^2/d.o.f=117.46/28$ and $\chi^2/d.o.f.=142.71/53$) and especially \citet{Genzel_96} ($\chi^2/d.o.f.=323.04/28$ and $\chi^2/d.o.f.=165.99/53$) fit especially our flux data better. 
Still their average $\chi^2/d.o.f.$ is worse than our average $\chi^2/d.o.f.$
when we fit both data sets with the same parameters
The reason is that the core radii of \citet{Genzel_96,Trippe_08} are small. 
The different profiles of \citet{Do_13b}, which are very similar to each other, do not fit our data at all, because the outer slope of the cluster is much too steep and the core too large.
These discrepancies with our fit are not surprising, since none of these works used such a large radial coverage as we. \citet{Schoedel_09a} did not fit the density model, it is a fixed input to their modeling. They were guided by recent literature. \citet{Trippe_08} only fit dynamic data.
\citet{Do_13b} fit density data together with dynamic data, but only inside 12$\arcsec$. That covers essentially only the core.
This shows that it is important to use density data over a scale larger than the core of the nuclear cluster since otherwise
the properties of the outer profile, which have influence on the core parameters, cannot be determined accurately.

\subsection{Flattening of the Cluster}
\label{sec:flattening}

Our profiles show that the axis ratio ($q=b/a$) of the nuclear cluster increases with radius, see Figure~\ref{fig:_surf_bri5b}.
We measure the flattening binwise, see Appendix~\ref{app:meas_flat}, and Table~\ref{tab:_flat1}. 
In the inner most bin,  $|l^*|<68\arcsec$, $q=0.80\pm0.04$. There is some indication in the data (Figure~\ref{fig:_surf_bri5b}), that the flattening is smaller around 40$\arcsec$ than around 20$\arcsec$. That dip is consistent with noise. We do
not find a relevant a systematic error source, see Appendix~\ref{app:flat_syst_err}.
The flattening increases further outside of our field, as it is visible in large scale IRAC data and in
 \citet{Launhardt_02}. 
They model the inner r$_{\mathrm{box}}=2^{\circ}$ of the GC and obtain an axis ratio of 0.2 at 
around $l^*=3100\arcsec$.

\begin{deluxetable}{ll} 
\tabletypesize{\scriptsize}
\tablecolumns{2}
\tablewidth{0pc}
\tablecaption{flattening profile \label{tab:_flat1}}
\tablehead{ $|l^*|$ range &  $q=b/a$ }
\startdata
 0 to 68$\arcsec$ & $0.80\pm0.04$\\
 68 to 130$\arcsec$ & $0.63\pm0.03$\\
 130 to 248$\arcsec$ & $0.58\pm0.05$\\
 248 to 473$\arcsec$ & $0.45\pm0.04$\\
 473 to 1030$\arcsec$ & $0.32\pm0.03$\\
\enddata
\end{deluxetable}

\begin{figure*}
\begin{center}
  \includegraphics[width=1.19 \columnwidth]{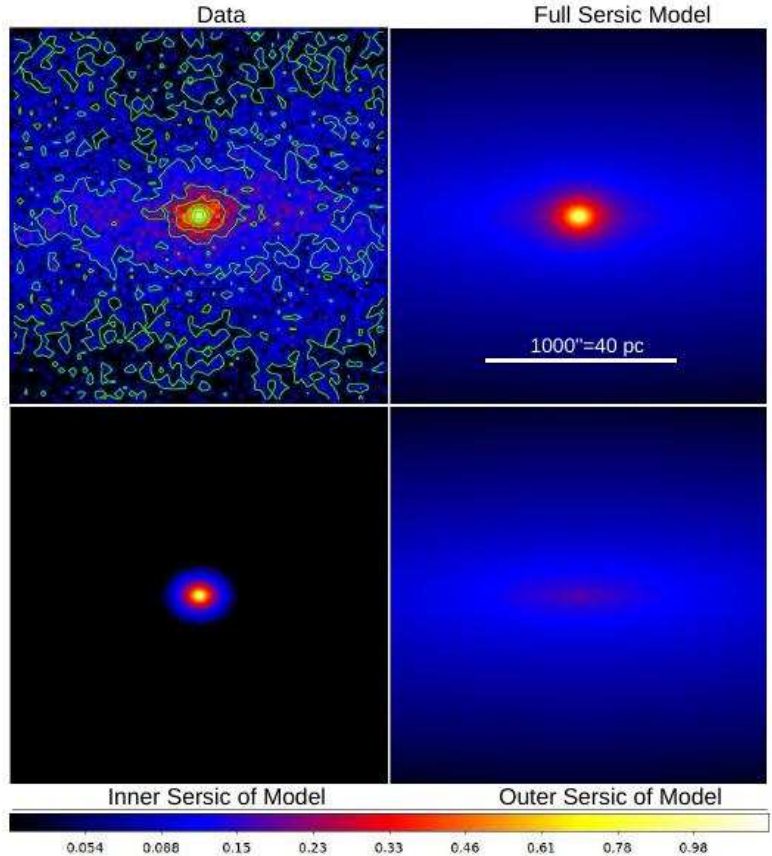} 
\caption{Map of the stellar density in the inner r$_{\mathrm{box}}=1000\arcsec$. All panels use the same
color scale. Upper left: stellar density from VISTA/WFC3/NACO star
counts, corrected for completeness and extinction. To the fitted surface brightness data we added smooth contours for illustration.
Upper right: GALFIT fit to the data. The fit consists of two components, which are
shown in the lower two panels. Lower left: the central component, a n$=1.46$ Sersic profile, with $q=0.80$ slightly flattened;
lower right: a Nuker profile ($q=0.26$), which is not well constrained since it extends well outside of our field of view.
} 
\label{fig:2dim-map}
\end{center}
\end{figure*}

We can use our two dimensional data to distinguish between nuclear cluster and nuclear disk, where the outer disk component is 
used to estimate the background for the cluster in the following manner.
In the central 68$\arcsec$ we use the 
density profiles shown in Figure~\ref{fig:_surf_bri5b} in the corresponding quadrants.
Further out, we use the VISTA data. Figure~\ref{fig:2dim-map} shows the resulting map.
For finding an empirical description we use GALFIT \citep{Peng_02}.
In the first fits we fix the centers to the known location, and enforce alignment of  the
 inner component with the Galactic Plane, but not of the outer component. (Because we measure the density only in two sectors inside of 68$\arcsec$ alignment with the Galactic Plane is enforced there by construction.) In this fit the outer component is aligned within 1.1$\pm1.3^{\circ}$. The uncertainty of that angle is obtained by four trials, in each we mask out the lower/upper half in l/b. In contrast to many other parameters that angle is robust. 
 Thus, the misalignment is not significant and we fix the angle to be $0^{\circ}$.
Our result is consistent with \citet{Schoedel_14} who obtain both for the nuclear disk, and the nuclear cluster
alignment with about 2$^{\circ}$ uncertainty.

We fit the data with Sersic \citep{Sersic_68} and Nuker profiles. Which of them is used is not important, similar sizes and flattening are  obtained with both.
However, it is difficult to disentangle the inner and the outer component. Small changes in $\chi^2/d.o.f.$ result in large differences in most parameters.
Mainly because the flattening has a local minimum around $40\arcsec$, a free fit results in a large Sersic index for the outer components. In that case the outer component contributes relevantly again in the center, which reduces the flattening of the inner component. Then the nuclear cluster has 
$q\approx0.91$. 
However, it is physically unlikely that outer component is more important in the very center than slightly further out. Since in the case of our $\gamma$-profiles (Section~\ref{sec:fitting_profile_gam}) the outer component does not contribute relevantly in the center, we assume that the inside 68$\arcsec$ measured flattening is identical with flattening of the nuclear cluster as a whole.  
To obtain this flattening we use for the nuclear disk a Nuker profile with a flat core, $\alpha=0.6$, $\beta=3$, $\Gamma=0$. This set of parameters approximately models a projected $\gamma$ model
in its outer and inner slope (Section~\ref{sec:fitting_profile_gam}). The fit obtained for the disk is q$=0.264$. 
In the center we use a Sersic model to enable comparison with the recent literature. 
Under these assumptions we obtain for the inner component $q=0.80\pm0.05$, n$=1.46\pm0.05$, R$_e=127 \pm 10\arcsec$, and an integrated count uncertainty of 10\%.   In contrast to many other parameters, the Sersic parameter is well constrained. It is in all cases between 1.4 and 1.6.
 In the very center the outer component contributes 13\% of the star counts of the inner component.
 The outer component dominates outside about 104$\arcsec$.
 
Our half light radius of 5.0 pc is slightly larger than the preferred value of \citet{Schoedel_14} (R$_e=4.2\pm0.4$). However, when the outer Sersic is also free, in their two Sersic fits they obtain $6\pm0.2$ pc. Our axis ratio is somewhat larger then their value of $q=0.71\pm0.02$ but within the uncertainties. Our Sersic index n is smaller than their of
n$=2\pm0.2$. Since they were not able to use the central parsec in their fit, it is likely that our fit is better in that region. The very center is important for the Sersic index, thus our n is probably better. 
 Overall from comparing our and their different fits it is probable that both works underestimate the systematic error in component fitting. The main reason for that fact is the existence of two not clearly separated components in the GC and the high extinction towards the region.

\subsection{Luminosity of the Nuclear Cluster}
\label{sec:total_luminosity}

To obtain the Ks-luminosity in the GC 
we integrate the flux of the old stars (Figure~\ref{fig:_surf_bri2}).
The total extinction-corrected flux within R$<100\arcsec$ is  $1052\pm200\,$Jy. The absolute error of 20\% contains the uncertainty of the extinction law toward the GC (\citealt{Fritz_11}: 11\%), the calibration uncertainty (7\%), and 14\% for the differences between
 the stellar density and flux density profiles. The latter we obtain from the scatter between the star counts profile and the flux profile scaled to each other. 
 
To estimate the total luminosity of the nuclear cluster, we need to estimate its size. We use again different methods, to estimate
the systematic error. Firstly, we use  two-dimensional decomposition
of the star counts in nuclear cluster and nuclear disk (Section~\ref{sec:flattening}).\footnote{We do not consider other Galactic components, they are very minor in the center. In the model of \citet{Launhardt_02} (their Figure~2) these contribute 0.35 mJy/''$^2$ before extinction correction, while we have 2 mJy/''$^2$ in total at R$=100\arcsec$. Extinction corrected, their contribution is even smaller.}
The fraction of star counts from the inner component is 67\% and the fraction the nuclear cluster which is within 100$\arcsec$ is 44\%.
That leads to a total luminosity of 1599 Jy. Using instead the one dimensional $\gamma$ decomposition of Section~\ref{sec:fitting_profile_gam} of the counts leads
to 5058 Jy. The $\gamma$-model fit of the flux implies 3420 Jy, see Table~\ref{tab:_surf_fit2}. 
The main reason for the different fluxes are the different models: a Sersic can, when it is as here close to exponential, decay fast outside its characteristic radius, while a $\gamma$-model decays only with its fixed power law of -3.

We obtain absolute luminosities using  R$_0=8.2$ kpc and  M$_\odot(Ks)=3.28$.
Thereby, we use the 3420 Jy as best estimate for the total flux and the other two values for the error range.
We obtain 
$7.4^{+3.5}_{-3.9} \times 10^7\,$L$_{\odot}$  
for the total nuclear cluster, consistent with the estimate of $ 6\pm 3 \times 10^7\,$L$_{\odot}$ of
\citet{Launhardt_02} and the $4.1\pm0.4 \times 10^7\,$L$_{\odot}$  at 4.5~$\micron$ of  \citet{Schoedel_14}.

The young O(B)-stars in the center, which are not included in our sample, add 25 Jy in the Ks-band. Although they are
irrelevant for the light in the Ks-band, this is different for bolometric measurement:
the bolometric luminosity of the young stars is  about L$_{UV}\approx10^{7.5}\,L_{\odot}$ and $M_{\mathrm{bol}}\approx-14.1$ \citep{Genzel_10,Mezger_96} 
and thus larger than what we obtain for the old stars, $M_{\mathrm{bol}}\approx-13.4$ within R$<100\arcsec$.
Also, the young stars are concentrated on a more than $\approx1000$ times 
 smaller volume than the old stars.

\section{Kinematic Analysis}
\label{sec:an_method}

We now use our kinematic data to characterize the properties of the data and to obtain a rough mass estimate from it.
In Section~\ref{sec:_anisotropy},~\ref{sec:fast_stars} and \ref{sec:rotation} we discuss anisotropy, 
fast stars, and rotation, respectively. In  Section~\ref{sec:binning} we explain and justify the binning
used in our Jeans modeling and in \citet{Chatzopoulos_14}.
In this work we use isotropic spherical symmetric Jeans modeling \citep{Binney_08} 
as illustrative models of what can be derived from our data (Section~\ref{sec:jeans_modelling}). 
With that relatively simple model we can also explore
many systematic error sources easily.
In \citet{Chatzopoulos_14} we use two-integral modeling with self-consistent rotation \citep{Hunter_93},
which allows us to include intrinsic flattening and rotation.

\subsection{Velocity Anisotropy}
\label{sec:_anisotropy}

The cluster could be anisotropic. One type of anisotropy is radial anisotropy, 
which would manifest itself as a difference between the dispersions in tangential and radial direction
 \citep{Leonard_89,Schoedel_09a}. 

We obtain an estimate of the anisotropy from the proper motions from $\beta'_{\mathrm{pm}}(R)=1-[\sigma_\mathrm{tan\,PM}(R)/\sigma_\mathrm{rad\,PM}(R)]^2$, where $\sigma_\mathrm{tan}$ and $\sigma_\mathrm{rad}$ are the radial and tangential dispersions of the proper motions, respectively.
$\beta'_{\mathrm{pm}}$ has the advantage that it follows directly from measured properties 
without modeling
and does not depend on $R_0$.
Uneven angular sampling of stars together with  the flattening  which causes $\sigma_{l^*}>\sigma_{b^*}$ (Figure~\ref{fig:_std_v}) can mimic anisotropy in the following way:
consider a radial bin that is only covered close to the $b^*$-axis. Then, $\sigma_{\mathrm{tan}}\approx 
\sigma_{l^*}$ and $\sigma_{\mathrm{rad}}\approx \sigma_{b^*}$. Since $\sigma_{l^*}>\sigma_{b^*}$, $\beta'_{\mathrm{pm}}<0$ implies tangential anisotropy.
The arguments also hold for uneven angular sampling.
\citet{Do_13b} had outside the center only covered close to the $b^*$ axis
the Galactic Plane. As expected $\sigma_{\mathrm{tan}}>\sigma_{rad}$ in these data. That is the reason that their fit 
prefers $\beta_\infty<0$.

\begin{figure}
\begin{center}
\includegraphics[width=0.70 \columnwidth,angle=-90]{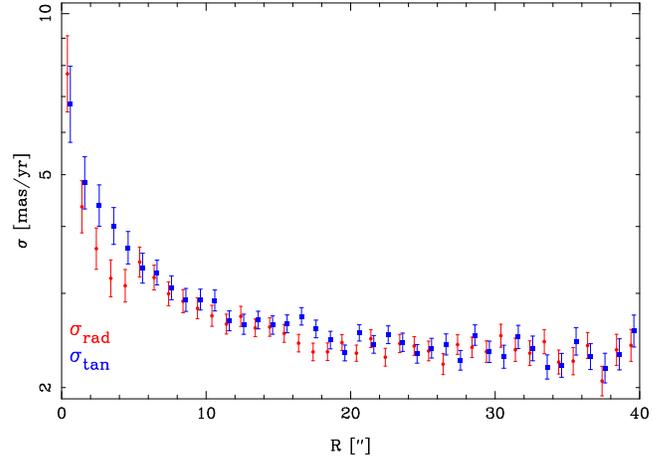} 
\caption{
Radial and tangential dispersions as function of the radius. The points are slightly offset in R from each other for better visibility.
} 
\label{fig:_disp_aniso}
\end{center}
\end{figure}

To avoid 
a spurious influence of the flattening on the anisotropy we firstly restrict the analysis here to 
r$<$40$\arcsec$ for having full (not necessarily even) angular coverage.
Secondly, to even out density fluctuations, we obtain the dispersions by taking the average of the dispersions in two angular bins: one within $\phi<45^{\circ}$ to the Galactic Plane and the other with $\phi> 45^{\circ}$. To even out coverage fluctuations, we give the two bins equal weight.
 We obtain  $\beta'_{\mathrm{pm}}=-0.040 \pm 0.022$ (Figure~\ref{fig:_disp_aniso}). 
The scatter between the bins is consistent with the Poisson errors. Similar to what was found by \citet{Schoedel_09a} we see that 
 $\sigma_{\mathrm{tan}}$ is somewhat larger in the center. 
However, since $\chi^2/d.o.f.=33.71/39$
 shows that this is not significant. 
\citet{Schoedel_09a} suggested that the increase in the center could be due to pollution with early-type stars, 
which are in average on more tangential orbits \citep{Genzel_00,Bartko_09,Bartko_10}. We test this hypothesis by using only stars with late-type 
spectra. Integrated over the full field this yields $\beta'_{\mathrm{pm}}=0.045 \pm 0.049$. 
That is smaller than for all stars, but not significantly different. 

Between 2'' and 5''  $\beta'_{\mathrm{pm}}=-0.463 \pm 0.170$
when using all stars and  $\beta'_{\mathrm{pm}}=-0.134 \pm 0.207$ when using only late-type stars. Since these two values
are again consistent, pollution is probably not important in this radial range. 

The anisotropy parameter $\beta$ is defined in 3D coordinates (r) \citep{Binney_08}. $\beta$ can be only estimated in full modeling which also needs to account for projection effects. 
In such model, $R_0$ would also need to be fit and it would also be necessary \citep{Chatzopoulos_14} to use a flattening model to account for the different effect of the flattening on v$_z$, v$_{l^*}$, and v$_{b^*}$. That is beyond the scope of this paper. 
The deprojected anisotropy parameter $\beta$ is more different from 0 than $\beta'_{\mathrm{pm}}$ \citep{Marel_10}. Due to the core-like density profile the difference between  $\beta'$ and $\beta_{\mathrm{pm}}$ can be large especially in the center. Because of projection effects, full modeling is required to constrain the
radial dependency of the anisotropy \citep{Marel_10}.
While the overall radial anisotropy is small, other deviations from prefect isotropy exist; see Section~\ref{sec:rotation}.

\subsection{Rotation and dynamic main axis}
\label{sec:rotation}

In Section~\ref{sec:flattening} we determined the major and minor axes of the
distribution of stars in the nuclear cluster. Here we use the new kinematic
data to find the rotation axis of the system. First, we use the
proper motion data.

\citet{Trippe_08} interpreted the difference between the velocity
dispersions $\sigma_l$ and $\sigma_b$ (Figure~\ref{fig:_std_v}) as a sign of rotation.
However, $\sigma_l>\sigma_b$ is globally required for any axisymmetric star
cluster flattened parallel to the Galactic plane. This is independent
of whether the extra kinetic energy along the Galactic plane is due to
net rotation or to higher in-plane velocity dispersions; reversing Lz
for any orbit does not change $\sigma_l$.  Furthermore, spherical clusters
with rotation \citep{Lynden-Bell_60} do not show $\sigma_l>\sigma_b$.  Therefore
the difference between $\sigma_l$ and $\sigma_b$ is ultimately due to the
flattening, even though most of the flattening of the nuclear cluster is
in fact generated by additional rotational kinetic energy
\citep{Chatzopoulos_14}.

However, in any case we can follow the approach of \citet{Trippe_08}
to find the kinematic major axis of the nuclear cluster. To this end, we bin the motions in angle (their Figure~7). The pattern is sinusoidal but with more variation close to the peaks. Thus, we use for fitting following function: \begin{equation} 
f(\theta)=a+b\times(|\theta-\phi|)^c\end{equation} 
Therein, theta is the angle relative to the line to the east;
$\phi$ the position of the maximum; $a$ is the constant floor, b the peak parameter; a big $b$ implies that the
maximum has small width than the minimum.
We obtain $b=3.97\pm0.49$ and $\phi=58.8\pm1.2^{\circ}$
consistent with the Galactic plane ($\phi=58.6^{\circ}$).

Lastly, we use the radial velocities, whose
 gradient in the mean radial velocity as a function of $|l^*|$ \citep{Trippe_08} can only be explained 
 by rotation. For fitting the Galactic plane we aim to find the angle for which the radial velocity is constant along the coordinate $x''$, which we obtain by the rotation. We assume cylindrical rotation, since we ignore $y''$. The angle is the rotation axis, rotated by 90$^\circ$. 
  The advantage of that angle compared to the rotation axis is that it is not necessary to fit at the same time for the possible complex rotation curve, because vertical to it the velocity is identical everywhere. We obtain $53.7\pm4.0^{\circ}$ broadly consistent with the Galactic plane.
 
\begin{figure}
\begin{center}
\includegraphics[width=0.70 \columnwidth,angle=-90]{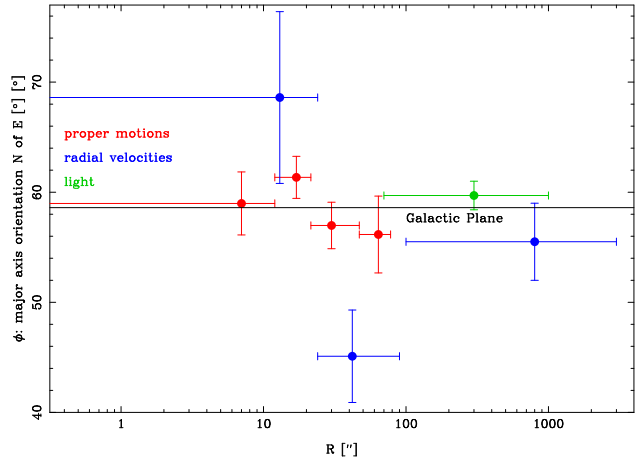} 
\caption{Orientation of the major (flattening/rotation) axis.
} 
\label{fig:_rot_axis}
\end{center}
\end{figure}

Due to the finding of \citet{Feldmeier_14} that radial velocity field is not only a function of l$^*$ and b$^*$ as expected, we divided our velocities in radial bins (Figure~\ref{fig:_rot_axis}). 
We can confirm their findings mostly. At medium radii (between 24 and 90$\arcsec$) the radial velocity plane follows an angle of $45.1\pm4.2^{\circ}$. The value is 3.2~$\sigma$ different from the Galactic Plane. That agrees well with the measurement of \citet{Feldmeier_14}. Further out and further in the rotation axis aligns with the Galactic plane within the partly large error. 
We do not find deviations of $\sigma_z$ in these bins.

\begin{figure}
\begin{center}
\includegraphics[width=0.70 \columnwidth,angle=-90]{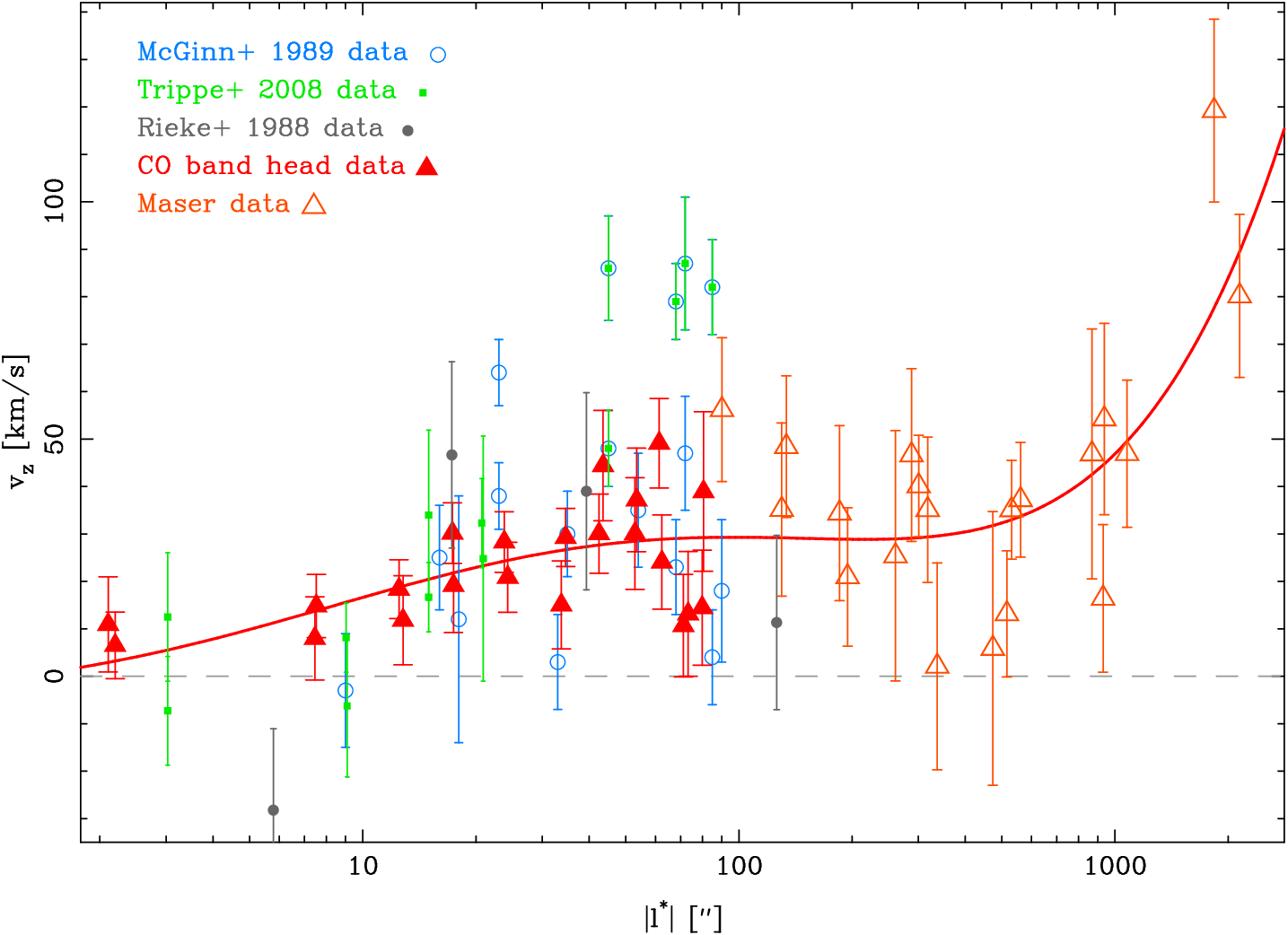}
\caption{Average radial velocities from our data and the literature.
 We assume symmetry of the rotation pattern and reverse the sign of the radial velocities
For the maser data we use the velocities from \citet{Lindqvist_92a} and \citet{Deguchi_04}. 
 The data from \citet{Trippe_08} overlaps within $|l^*|<27\arcsec$ largely with our data. Further out \citet{Trippe_08} utilized 
a subsets of the dataset of \citet{McGinn_89}. We also plot the data of \citet{Rieke_88}.
We fit the good (reddish) data without binning by a polynomial for illustration.
} 
\label{fig:_v_rad_fit1}
\end{center}
\end{figure}

Although it is likely that the rotation is not a function of l$^*$ only, we assume it here to ease comparison with 
most of the literature.
For illustration we bin the radial velocity data (Section~\ref{sec:rad_vel}) in a less crowded way (Figure~\ref{fig:_v_rad_fit1}).

Inside of  27$\arcsec$ our velocities are consistent with the velocities of \citet{Trippe_08}, since the data set is largely identical.
Outside of 27$\arcsec$ our new SINFONI radial velocities are on average smaller than 
the velocities of \citet{Trippe_08}, who used only a subset of the velocities in \citet{McGinn_89}, a problem already pointed out
by \citet{Schoedel_09a}. Surprisingly, our new high resolution data do not yield the average of the velocities reported by \citet{McGinn_89},
but agree roughly with the lower end of values found. These lower end of values agrees with the 
 velocities of single bright stars by \citet{Rieke_88}.

 The maser data of \citet{Lindqvist_92a} and \citet{Deguchi_04} agree with the lower velocity data of \citet{McGinn_89} and our CO band head
velocities. \citet{Schoedel_09a} suggested that the differences in the radial velocities in the literature could be a sign of two populations in the GC.
However, in our data we find no sign for any population dependence of the rotation pattern.
Possibly, the difference in radial velocities in \citet{McGinn_89} and \citet{Rieke_88} is an indication that the velocity calibration
of these old CO band head measurements was more difficult than assumed back then. 
Overall we are confident that our smaller rotation of the cluster compared to \citet{Trippe_08} is correct and is not population dependent.

\subsection{Binning}
\label{sec:binning}

\begin{figure}
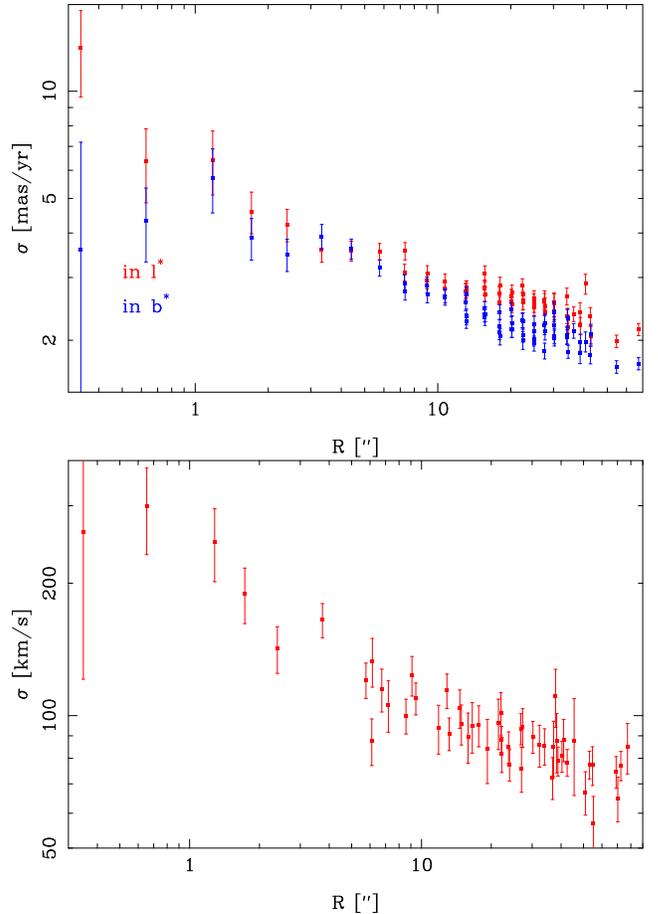

\begin{center}
\includegraphics[width=0.70 \columnwidth,angle=-90]{f10a.eps} 
\includegraphics[width=0.70 \columnwidth,angle=-90]{f10b.eps} 
\caption{Binned velocity dispersion used for Jeans modeling. The upper panel presents the proper motion data, 
the lower one the radial velocity data. 
} 
\label{fig:_std_v}
\end{center}
\end{figure}

For simplicity, we choose to bin our data. The loss of information  \citep{Merritt_94,Feigelson_12,Scott_92}  is small, since we use
a  large amount of data and the variations between the bins are smaller than the errors.
We assume symmetry relative to the Galactic plane, as supported by most observations, see Section~\ref{sec:rotation}.
There might be little deviations in the radial velocities. However, since we fit second moments and not velocities and dispersions the impact is very small. 
An edge-on flattened system has different symmetry properties in proper motion and in radial velocity. That is another reason for our different binnings which is explained in 
 Appendix~\ref{app_binning}. The same bins are also used by \citet{Chatzopoulos_14}.
We have tried also different binnings, which bin only in r and bin proper motions and radial velocities together. 
Our test includes bins of nearly equal size in r, log(r) and nearly equally populated bins.
The Jeans masses varies usually by less than the formal fitting errors, sometimes slightly more.
Since systematic errors are much larger than the fitting error (thus also larger than the binning error) it is therefore not necessary to include binning terms in the errors.
We show in Figure~\ref{fig:_std_v} the binned dispersion data in all three dimensions (l$^*$, b$^*$, z).

\subsection{Jeans Modeling}
\label{sec:jeans_modelling}

It is obvious from the dispersion difference (Figure~\ref{fig:_std_v}) in the two proper motion axes
that the nuclear cluster is not a spherical, isotropic system, which would have the dispersions in all directions. 
As shown in \citet{Chatzopoulos_14}, this difference is caused by
the flattening of the NSC; their Figure~10 shows how $\sigma_l>\sigma_b>\sigma_z$
over the whole range of radii for their favored axisymmetric models.
While we cannot include anisotropy totally, it is very likely minor given our constrains of Section~\ref{sec:_anisotropy} and the fact that an isotropic rotator fits the data well \citep{Chatzopoulos_14}.
Thus, anisotropic spherical modeling is at most only a small improvement compared to isotropic spherical modeling. 
Therefore, we use the most simple kind of Jeans modeling \citep{Binney_08} assuming isotropy and spherical symmetry for mass and light.
That simple model is mainly used for illustration of what can be derived from the data. In a simple model,
tests for other effects are also easier and faster, in contrast to more complex models.

 Two variants of isotropic Jeans modeling were used in the past for the GC:
\begin{itemize}
\item \citet{Genzel_96} and \citet{Trippe_08}  parametrized the deprojected dispersion.
\item \citet{Schoedel_09a} used a direct mass parametrization.
\end{itemize}
We follow here \citet{Schoedel_09a}. 
Since the existence of the SMBH is shown with orbits \citep{Schoedel_02}, is
advantageous when it can be parametrized directly.

\subsubsection{Relation of dispersion and mass}

To relate the measured dispersion to the mass we use the following equation derived from the Jeans equation \citep{Schoedel_09a,Binney_08}:

\begin{equation}
\begin {split}
\sigma_P^2(R)/G=\frac{ \int^{\infty}_R dr\, r^{-2} (r^2-R^2)^{1/2} n(r) M(r)}   {
\int^{\infty}_R dr\, r (r^2-R^2)^{-1/2} n(r) \,\, ,
}
\end{split}
\label{eq:jeans_par_is}
\end{equation}

and one needs to choose a parametrization for $M(r)$. The assumption which has
the fewest degrees of freedom is a constant mass to light ratio, 
 or a theoretically predicted profile of that 
ratio \citep{Luetzgendorf_12}. 
The GC data sets have been rich enough to 
leave the shape of the extended mass
distribution as a free parameter \citep{Schoedel_09a}.  
Given that the presence of the central SMBH is well established in the GC, one
can add the SMBH mass explicitly ($M_{\bullet}$) to $M(r)$, as done in \citet{Schoedel_09a}.

\subsubsection{Projection}

It is necessary to deproject the observed radii $R$ into
true 3D radii $r$, for which one needs to know the space tracer density distribution $n(r)$, see Section~\ref{sec:fitting_profile}. 
This is done by two projection integrals, namely equation~(\ref{eq:abel_1}) and the following integral:

\begin{equation}
\begin {split}
\Sigma(R) \sigma_P(R)^2=2 \int_{R}^{\infty} n(r) \sigma(r)^2 r dr/\sqrt{r^2-R^2} 
\end{split}
\label{eq:abel_2}
\end{equation}

\subsubsection{Averaging the 3D-dispersions}

In the Jeans equation a one-dimensional dispersion $\sigma$ is used. 
We have data in all three dimensions and thus need to average them. We use proper motions and radial velocities separately
since they cover different areas and have different errors. 
\begin{itemize}

\item 
For the radial velocities, there is significant rotation at large radii (Figures~\ref{fig:_v_rad_fit1}). 
To roughly take care of rotation we use instead  of  $\sigma_z$,  $\langle v^2_{z}\rangle^{1/2}$ (see e.g. \citet{Tremaine_02,Kormendy_13}) in Equation~\ref{eq:jeans_par_is}. 
Our approach is an approximation to axisymmetric modeling, which we use in \citet{Chatzopoulos_14}.

\item By definition of our coordinates, there
is no rotation term in the proper motions. Still a non-zero mean velocity can appear locally. 
Thus we also use $\langle v^2\rangle^{1/2}$.
The impact of using this measurement instead of $\sigma$ is very small.
We combine both dimensions in each bin to 
$\langle v^2_{\mathrm{pm}}\rangle=1/2\times(\langle v^2_{l}\rangle+\langle v^2_{b}\rangle)$  
to reduce the impact of the flattening. 
\end{itemize}

\subsubsection{Errors of the Dispersions}

Calculating the error of the dispersion with $\delta\sigma=\sigma\times 1/\sqrt{2N}$ is problematic when there
are only a few stars in a bin (especially as in our maser radial velocity data), since the weight of the individual bins will scale inversely with the 
dispersion value. Thus, low dispersion values have too large weights. 
To obtain errors which do not bias the result, we determine the errors with an iterative procedure: we fit the binned profile $\sigma(r)$ with an empirical function 
(a fourth order log-log-polynomial). In the first iteration we weight the points according to the observed $\sigma\times 1/\sqrt{2N}$.
In the further iterations we use the value of the polynomial fit ($\sigma_\mathrm{fit}$) to obtain the errors: $\delta\sigma=\sigma_\mathrm{fit}\times 1/\sqrt{2N}$. 
 With these errors we repeat the fit and get refined error 
estimates. After four iterations the procedure converges.  
We use the same method also for the proper motions. Due to the higher star numbers the effect of this correction
is smaller there.

As result our errors are slightly correlated between the different bins. The effect is small.
An estimate of its size can be obtained by comparing our masses when using different binning.
These differences are smaller than the formal fitting errors, and thus negligible compared to the
systematic errors.

\subsubsection{Tracer Distribution Profiles}

We fit our data using three different types of profiles for the three-dimensional tracer distribution:
\begin{itemize}
\item For most causes we fit a double $\gamma$-profile (Section~\ref{sec:fitting_profile}).
\item For checking the robustness of these, we also use a single $\gamma$-profile which we fit only to our inner density data. 
\item For comparison with the literature, we also use the single component tracer models of
\citet{Genzel_96,Trippe_08};
 and \citet{Schoedel_09a}. 
\end{itemize}

\subsubsection{Mass Parametrization}

Our mass model contains the point mass SMBH at the center and an extended component made of stars. It is justified to ignore gas clouds, since even the most
 massive structure,
the circumnuclear disk, has a mass of only a few times $10^4$ M$_{\odot}$ \citep{Mezger_96,Launhardt_02,Requena_12}.
We use two different ways of 
parametrization for the extended
mass:\begin{itemize}
\item We use a power law, similar to \citet{Schoedel_09a}: 
\begin{equation}
\begin{split}
M(r)=M_{\mathrm{100}\arcsec}\times(r/100\arcsec)^{\delta_\mathrm{M}}
\end{split}
\label{eq:mass_param1}
\end{equation}
The fact that the integrated mass is not finite for $r\rightarrow
\infty$ is not a problem, since our tracer profile $n(r)$ falls more rapidly than $M(r)$. 
\item We use a constant mass to light ratio for the extended mass:
\begin{equation}
\begin{split}
M(r)=M/L\times L(r)
\end{split}
\label{eq:mass_param2}
\end{equation}
The light is either flux or the star counts (Section~\ref{sec:lum_prop}).
\end{itemize}

For normalization of each 
we choose 100$\arcsec$, since that mass is well determined from our data.

\subsubsection{Fitting}

We fit simultaneously the surface  density and the projected dispersion data. 
To the surface density we fit the density model, the double $\gamma$ model (Section~\ref{sec:fitting_profile_gam}). 
The dispersion data depends on both the mass model and the density model, which in that case is the tracer model. Since we fit both at once both model components are constrained by both data sets.
When we use for the mass the power law model, the fit values of the density model are consistent
with the fits which only use the surface density (Table~\ref{tab:_surf_fit2}).
This confirms that our fits converge well. For constant M/L the parameters for the best fitting density model
differ often  by more than 1$\,\sigma$ from the ones when we fit only the surface density.  The reason is, that in the second case the density model depends more on the dispersions, because due to constant M/L the influence of the dispersion on the density model is grater than in the other case.
 Usually, the density model is less concentrated in the full fit than in the density only fit. That is an expression of the low dispersion problem in the center
(Section~\ref{sec:mass_smbh}).
However, the density parameter differences are not very big and do not influence the obtained masses relevantly. 
In one of our fits the central slope ($\gamma_{\mathrm{inner}}$) of the inner component of the tracer density is slightly (by only 0.02)
smaller than 0.5. This is problematic, because a slope smaller than 0.5 is not possible in the spherical isotropic case
when a central point mass is dominating \citep{Schoedel_09a}. 
To retain self-consistency in our simple isotropic models we fix the smaller slope to 0.5 in that case. The fix has no influence on the obtained masses. Also, the inner slope of the outer component is not well constrained and
we fix it usually to 0. 
Both restrictions have no relevant influence on the masses obtained.
In the fitting we optimize $\chi^2$ \citep{Press_86}: 
\begin{equation} 
\chi^2=\sum\limits_{i=1}^n (\frac{x_i-\mu_i}{\sigma_i})^2
\end{equation}
Therein $\mu_i$ is the function which we optimize; it consists of two function, one for density and one for the dispersion. The density function consists of two terms of Equation~\ref{eq:eta}  projected with Equation~\ref{eq:abel_1}, while the dispersion function consists of Equation~\ref{eq:abel_2} using one of our two types of extended mass parameterizations (power law or constant M/L). In both cases, the dispersion function is projected with Equation~\ref{eq:jeans_par_is}.
$x_i$ and $\sigma_i$
are the values and errors of the observables which are obtained in bins. 
We assume Gaussian errors.

We present the fits corresponding to the various choices of how to set up the Jeans model of the nuclear cluster in Table~\ref{tab:_mass_iso1}. We do not show the cluster properties since they do not differ by much. The
 robustness of the results can be assessed by comparing the different fits. For the fitting itself we first use routine mpfit \citep{Markwardt_09}. 
Secondly, we also fit the data by using Markoff-Chain Monte Carlo (MCMC) simulations following the method of \citet{Tegmark_04,Gillessen_09}. Our modification is that we start in the previously found minimum. We never find a better minimum
 than the starting point. 
 The errors are usually rather similar with both methods. There is usually some asymmetry, i.e. the errors
 are larger on the positive side. For mass properties that asymmetry is small. It is larger for some
 of the less well defined tracer density parameters. Such asymmetry cannot be found by mpfit.
 In Table~\ref{tab:_mass_iso1} we give as best value the median value of the accepted MCMC values. As error we give half of the difference between the 84.1\% and the 15.9\% quantile. Such errors encompass the central 1~$\sigma$ range. 
 
Initially, we fit with a free SMBH mass (Rows 1 and 2 in Table~\ref{tab:_mass_iso1}). However, especially for the power law mass model 
 which has one parameter more, the cluster shape, the resulting mass of the SMBH is smaller than the direct mass measurements 
by means of stellar orbits \citep{Ghez_09,Gillessen_09}. 
 We discuss the reason for the small SMBH mass in Section~\ref{sec:mass_smbh}. The direct cause 
for the too small SMBH mass is that models with a realistic black hole mass (M$_\bullet=4.17 \times 10^6M_{\odot}$) predict a larger dispersion  within $r<10\arcsec$ than we measure. That implies that our model is in some aspect incomplete near the black hole.
In the following we fix the central mass to M$_\bullet=4.17 \times 10^6M_{\odot}$, corresponding to our fixed distance of $R_0=8.2\,$kpc, and we neglect the small distance independent uncertainty of 1.5\% \citep{Gillessen_09}.

\begin{deluxetable*}{llllllllll} 
\tabletypesize{\scriptsize}
\tablecolumns{10}
\tablewidth{0pc}
\tablecaption{Jeans model fits \label{tab:_mass_iso1}}
\tablehead{No.& mass & tracer  & tracer &tracer  & dynamics   & M$_\bullet$ &  M$_{100\arcsec}$& $\delta_\mathrm{M}$&$\chi^2$/d.o.f \\
                          &  model &  model & source &range &  range & [$10^6 M_{\odot}$] &  [$10^6 M_{\odot}$] & &}

\startdata
1  & power law & double $\gamma$ & stars&all  & all R &  $ 2.26\pm 0.26$ &  $9.28 \pm 0.48 $ & $ 0.92 \pm 0.04 $&185.43/182 \\ 
2  &   M/L$=$const & double $\gamma$ & stars&all  & all R &   $ 4.37\pm 0.13$ &  $5.17 \pm 0.24 $ & & 276.16/184 \\ 
\hline
\bf 3 &\bf power law &\bf  double $\bf\gamma$ &\bf stars&\bf all &\bf 10$\arcsec<$R$<$100$\arcsec$ &\bf  4.17  &\bf  5.81 $\pm$ \bf 0.26 &\bf 1.21 $\pm$ \bf 0.05 &\bf145.83/137 \\ 
\bf 4  &\bf power law &\bf double $\gamma$ &\bf  flux &\bf all  &\bf  10$\arcsec<$R$<$100$\arcsec$ &  \bf 4.17  & \bf 6.17 $\pm$ \bf 0.23 & \bf 1.25 $\pm$ \bf 0.04  &\bf 349.51/286\\
\bf 5 &\bf   M/L$=$const &\bf double $\gamma$ &\bf stars &\bf all  &\bf  10$\arcsec<$R$<$100$\arcsec$ & \bf 4.17 &\bf  5.62 $\pm$ 0.17  &&\bf 156.97/138   \\ 
\bf 6 & \bf M/L$=$const &\bf  double $\gamma$ &\bf flux &\bf all &\bf  10$\arcsec<$R$<$100$\arcsec$ &\bf  4.17  &\bf  6.81 $\pm$ 0.16  &&\bf 342.58/287   \\
\hline
7  & power law & double $\gamma$ &stars& all  & R $<100\arcsec$ &  $4.17 $ &  $4.98 \pm 0.29 $ & $1.33 \pm 0.05 $&199.69/163 \\ 
8  &  power law &double $\gamma$ &  flux &all  & R $<100\arcsec$ &  $4.17 $ &  $5.35 \pm 0.23 $ & $1.37 \pm 0.04 $ &435.79/312\\
\hline 
9&  M/L$=$const &  double $\gamma$ &stars& all & R $<100\arcsec$ &  $4.17 $ &  $5.48 \pm 0.16$ &&195.82/164    \\ 
10  &  M/L$=$const &  double $\gamma$ &  flux &all & R $<100\arcsec$ &  $4.17 $ &  $6.70 \pm 0.18$ &&414.04/313  \\ 

11  & power law & double $\gamma$ &stars& all  &  R$>$10$\arcsec$ &  $4.17 $ &  $6.14 \pm 0.18 $ & $1.12 \pm 0.03 $&182.36/157 \\ 
12 & power law & double $\gamma$ &  flux &all  &  R$>$10$\arcsec$ &  $4.17 $ &  $6.57 \pm 0.17 $ & $1.12 \pm 0.03 $&399.37/306 \\ 
13  &  M/L$=$const &  double $\gamma$ &stars& all  & R$>$10$\arcsec$ & $4.17 $ &  $5.58 \pm 0.15 $ &&240.18/158  \\\ 
14  &  M/L$=$const &  double $\gamma$ &  flux &all & R$>$10$\arcsec$ &$4.17 $ &  $6.65 \pm 0.16 $ &&433.36/307  \\ 
15  & power law & double $\gamma$ &stars& all  & all R & $4.17 $ &  $5.59 \pm 0.17 $ & $1.19 \pm 0.03 $&247.59/183 \\ 
16  & power law & double $\gamma$ &  flux &all & all R &  $4.17 $ &  $6.03 \pm 0.17 $ & $1.19 \pm 0.03 $&507.08/332 \\ 
17  &  M/L$=$const &  double $\gamma$ &stars& all  & all R & $4.17 $ &  $5.47 \pm 0.14 $ &&277.14/185 \\ 
18 &  M/L$=$const &  double $\gamma$ &  flux &all  & all R &  $4.17 $ &  $6.56 \pm 0.15 $ &&500.64/333   \\ 

\hline
19 &power law & single $\gamma$ & stars&R$<220\arcsec$ & 10$\arcsec<$R$<100\arcsec$ &   $4.17 $ &  $5.91\pm 0.21$ & $1.32 \pm 0.08 $&126.27/108  \\ 
20 &power law & single $\gamma$ & flux &R$<220\arcsec$ &  10$\arcsec<$R$<100\arcsec$  & $4.17 $ &  $6.27\pm 0.18$ & $1.36 \pm 0.07 $ &207.12/132 \\
21 & M/L$=$const &  single $\gamma$ & stars&R$<220\arcsec$ &  10$\arcsec<$R$<100\arcsec$  &  $4.17 $ &  $6.23\pm 0.16$ &&130.58/108  \\ 
22  &  M/L$=$const & single $\gamma$ & flux & R$<220\arcsec$ &  10$\arcsec<$R$<100\arcsec$ &   $4.17 $ &  $6.92\pm 0.15$  &&205.30/134  \\
\hline
23  & power law & Sch\"odel+ 2009 & stars& R$<220\arcsec$ &  10$\arcsec<$R$<100\arcsec$ &   $4.17 $ &  $5.79\pm 0.25$ & $1.16 \pm 0.05 $&195.97/110\\
24  & power law & Trippe+ 2008 & stars& R$<220\arcsec$ &  10$\arcsec<$R$<100\arcsec$ &    $4.17 $ &  $7.22\pm 0.17$ & $1.20 \pm 0.04 $&453.40/110\\
25  &power law & Genzel+ 1996 & stars& R$<220\arcsec$ & 10$\arcsec<$R$<100\arcsec$ &    $4.17 $ &  $6.75\pm 0.18$ & $1.12 \pm 0.04 $&237.05/110\\
26 &  M/L$=$const & Sch\"odel+ 2009 & stars&R$<220\arcsec$ &   10$\arcsec<$R$<100\arcsec$&   $4.17 $ &  $4.66\pm 0.09$& &218.72/111  \\
27 & M/L$=$const & Trippe+ 2008 & stars&R$<220\arcsec$ &  10$\arcsec<$R$<100\arcsec$ &   $4.17 $ &  $7.56\pm 0.15$& &444.73/111  \\ 
28  &  M/L$=$const & Genzel+ 1996 & stars& R$<220\arcsec$&  10$\arcsec<$R$<100\arcsec$  &   $4.17 $ &  $6.19\pm 0.12$ &&243.69/111 \\
\hline
29&  M/L$=$const &  double $\gamma$ &stars& all & R $<27\arcsec$ &  $3.37\pm0.16 $ &  $7.31 \pm 0.42$ && 91.93/120 
\enddata
\tablecomments{
Jeans model fitting of our dynamics and density data, assuming different mass and tracer models, and different selections for the data. For the fitted surface density (tracer) we use two data sources and restrict us sometimes to
a subset of the available data range (Column 3 and 4). The dynamics data consists of $\langle v^2\rangle$ in all three dimensions. We restrict it partly radially.
The mass model includes in all cases a central point mass.  M$_{100\arcsec}$ is the nuclear cluster mass within 100$\arcsec$. 
If no error is given for a parameter it is fixed. 
The literature tracer models are from
\citet{Trippe_08,Schoedel_09a},
 and \citet{Genzel_96}.
}
\end{deluxetable*}

\subsubsection{Results}

 The fits which meet our assumption best (Table~\ref{tab:_mass_iso1}) are in the rows 3, 4, 5, and 6. They use
a range of 10$\arcsec<R<$100$\arcsec$ for the dynamics data. The surprisingly low dispersion in the center impacts the cluster mass in case of
the fits in rows 7 and 8. These cases
use $R<100\arcsec$ for the dynamics data (i.e. including the central region) and allow the cluster shape to vary with  the power law.
All other fits are less influenced by the central dispersion and yield masses $M_{\mathrm{100}\arcsec}$ between 5.5 and $6.8 \times 10^6\,$M$_{\odot}$. The fact that the mass does not strongly depend ($\delta M<15$\%) on whether we include the central 10$\arcsec$ shows
that the influence of the too small SMBH mass on the cluster mass is much smaller than the effect of the neglected flattening, see Section~5.4.8.

\begin{figure}
\begin{center}
\includegraphics[width=0.70 \columnwidth,angle=-90]{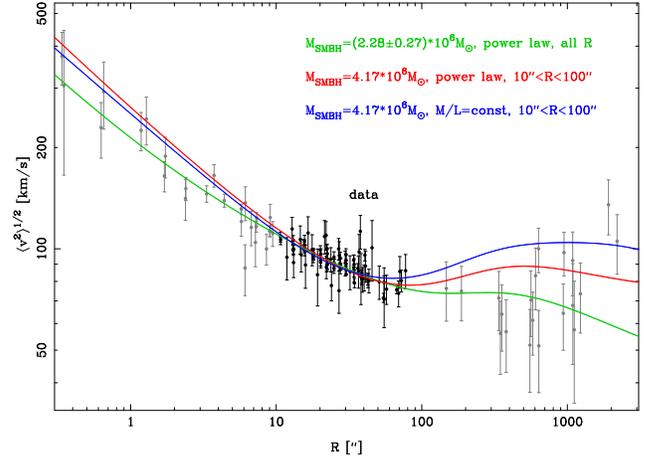}
\caption{Data and fits of the Jeans modeling. The data within 10$\arcsec$ and outside 100$\arcsec$ (gray dots) are somewhat less consistent with 
our simple model; the other data is plotted as black dots. The curves show the fits from rows 1 (green), 3 (red), and 5 (blue) of 
Table~\ref{tab:_mass_iso1}. 
}
\label{fig:_fits_iso3}
\end{center}
\end{figure}

\begin{figure*}
\begin{center}
\includegraphics[width=1.999 \columnwidth,angle=-90]{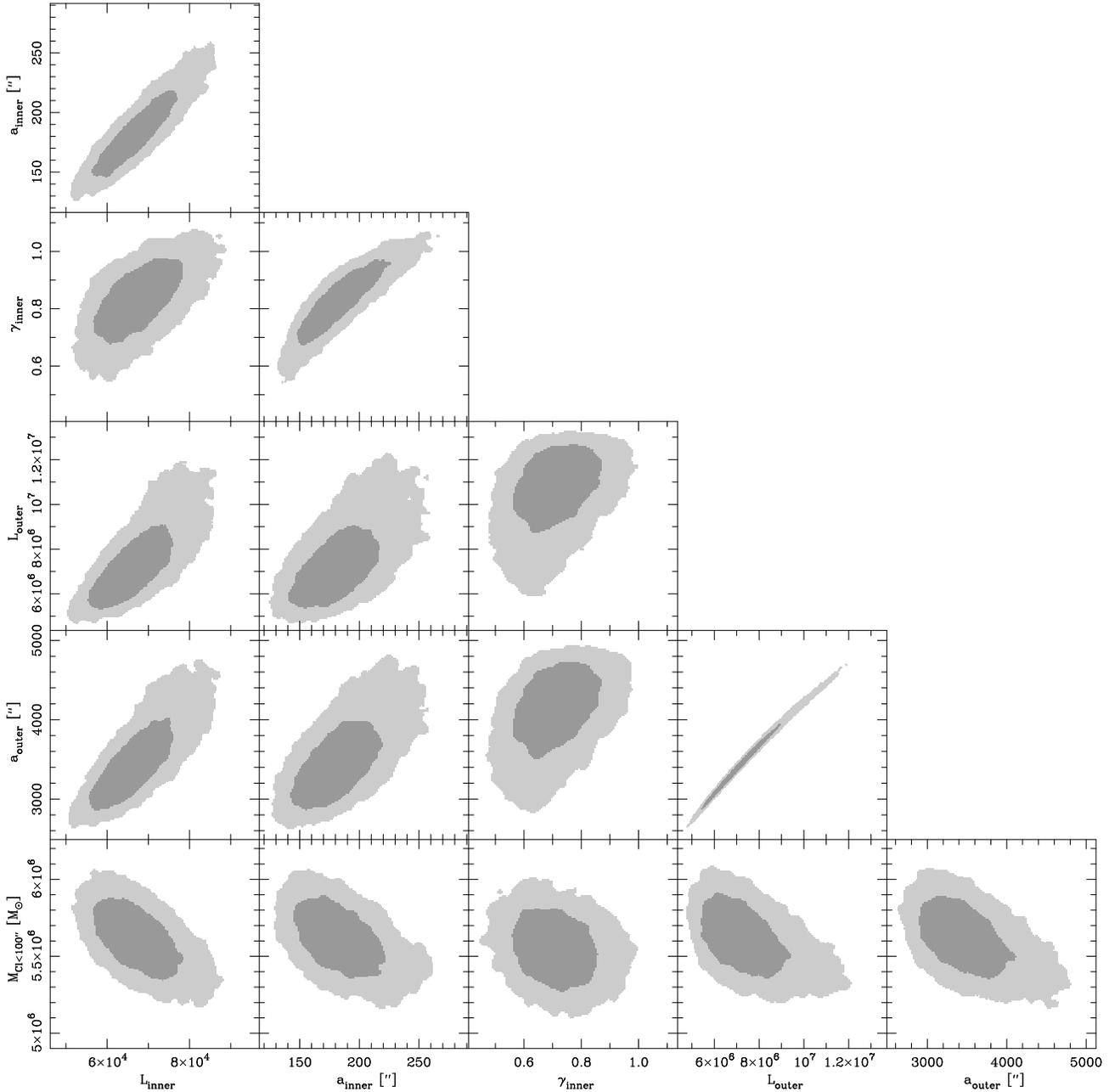} 
\caption{Parameter correlations for model 5 (Table~\ref{tab:_mass_iso1}). The parameters are the `luminosity' (in star counts), the inner slope parameter and the 'core' radius for the two $\gamma$ components and the extended mass at 100$\arcsec$. The black hole mass
is fixed here. The dark gray corresponds to the 1 $\sigma$ area, the light gray to the 2 $\sigma$ area. 
The parameter correlations are obtained from a Markoff-Chain Monte Carlo simulation.
} 
\label{fig:corr_plot1}
\end{center}
\end{figure*}

Figure~\ref{fig:_fits_iso3} illustrates the fits of rows 1, 3, and 5. 
In this figure it is visible that our models with the right SMBH mass have higher dispersions than the measurements at $R<10\arcsec$. Probably the models are not correct there. 
In Figure \ref{fig:corr_plot1} we show all correlations of the fit in row 5. Some (outer) tracer density parameters are strongly correlated with each other. The cluster mass itself shows only weak correlations with other parameters. That explains the small fit errors for the cluster mass.

 Outside of $100\arcsec$ the fits are also not very good. The reason for that is probably:
\begin{itemize}
\item  We don't have density data outside of 1000$\arcsec$, and hence cannot constrain the tracer profile there.
The outer slope of the tracer profile is thus not well-constrained.
\item The fixed outer slope in the $\gamma$-model does not allow for variability in the outer slope.
\item The assumption of spherical symmetry is a poor approximation outside of $\approx 300\arcsec$.
\end{itemize}
Therefore, the mass outside the central $\approx100\arcsec$ is less reliable.

\begin{figure*}
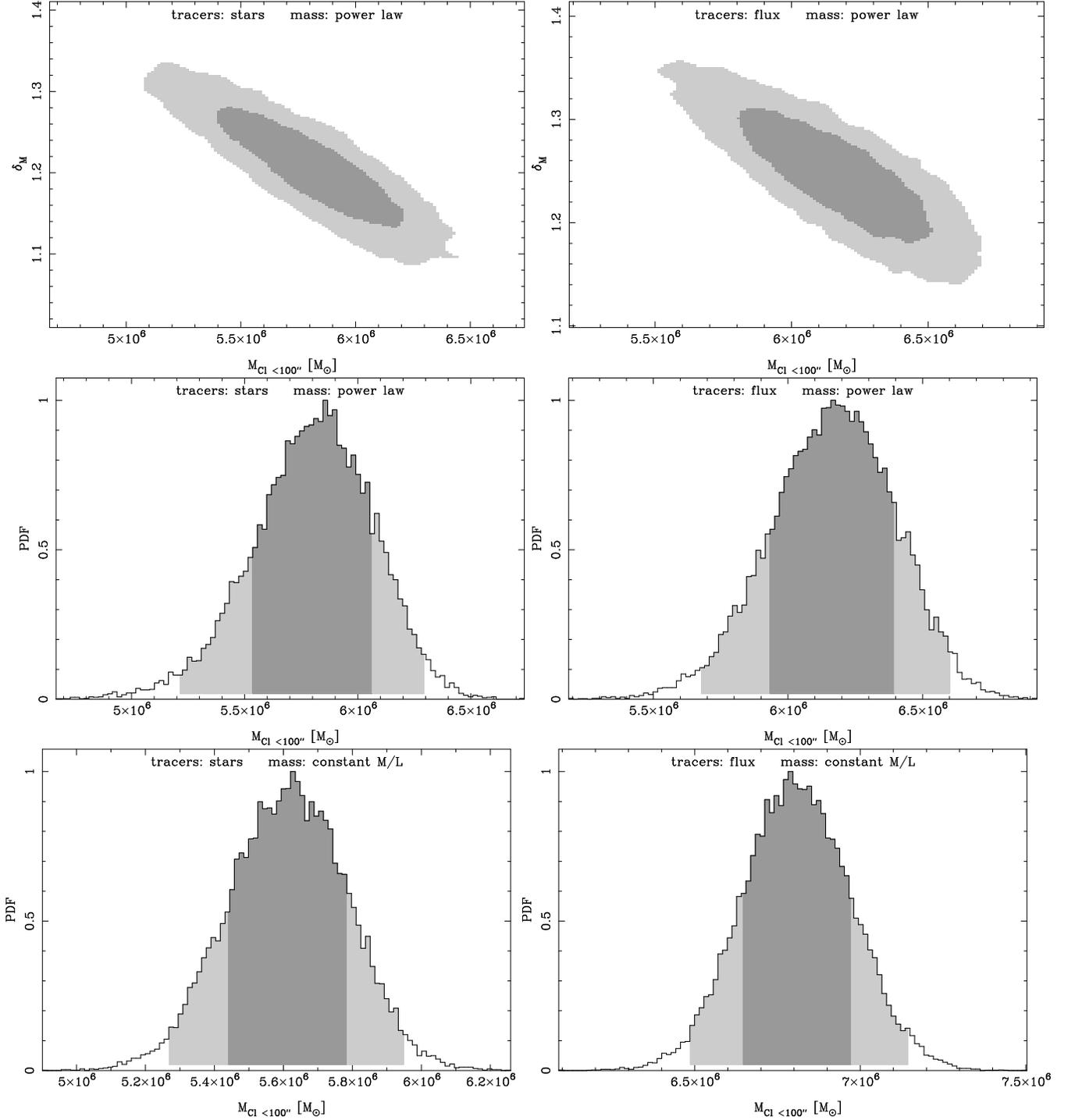

\begin{center}
\includegraphics[width=0.74 \columnwidth,angle=-90]{f13a.eps} 
\includegraphics[width=0.74 \columnwidth,angle=-90]{f13b.eps} 
\includegraphics[width=0.73 \columnwidth,angle=-90]{f13c.eps} 
\includegraphics[width=0.73 \columnwidth,angle=-90]{f13d.eps} 
\includegraphics[width=0.73 \columnwidth,angle=-90]{f13e.eps} 
\includegraphics[width=0.73 \columnwidth,angle=-90]{f13f.eps} 
\caption{Cluster mass correlations and probability distributions for model 3, 4, 5 and 6 (Table~\ref{tab:_mass_iso1}).
 These four cases use the same dynamic data range ($10\arcsec<R<100\arcsec$) but use different tracer distributions and extended different mass models. The ones in the left column use the star density as tracer, the ones in the right the flux density. For the two top rows a power law is fitted to nuclear cluster. In the bottom row a constant mass to light ratio is assumed.
For all the SMBH is fixed to M$_{\mathrm{SMBH}}=4.17\times10^6\, M_{\odot}$. The dark gray marks the 1$\,\sigma$ area from MCMC,  the light gray the 1$\,\sigma$ area.
} 
\label{fig:corr_plot2}
\end{center}
\end{figure*}

Apart from the fits in rows 7 and 8, the selection of the range for the dynamics data is less important
than the choice of the tracer source for the obtained mass. The latter is particularly true when choosing a 
constant mass-to-light ratio as the mass model. Choosing the stellar number counts or the 
flux as the tracer results in a difference of up to 20~\% in $M_{100\arcsec}$ (compare rows 5 and 6 in table). 
The smaller value occurs for the star density profile. The main reason is that in this profile the inner component has a larger core radius than in case of the flux profile. With a smaller core a bigger part of the mass which causes the dispersion need to be close to the black hole.
In Figure~\ref{fig:corr_plot2} we show the cluster mass probability distribution of row 3 to 6, i.e. the ones which all use the same, likely best, dynamic data. It is also visible that there is a clear anticorrelation between cluster mass and cluster mass slope. 
Still, the best values of both are well defined, in contrast to \citet{Schoedel_09a}, thanks to our larger radial coverage.

Looking at the $\chi^2$-values in the table seems to indicate that the fits using the flux density as a tracer profile are worse than 
the fits using the stellar number counts. 
However, the difference is due to the density data, where because of our error calculation 
(Appendix~\ref{sec:obt_lum}) the $\chi^2$ of the flux density data is worse (Section~\ref{sec:fitting_profile}). 
Since we identify no reason to prefer one of the two data sets (Section~\ref{sec:number_density}), we give of both of them equal weight in the following.

From the sample we exclude rows 7 and 8, because the obtained slope $\delta_\mathrm{M}$ are significantly bigger than the slope of stars counts
and flux profiles. That is unrealistic and is caused by the dispersion problem in the center. Thus, we consider a sample, that contains the 14 fits in the rows 3 - 6 and 9 - 18 
which we all give equal weight. The average mass is
\begin{equation}
M_{100\arcsec}=(6.09 \pm 0.53) \times \,10^6\,M_{\odot} \,\, .  
\end{equation}
As \citet{Gillessen_09} we calculate the systematic error from the scatter between the different fits.
That errors dominates over the smaller fit errors. That way of error calculation has the advantage that
it uses realistic uncertainties for the extended mass and star distribution. In the case of the mass, it explores the presence and absence of a dark cusp. In the case of the star distribution, it enables a realistic assessment of the uncertainty in the profile, which a single profile cannot provide. 
Using only rows 3 - 6 yields a very similar result. 
 According to Table~\ref{tab:_mass_iso1}, the main uncertainty is whether the star counts or the flux is used. They have different density profiles which results in different masses. By comparison, the other assumptions (like the mass profile) play a minor role.

The cluster mass slopes are between 1.12 and 1.25 for the cases used. The
mean is $\delta_\mathrm{M}=1.18 \pm0.06$, where the error again is dominated by the scatter between the different fits. This number is in 
reasonable agreement with the tracer profile, which has a slope of 1.18 (stellar number count based) or 1.06 (flux based) in the 
range 50$\arcsec< R <$ 200$\arcsec$.

In order to check how important our two-component model is we also fit single $\gamma$ models to our data (Rows 19 - 22). To ensure, that a single component model is a reasonable fit to the data, we restrict the
range of the tracer density data inside the break to the nuclear disk, i.e. $R < 220\arcsec$.  
With this restriction, the masses are slightly larger than before. The main reason is that the core of the inner $\gamma$ component, which is now the only $\gamma$,
is larger in this case.
Using the literature profiles of
\citet{Genzel_96,Trippe_08}; and 
\citet{Schoedel_09a} the mass range is larger, from 
$4.66\times 10^6$ to $7.56\times 10^6\,M_{\odot}$. The main reason for the mass trend between them is the
effective outer slope. When it is steep as in \citet{Trippe_08} the expected dispersion at large radii gets smaller.

Our data set would allow us to determine the distance to the GC, $R_0$, by means of a statistical parallax \citep{Genzel_00,Eisenhauer_03,Trippe_08}.
To be accurate in $R_0$ it is necessary to have a model which accounts for differences between $\sigma_l$, $\sigma_b$ and $\sigma_z$. Our spherical model cannot provide that, and we
defer this to the work of \citet{Chatzopoulos_14}, where we present self-consistent flattened models.

We now obtain the mass uncertainty due to the distance uncertainty. 
We use the latest results for the orbit of the S-stars of $R_0=8.2\pm 0.34$ kpc \citep{Gillessen_13}. Due to the small distance-independent SMBH mass error of 1.5\% \citep{Gillessen_09}, which is vanishingly small compared to our other uncertainties, it is sufficient
to derive the mass dependency on distance.
To include the distance error in the error budget we repeat the Jeans modeling 
for $R_0$ of 7.86 and 8.54 kpc. Thus we assume that the distance error is Gaussian, which seems to be true approximately, see Figure~15 in \citet{Gillessen_09}. We perform full $\chi^2$ adding in \citet{Chatzopoulos_14}. We determine the SMBH mass 
from the M$_\bullet-R_0$ relation of
\citet{Gillessen_09}: $M_\bullet \propto R^{2.19}$. 
 We find following mass distance relation: 
\begin{equation}
M_{100\arcsec}=6.09 \cdot 10^6
\times \left(\frac{R_0} M_{\odot} {8.2 
\mathrm{kpc}}\right)^{3.83}  
\end{equation}
The distance error of \citet{Gillessen_09} leads to following
expression
\begin{equation}
M_{100\arcsec}=(6.09 \pm 0.53|_{\mathrm{fix} R_0}\pm  
0.97|_{R_0} ) \times 10^6  M_{\odot}\,\, . 
\end{equation}
Thus, the distance-induced error is larger than the other considered errors. 
However, we do not consider all error sources in our work. Our model does not include anisotropy and flattening. The inclusion of both can introduce important mass changes, see e.g. \citet{Marel_10,Dsouza_13}. 
Since we detect signs for flattening but not for anisotropy, flattening is more important. We use a flattened model in \citet{Chatzopoulos_14}. Comparing with that work it can be clearly seen that the cluster mass is 47\% bigger when flattening is included. Thus, it is important to include the flattening in nuclear cluster models. Still, our relative errors are useful as an estimate of some systematic uncertainties. While the distance dependent error may improve with a better distance from e.g. the statistical parallax \citep{Chatzopoulos_14}, it is more difficult to reduce contributions of the light and mass profile uncertainty. They cause an error of about 8.5\% in our modeling which would be rather similar also in more complex models. That error is larger than the statistical error of 3.5\% of \citet{Chatzopoulos_14}, which includes also the distance uncertainty. It is however smaller than their systematical error of 10.1\%. Their systematic error does not include the tracer profile uncertainty, which is dominant in our modeling, but is dominated by the flattening uncertainty. Overall, our tests show that the uncertainty which is caused by the tracer profile is not the dominant error contribution.

Finally, we obtain an estimate for the total mass of the nuclear cluster by using the inner components of the fits with constant M/L. 
From the average and scatter we obtain \begin{equation}
M_\mathrm{NC}=(4.22 \pm 0.50|_{\mathrm{fix} R_0}\pm 0.67|_{R_0}) \times 10^7  M_{\odot} \,\, .
\end{equation}
Since the outer cluster slope is more uncertain than the mass at 100$\arcsec$ the mass uncertainty is now larger.
The full uncertainty is larger still. It is not present here because the outer slope
is fixed in $\gamma$-models. The true total mass uncertainty is probably as in the case of the luminosity 
(Section~\ref{sec:total_luminosity}) around $\pm$50\% of the mass. In contrast, the extended mass within 100$\arcsec$ is not affected by that uncertainty, because within $\approx100\arcsec$ the data are better.

\subsection{Velocity Distribution and Fast Stars}
\label{sec:fast_stars}

The Jeans modeling only used the first two moments of the velocity distributions, but they contain information beyond that. In particular, stars with
high velocities in the wings of the distributions are interesting. The unbiased space (three-dimensional) velocity for each star is \citep{Trippe_08}
\begin{equation}
\begin {split}
v_{\mathrm{3D}}=\sqrt{v_x^2+v_y^2+v_z^2-\delta v_x^2-\delta v_y^2-\delta v_z^2} \,\, .
\end{split}
\label{eq:vel_3d}
\end{equation}

\begin{figure}
\begin{center}
\includegraphics[width=0.70 \columnwidth,angle=-90]{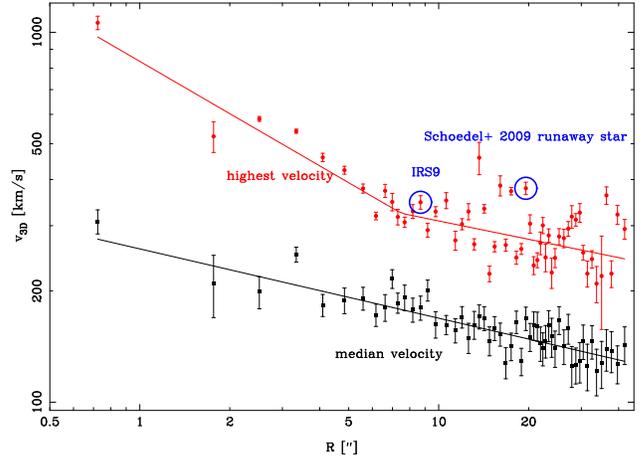}
\caption{
Binwise median and maximum three-dimensional velocity. We describe the data with (broken) power laws. 
The error indicates the velocity error of the median velocity stars and the velocity 
error of the fastest stars, respectively. Fast stars already discussed by \citet{Reid_07} or \citet{Schoedel_09a} are marked with open blue circles. 
} 
\label{fig:_max_vel}
\end{center}
\end{figure}

Since the three-dimensional velocity is, like the dispersion positive by definition we subtract the errors to get unbiased velocities.
The errors of $v_{\mathrm{3D}}$ do not depend on the value, and in particular there is no indication (especially inside 20'') that the higher velocity stars are caused by measurement problems.
We divide the sample into radial bins of 32 stars each, and determine in each the maximum and the median 3D-velocity 
(Figure~\ref{fig:_max_vel}). The maximum velocity cannot be described by a single power law. Inside of 7.5$\arcsec$
it follows a power law slope of $-0.47 \pm 0.04$, 
close to a Keplerian slope of -0.5. Other high quantiles like the second fastest star follow similar slopes in the center.  Outside of 7.5$\arcsec$ these slopes 
appear to be consistent with
the single slope of the median velocity, of around $-0.19$.

\subsubsection{Minimum Binding Mass}

\begin{figure}
\begin{center}
 \includegraphics[width=0.70 \columnwidth,angle=-90]{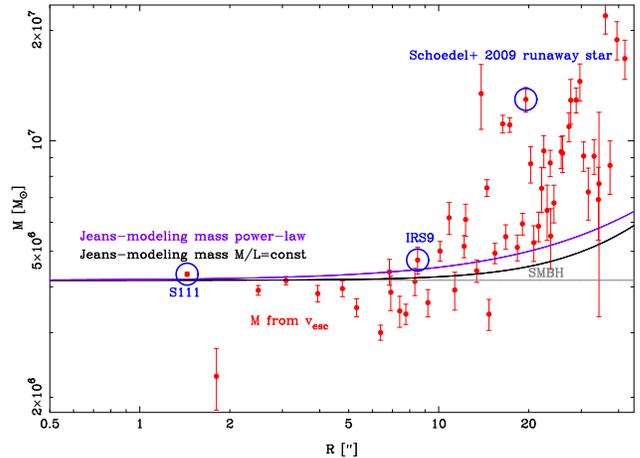} 
\caption{
Maximum of the minimum binding mass in bins. The mass errors 
follow from the 1$\,\sigma$ velocity errors We compare these masses with the SMBH mass of 
\citet{Gillessen_09} and two Jeans-models, model A and C from Table~\ref{tab:cum_mass}. 
The fast stars from \citet{Reid_07,Trippe_08}
and \citet{Schoedel_09a} are indicated with open blue circles. 
} 
\label{fig:_esc_mass}
\end{center}
\end{figure}

We now calculate for all stars the minimum binding mass, i.e. we assume that each star is in the plane of the sky and on a parabolic orbit
around a point mass. 
 In each radial bin we thus can determine the highest mass (Figure~\ref{fig:_esc_mass}). Inside of 8.3$\arcsec$ the escape mass 
is close to the SMBH mass. The average escape mass there is $3.67\times10^6\,$M$_{\odot}$.
The most significant mass larger than that of the SMBH occurs for the star S111 with 
$4.33 \pm 0.05 \times10^6\,$M$_{\odot}$, as was noted already by \citet{Trippe_08} and \citet{Gillessen_09}.
High escape masses indicate that the stars are close to the SMBH not only in projection, but also in 3D. 
These fast stars can be used as additional constraints on the late-type density profile close to the SMBH in advanced 
dynamic modeling.
\subsubsection{A Distance Estimate}

\begin{figure}
\begin{center}
\includegraphics[width=0.70 \columnwidth,angle=-90]{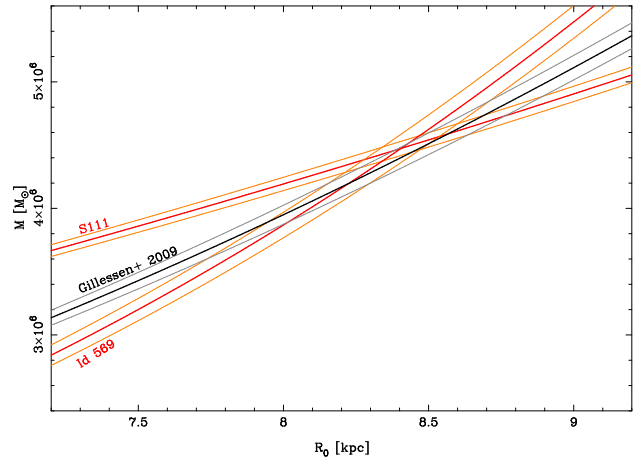}
\caption{Distance constraints from stars, that are close to escaping. The figure shows the mass distance relations for the SMBH from 
\citet{Gillessen_09}.
The mass distance relations for the two stars which are closest to escape is derived from the minimum binding mass. 
Red and black lines mark the value; orange and gray lines the respective 1$\,\sigma$ error range.
} 
\label{fig:_dist_esc}
\end{center}
\end{figure}

The number of stars that appear unbound depends on the assumed distance $R_0$. Stars whose velocity is dominated by the
radial velocity are bound for a large $R_0$, while stars with a large proper motion are bound for small $R_0$. 
In the following we restrict the used data to r$\leq8.45\arcsec$, the area dominated by the SMBH. 
Our sample contains nine stars, which are unbound for some distances between 7.2 and 9.2 kpc. 
From these stars we calculate a $\chi^2$ as a function of $R_0$ summing up in squares the significances of the differences from
the minimum binding mass minus the mass inside that radius. We only consider stars when they appear unbound, however not when the minimum binding is smaller than the mass inside that radius. The mass is dominated by the SMBH, for which we use 
M$_{\bullet}=(3.95 \pm 0.06 )\times 10^6 \,(R_0/(8.0 \mathrm{kpc})^{2.19}\,$M$_{\odot}$ \citep{Gillessen_09}. 
That means we use from that paper only the distance independent mass error, not the distance dependent component.
For the extended mass we use our two preferred models, a power-law and a constant M/L case, see Section~\ref{sec:cum_mass}.
The mass of the SMBH dominates, the extended mass adds at most 10$^5\,M_{\odot}$.
The error on the SMBH mass we add by exploring the Gaussian 1.5\% error. In case of the extended mass we give the two cases equal weight and get the error from the difference.
Minimizing the $\chi^2$ as a function of $R_0$ yields  $R_0=8.53^{+0.21}_{-0.15}\,$kpc. 
The main constraint  comes from 2 stars: S111 and a star with high proper motion, Id 569
(Figure~\ref{fig:_dist_esc}). 

This estimate relies on some assumptions: 
\begin{itemize}
\item We assume that all stars are bound. This is justified, since the chance to see an escaping star from a Hills-event \citep{Hills_88} 
is very low \citep{Yu_03,Perets_07}. Also the Hills mechanism, which causes stars faster than the local escape velocity at $\gtrsim 10\arcsec$ is unlikely to play 
a role, as suggested
by the break around 8$\arcsec$ visible in Figure~\ref{fig:_max_vel}).
\item The method relies on the mass-distance scaling of the SMBH mass from stellar orbits.  Using the relation from 
\citet{Ghez_09} we obtain R$_0=8.38^{+0.23}_{-0.25}\,$kpc.
\item We assume that we have correctly debiased the velocities. Since the total velocity error of the two most 
important stars is less than 1.5$\,$\% of their velocity
this assumptions seems to be uncritical.
\end{itemize}
A lower limit on the SMBH mass independent of the M$_{\bullet}-R_0$ can be obtained by using the smallest recent R$_0$ measurement of 7.2 kpc \citep{Genzel_10,Bica_06}.
For that distance S111 gives $3.6\times 10^6 M_{\odot}$, 
consistent with the orbit-based estimates, and higher than most Jeans-model estimates.

\subsubsection{Fast Stars at $R>10\arcsec$}

Outside of 10$\arcsec$ most fast stars are unbound to SMBH and nuclear cluster mass (Figure~\ref{fig:_esc_mass}). These stars are detected in all three velocity dimensions: we have 5, 6, 5, with $|v|>275$ km/s in either/or z, $l^*$ and $b^*$, respectively among the stars with all velocity components measured outside of 7$\arcsec$ in the central and extended sample.
Some of these stars were already discussed in 
\citet{Reid_07} and \citet{Genzel_10}. Our improved mass estimate for the nuclear cluster reinforces the statement that these stars
are not bound to the GC. Note that placing a star out of the plane of the sky enlarges the discrepancy \citep{Reid_07}, 
e.g. for a star  at 40$\arcsec$ the escape velocity decreases out to $z=200\arcsec$ and then increases  only very slowly to about 165 km/s at 3600$\arcsec$, 
which is still less than in the plane of the sky. The fact that the maximum velocity decreases for $R>8\arcsec$ in the same
way as the median velocity (Figure~\ref{fig:_max_vel}) excludes that the fast stars are foreground objects, for which 
the velocity would not depend on radius. The extinction appears to be normal for the fast stars.

\citet{Reid_07} also discussed binaries as a solution for the high velocities. This is excluded by our data set since for some stars the high velocity 
is dominated by long-term proper motions measurements, which cover much more time than what one orbital period would need to be.

For testing whether the Hills-mechanism is important, the directions of motion of the fast stars can help. We use therefore  the two sided K-S~test on the distribution of proper motion vectors in polar coordinates comparing fast stars with all stars. Fast stars are again selected in radial bins to avoid influence of radial trends.
We obtain for different selections borders within the fastest 1\% that the two distributions are identical, probability usually 0.5, in one case 0.07.  Thus the fast stars are not preferentially moving away from the black hole.
That makes it unlikely that the Hills mechanism is the main mechanism.
The comparably large number of fast stars makes that also unlikely.
Probably as already advocated by \citet{Reid_07}, the best solution is that these stars are on very eccentric orbits in the large scale potential.
Therein they can obtain higher velocities than the escape velocity determined from the local mass distribution.

Since the database of  \citet{Genzel_10} contained many bright, unbound stars, these authors suggested that preferentially young, 
relaxed, bright TP-AGB stars are on these unbound orbits. This finding might be affected by low number statistics 
and a bias. Bright stars have smaller velocity
errors and are therefore easier to identify as significantly unbound. In our sample and excluding the database of \citet{Genzel_10}, 
we see no evidence that bright stars   
are dynamically distinct from the other stars. This also holds for the subsample of medium old TP-AGB stars from \citet{Blum_02}. Also 
\citet{Pfuhl_11} found that their two samples of younger and older giants show consistent dynamics.

\subsubsection{Velocity histograms}

\begin{figure}
\begin{center}
 \includegraphics[width=0.70 \columnwidth,angle=-90]{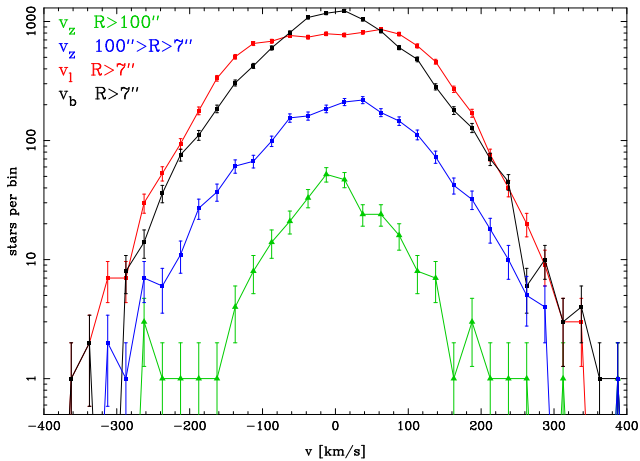}
\caption{Velocity histograms of our dynamics sample.} 
\label{fig:_velocity_his}
\end{center}
\end{figure}

In Figure~\ref{fig:_velocity_his} we show the velocity histograms for the three dimensions. In case of the proper motion samples we show
$R>7\arcsec$. For the radial velocities, we show two bins,  7$\arcsec<R<100\arcsec$ and $R>100\arcsec$. In the latter
bin (consisting of the maser stars) we have subtracted off the rotation.  

 We see deviations from Gaussian distributions in many aspects:

\begin{itemize}
\item The central part of distributions in $l^*$ and $b^*$ are not as peaked as Gaussians (Figure~\ref{fig:_velocity_his}).
In $l^*$ the distribution has a flatter peak than a Gaussian \citep{Trippe_08,Schoedel_09a}.
In $b^*$ the distribution has a slightly, but significantly ($>5\,\sigma$), steeper peak than a Gaussian.
The flatter peak in $l^*$   is actually a signature of the flattening of the cluster, as shown in \citet{Chatzopoulos_14}.
\item The varying ratio of maximal and median velocity (Figure~\ref{fig:_max_vel}) is indicative of non-Gaussian wings
within 7$\arcsec$. 
\citet{Trippe_08} did not see that
because they did not radially subdivide their sample. Furthermore our sample contains
twelve more stars in the central 2$\arcsec$ with v$_{\mathrm{3D}}>460$ km/s than the one from \citet{Trippe_08}. 
In that work the only such fast star was S111. 
The high-velocity wings in all three dimensions are mainly caused by the presence of the SMBH.
 Therefore, because of the higher number of fast stars S111 is less special than discussed in \citet{Trippe_08}. 
\item From the 274 stars in the $R>100\arcsec$-bin, roughly 13 are in the high (radial) velocity wings of the distribution  (Figure~\ref{fig:_velocity_his}). 
Most of these outliers were already noted by \citet{Lindqvist_92b} and \citet{Deguchi_04}.
\end{itemize}

\begin{deluxetable*}{lllllllll}
\tabletypesize{\scriptsize}
\tablecolumns{9}
\tablewidth{0pc}
\tablecaption{Fast stars \label{tab:_fast star}}
\tablehead{ID & R.A. [$\arcsec$] & Dec. [$\arcsec$]  & v$_{\mathrm{R.A.}}$ [mas/yr]&v$_{\mathrm{Dec.}}$ [mas/yr]  & v$_{\mathrm{z}}$ [km/s]  &v$_{\mathrm{3D}}$ [km/s] & v$_{\mathrm{esc}}$ [km/s]  & Comment}
\startdata
770 & -1.11 & -0.91 & $-2.78 \pm 0.02$ & $-7.72 \pm 0.01$ & $-741 \pm 5$ &$807 \pm 5$ & 792 & S111\\
569 & 0.13 & 3.08 & $13.22 \pm 0.04$ & $-3.6 \pm 0.05$ & $-92 \pm 6$ &$540 \pm 7$ & 541 & \\
4 & 5.68 & -6.33 & $2.96 \pm 0.09$ & $2.58 \pm 0.08$ & $-312 \pm 14$ &$347 \pm 14$ & 326 & IRS 9\\
4258 & 2.66 & 13.61 & $-8.48 \pm 0.15$ & $-1.67 \pm 0.09$ & $-315 \pm 45$& $458 \pm 46$ & 255 & fastest v$_{\mathrm{3D}}$, R$>7\arcsec$ \\
899 & -8.37 & -12.21 & $-1.98 \pm 0.15$ & $11.03 \pm 0.24$ &             &$435 \pm 11$ & 247 & fastest v$_{\mathrm{2D}}$, R$>7\arcsec$\\  
903 & 14.11 & 7.45 & $-1.58 \pm 0.04$ & $-0.98 \pm 0.13$ & $379 \pm 24$ &$385 \pm 24$ & 238 & only high v$_{\mathrm{z}}$\\
787 & -6.8 & 18.35 & $-5.96 \pm 0.13$ & $7.36 \pm 0.2$ & $-91 \pm 11$ &$379 \pm 15$ & 215 & runaway candidate\\
 & &  &  &  & & &  &  \citep{Schoedel_09a}\\
\enddata
\tablecomments{Extraordinarily fast stars sorted by their projected distance from Sgr~A*. The errors of
the positions are smaller than 5 mas. v$_{\mathrm{esc}}$  assumes $R_0=8.2\,$kpc.
}
\end{deluxetable*}

Table~\ref{tab:_fast star} gives an overview of the unusually fast stars (R$\leq20\arcsec$, further out extremely fast stars are not reliably measured) in our sample. ID 787 is the star which 
\citet{Schoedel_09a} called a runaway candidate.
We find a proper motion that is 3.8$\,\sigma$ smaller, and derive a rather normal radial velocity from its late-type spectrum, 
making this star considerably less noteworthy. 
The difference in proper motion is probably due to our better distortion correction at the edge of the field of view
(Section~\ref{sec:mass_smbh}). 
The star with the highest 3D velocity is the giant ID 4258, but it is (projected) close to another bright late-type star, such that
the radial velocity error might well be larger than indicated.
The outlier fraction appears to be smaller than in the maser sample (Figure~\ref{fig:_velocity_his}), but at smaller radii the dispersion is higher and thus
the identification of outliers is more difficult there.

\section{Discussion}
\label{sec:discussion}

We discuss our results in the context of previous results for the nuclear cluster of the Milky Way and of other galaxies.

\subsection{Central Low Dispersion Problem and SMBH Mass}
\label{sec:mass_smbh}

Row~1 in Table~\ref{tab:_mass_iso1} shows that fitting for the SMBH mass with a power law profile yields a mass which
is much smaller than the estimates from stellar orbits \citep{Gillessen_09,Ghez_09}. 
We now look closer at that problem.
In order to reduce rotation or flattening influence in the result, 
 we use here only the data in the central 27$\arcsec$, where the assumption of 
spherical symmetry is fulfilled approximately. 
The same effect also occurred in 
earlier works \citep{Genzel_96,Trippe_08}. However, recently \citet{Schoedel_09a} and \citet{Do_13b} obtained 
higher masses from Jeans modeling of old stars.

\citet{Do_12} obtained from the three-dimensional motions
of 248 late-type stars in the central 12$\arcsec$  masses  between M$_\bullet=3.77^{+0.62}_{-0.52}\times 10^6 M_{\odot}$ and  M$_\bullet=5.76^{+1.76}_{-1.26}\times 10^6 M_{\odot}$, consistent with
the orbit-based estimates considering the R$_0$ of the fits. They obtained higher values with anisotropic spherically symmetric Jeans modeling than with isotropic spherically symmetric Jeans modeling. These fits obtain isotropy in the center, but tangential anisotropy  further out. The latter is probably spurious, see Section~\ref{sec:_anisotropy}.
The lowest value is obtained in isotropic modeling.

The break radius of \citet{Do_13b} is large and inconsistent with our data, see Figure~\ref{fig:_densities_1}.
We test its influence on the obtained mass.
We therefore select from our data the bins inside of 12$\arcsec$ and fit the SMBH while fixing all other parameters to fit number 1 of \citet{Do_13b}. We obtain  M$_\bullet=3.77\pm0.09\times 10^6 M_{\odot}$ fully consistent with \citet{Do_13b}. (Our error is smaller because we had to fix most parameters.) Thus, our motions are consistent with the motions of \citet{Do_13b}. Also the binning has no relevant influence. 
The SMBH mass increases to about $4.5\times10^6 M_{\odot}$ when fitting our profiles within 220$\arcsec$ with a single Nuker profile. In this and the previous fit the cluster mass is zero. When the cluster is included with constant M/L we obtain  M$_\bullet=3.33\pm 0.00|_{\mathrm{tracer\,profile}} \pm0.34|_{\mathrm{dispersion}}\times 10^6 M_{\odot}$ with the Do profile and 
 M$_\bullet=2.86\pm0.10|_{\mathrm{tracer\,profile}}\pm0.23|_{\mathrm{dispersion}}\times 10^6 M_{\odot}$ 
 with our density data. Note that for checking whether the tracer profile is important, only the error which is caused by the tracer profile is relevant. Thus, the reason for the high mass
 of \citet{Do_13b} is twofold: the tracer profile and the extended mass. 
 In conclusion, not only high quality dynamic data is important but also good data of the surface density profile. Further, all components, also the extended mass need to be fit, also in the inner 12$\arcsec$ due to projection effects.

\begin{figure*}
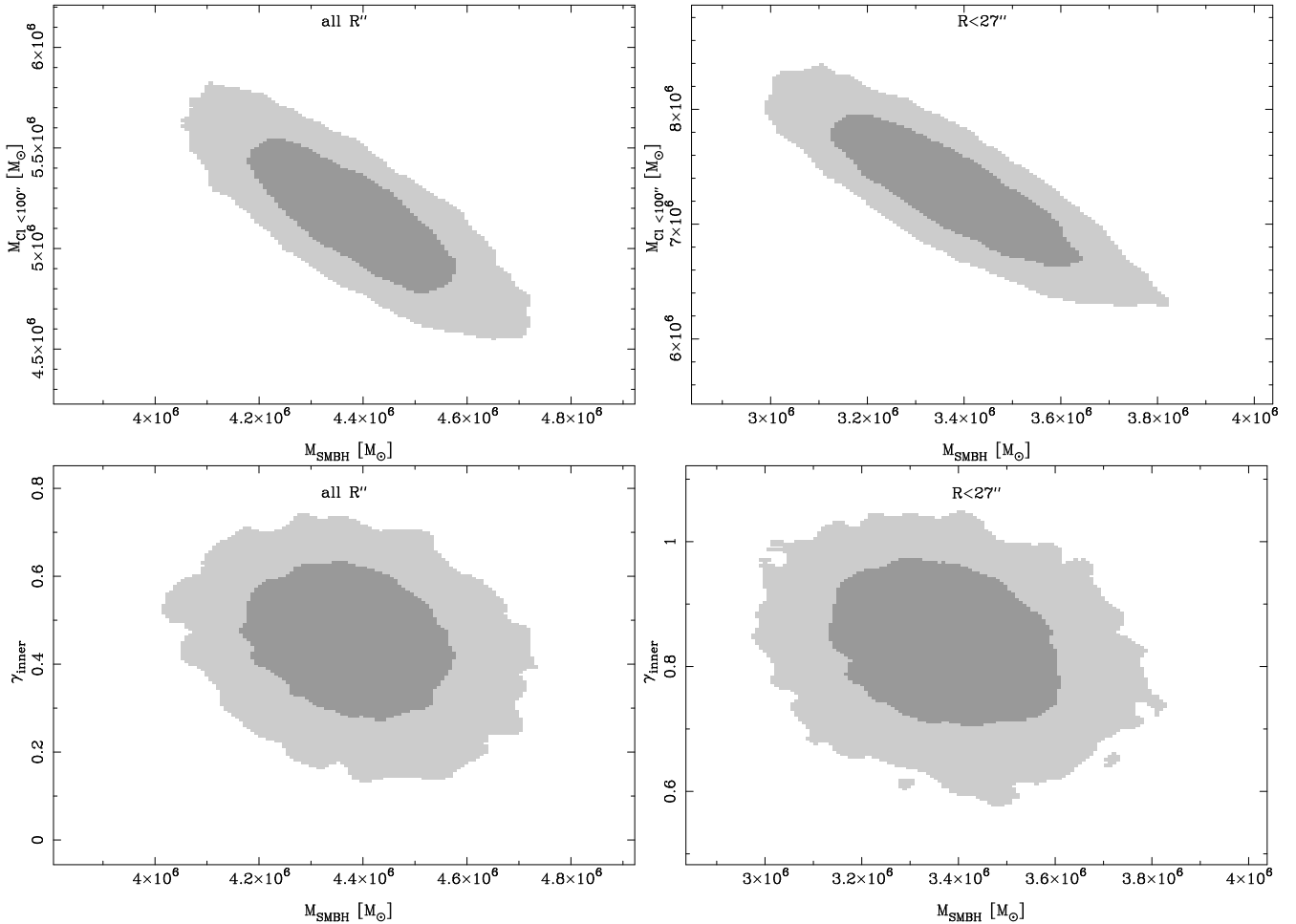

\begin{center}
\includegraphics[width=0.74 \columnwidth,angle=-90]{f18a.eps} 
\includegraphics[width=0.74 \columnwidth,angle=-90]{f18b.eps} 
\includegraphics[width=0.74 \columnwidth,angle=-90]{f18c.eps} 
\includegraphics[width=0.74 \columnwidth,angle=-90]{f18d.eps} 
\caption{M$_{\mathrm{SMBH}}$ correlation for model 2 and 29 (Table~\ref{tab:_mass_iso1}).
The top row shows the correlation with cluster mass, the bottom the correlation with  the inner slope $\gamma_\mathrm{in}$. 
Both models use the star counts as tracer density and assume a constant mass to light ratio. The left (model 2) use the full radial range
of the dynamics. The right (model 29) use only the data inside of 27$\arcsec$.
 The dark gray shows the 1$\,\sigma$ area from MCMC,  the light gray the 1$\,\sigma$ area.
} 
\label{fig:corr_plot3}
\end{center}
\end{figure*}

\citet{Schoedel_09a} obtained from isotropic Jeans modeling $M_\bullet=3.55 \times 10^6 \,M_\odot$, larger than 
our estimate and the earlier works. For their assumed distance of $R_0=8\,$kpc the SMBH mass of \citet{Gillessen_09} 
is within the 90\% probability interval of \citet{Schoedel_09a}, but in their anisotropic modeling it is excluded by 99\%. It is interesting to clarify the differences between our analysis and \citet{Schoedel_09a}, using their publicly available data set.  
\begin{itemize}
\item Using the \citet{Schoedel_09a} data, their tracer profile and our two mass parameterizations we get results consistent with what these authors found. For example with the power law and R$_0=8\,$ kpc we obtain M$_\bullet=3.33\pm0.56\times 10^6 M_{\odot}$. With our normal R$_=8.2$ kpc the mass increases to M$_\bullet=3.59\pm0.60\times 10^6 M_{\odot}$.
Hence, the details of the modeling only play a minor role.  
\item 
Using our dynamical data within 27$\arcsec$ and the tracer profile of \citet{Schoedel_09a} we retrieve $M_\bullet=2.88\pm0.48\times10^6 M_{\odot}$ repeating the previous fit with R$_0=8.2$ kpc. That decrease in mass is not because we add also radial velocities, in the opposite: 
when we use only our proper motion the mass decreases slightly to $M_\bullet=2.63\pm0.60\times10^6 M_{\odot}$. When we fit our density data inside the break with a single Nuker profile, in which $\gamma=0.5$ but the rest is free, we get $M_\bullet=2.34\pm0.60\times10^6 M_{\odot}$. 
Thus, the use of the \citet{Schoedel_09a} tracer profile increases that SMBH mass by about $0.5 \times10^6$ $M_{\odot}$.
The use of different proper motion data sets changes the mass by about $0.7 \times10^6$ $M_{\odot}$.
Therefore, differences in the tracer profiles are besides the dispersion data
the main reason for the differences in the mass obtained between our work and \citealt{Schoedel_09a}.
\end{itemize}

We now investigate the most important reason, namely the higher dispersions in the data of  \citealt{Schoedel_09a}.
 
\begin{enumerate}
\item The largest relative differences occur within $R<2.5\arcsec$. There, \citet{Schoedel_09a} still have a few early-type stars
contaminating their sample, since they used the spectroscopy-based, but slightly outdated star list in \citet{Paumard_06} and 
the photometric identifications of \citet{Buchholz_09}. Four certainly early-types are contained in their sample.
For example, the well-known fast  early-type
star S13 \citep{Eisenhauer_05}, for which an orbit is known, is part of their sample. In contrast, we use in the central 2.5$\arcsec$  
only stars which we positively identify spectroscopically as late-type stars. The issue is critical in the center,
since the early-type stars are more concentrated toward the SMBH than the late-type stars and thus
show a higher dispersion in the center. We obtain inside of 1.2$\arcsec$ $\sigma=9.44\pm1$ mas/yr using the data of \citet{Schoedel_09a} and $\sigma=6.92\pm0.69$ mas/yr from our proper motions.
Removing the early-type stars from the \citet{Schoedel_09a} sample
yields a central dispersion consistent with our value.
\item At R$>$15$\arcsec$ differences occur again in the proper motion data. This could be caused by differences in the distortion correction which is more important at large radii.
An imperfect distortion correction enlarges dispersions artificially, and thus a smaller measured value is more likely to be correct. 
\end{enumerate}
Overall, we believe that our dispersions are more reliable  than the ones of \citet{Schoedel_09a}. This is also supported by the fact that  we get an extended mass of $M_{100\arcsec}=5.65\pm 2.0 \times10^6 M_\odot$, consistent with our best value, 
when we restrict our dynamics data to the range R$<$27$\arcsec$ (for fixed SMBH mass with the count profile and constant M/L) in contrast
to \citealt{Schoedel_09a} (Section~\ref{sec:oth_GC}).

 Why do our and other attempts of Jeans-modeling fail to recover the right SMBH mass? 
The direct cause is that our measured dispersion within 10$\arcsec$ is smaller than the dispersion in our (isotropic) models
which obtain the right black hole, see Figure~\ref{fig:_fits_iso3}. 
What is the reason for the dispersion difference and thus of the low mass?

\begin{itemize}
\item Neglecting anisotropy is probably not the reason. While \citet{Do_13b} obtain higher SMBH masses in that case, that is mainly caused by the R$_0$-M$_{\bullet}$ relation. \citet{Do_13b} obtain R$_0=8.92$ kpc in their free fit, a distance larger than all recent measurements \citep{Genzel_10}. While the data of \citet{Schoedel_09a} is problematic in some radial ranges like the very center, we agree about anisotropy: we both get that the cluster is in the inner part slightly (not significant)  tangential anisotropy. Central tangential anisotropy decreases mass estimates, see e.g. \citet{Genzel_00}. Therefore,  \citet{Schoedel_09a} obtain a slightly smaller SMBH mass with anisotropic modeling. We therefore assume that the inclusion of anisotropy likely would decrease the obtained SMBH mass slightly.

\item The binning and the dispersion errors are probably not the reason. Our different binnings obtain all masses between $3.25\times10^6$ and  $3.38\times10^6$ M$_{\odot}$ for the SMBH using the count profile, a constant mass to light ratio and dynamics within 27$\arcsec$.
That makes it also less likely that we underestimated the dispersion errors, since in each of the binnings we use a slightly different
bias correction, but the results are consistent. The errors are easy to calculate, especially in the center, which is most relevant for the SMBH mass, since measurement errors are significantly smaller than
 Poisson errors.

\item For checking whether we can recover the correct SMBH mass for another $R_0$, we use, as previously, the SMBH mass-$R_0$ relation of \citet{Gillessen_09} for another $R_0$. Even for an unrealistically large value of
$R_0=9\,$kpc the SMBH mass-$R_0$ relation is not recovered. Thus, the SMBH mass underestimation is largely independent of the distance.

\item  The power law  extended mass parametrization 
 yields a smaller mass than using a constant M/L: 
$M_\bullet=(2.68\pm 0.51) \times 10^6~M_\odot$  for a free power law slope and $M_\bullet=(2.98 \pm 0.17) \times 10^{6}~M_\odot$ for a Bahcall-Wulf cusp of $\delta_M=1.25$.
While a constant mass to light yields  $M_\bullet=(3.37 \pm 0.16) \times 10^{6}~M_\odot$. Hence, a core in the stellar mass
instead of a Bahcall-Wulf cusp may be part of the solution.

\item \citet{Trippe_08} and \citet{Genzel_10} argued that a central core-like structure introduces a bias toward low SMBH masses.
This, however is only applicable when a tracer profile without core is used for cored data. If the correct profile contains a core and
is modeled as such, the central core only increases the error on the central mass. In our models such an uncertainty is included via the free 
density profile, yet the SMBH mass falls outside of the error band. In Figure~\ref{fig:corr_plot3}
it is visible that the correlation of $\gamma_{\mathrm{in}}$ and the SMBH mass is weak. It is not possible to obtain the right mass by changing $\gamma$ within our model.
Still, the fact that our flux and star count profiles are not consistent indicates that their errors are probably underestimated. 
Neither produces the right mass, but maybe the true profile is none of them.  A profile with a large enough core radius might
produce the right mass. Another possibility is that our functional form is inadequate. Perhaps a profile works with a large radial transition region, in which the profile has a constant slope, 
somewhat steeper than in the center.

\item   
Introducing a flat true core is unlikely to solve the issue. It is not only inconsistent with the apparent isotropy, 
it also under predicts the number of late-type stars with orbits. 

\item Also the flattening influences the SMBH mass, both by the tracer profile, and by reducing the dispersion in most dimensions,
see  \citet{Chatzopoulos_14}. Also well inside of 27$\arcsec$ the dispersion is different in l and b, indicating that 
flattening is important also there.
We test that case by fitting the data within 27$\arcsec$ with the flattened 2-integral model of  \citet{Chatzopoulos_14}.
We obtain $M_\bullet=(3.54\pm 0.18) \times 10^6~M_\odot$. This error also includes the distance uncertainty.
\end{itemize}

Summarizing, we have tested several potential solutions to the mass bias. For some we have shown quantitatively that they alone 
cannot correct the bias. Others would requires modeling which goes beyond this work. 
Likely, several together are necessary
for a solution.
The absence of a dark cusp and another more complex tracer profile
are maybe the most likely solution.
 Perhaps maximum likelihood modeling
is needed for our discrete data set \citep{Dsouza_13}, in particular for properly incorporating the  information the fast stars carry.

\subsection{Comparison with the M$_\mathrm{cluster}$ Literature}
\label{sec:oth_GC}

Overall, most M$_\mathrm{cluster}$ values from the literature are similar to our values. 
The comparison with other works needs some care, since different values for $R_0$ have been used, and hence
we scale to $R_0=8.2\,$kpc.
 We correct the masses in the literature to our distance using a mass scaling with exponent 1 for purely radial velocity based masses, and 
exponent 3 for purely proper motion based ones. 
We extrapolate from our best estimate $M_{100\arcsec}$ in two ways: 
 on the one hand we use a broken power law with break radius 100$\arcsec$. The inner slope is $\delta_\mathrm{M}=1.232$, the 
outer is $\delta_\mathrm{M}=1.126$. (That implies we use inside of 100$\arcsec$ the average $\delta_\mathrm{M}$ of row~3 and 4 of Table~\ref{tab:_mass_iso1} and outside the average $\delta_\mathrm{M}$ of row~11 and 12. The first uses dynamics between 10 and 100$\arcsec$, the second dynamics outside of 10 $\arcsec$.)
 On the other hand, we use one of the preferred fits with constant M/L (Row~5 in Table~\ref{tab:_mass_iso1})
Most other mass profiles are similar to one of these cases. 
We quantify the quality of the comparisons only when errors are given in literature and there is not more than one value in a source without clear preference for one.

\subsubsection{\citet{Schoedel_09a}}

The possible mass range of \citet{Schoedel_09a} within $1\,$pc is $(0.5-2.2)\times10^6M_{\odot}$, similar to our result. 
However, when assuming a constant M/L and isotropy they obtain a mass of $1.6 \times 10^6 M_{\odot}$,
larger than our value of $0.6\times 10^6\,$M$_{\odot}$ for the same assumptions. At 100$\arcsec$ their profile yields a
rather large mass of $16 \times 10^6\,$M$_{\odot}$. 
The main reason for the difference is their larger dispersion outside of 15$\arcsec$ (Section~\ref{sec:mass_smbh}).

\subsubsection{\citet{Trippe_08}}
At 100$\arcsec$ \citet{Trippe_08} found a mass that is roughly a factor three larger than ours. 
The difference is due to the high rotation at large radii as a result of their selective use of data from \citet{McGinn_89}.
At $1\,$pc where rotation is negligible they found a value of $1.2 \times 10^6 M_\odot$ (their Figure~14, gray dashed curve),
similar to our mass in the power law case.

\begin{figure*}
\begin{center}
\includegraphics[width=1.50 \columnwidth,angle=-90]{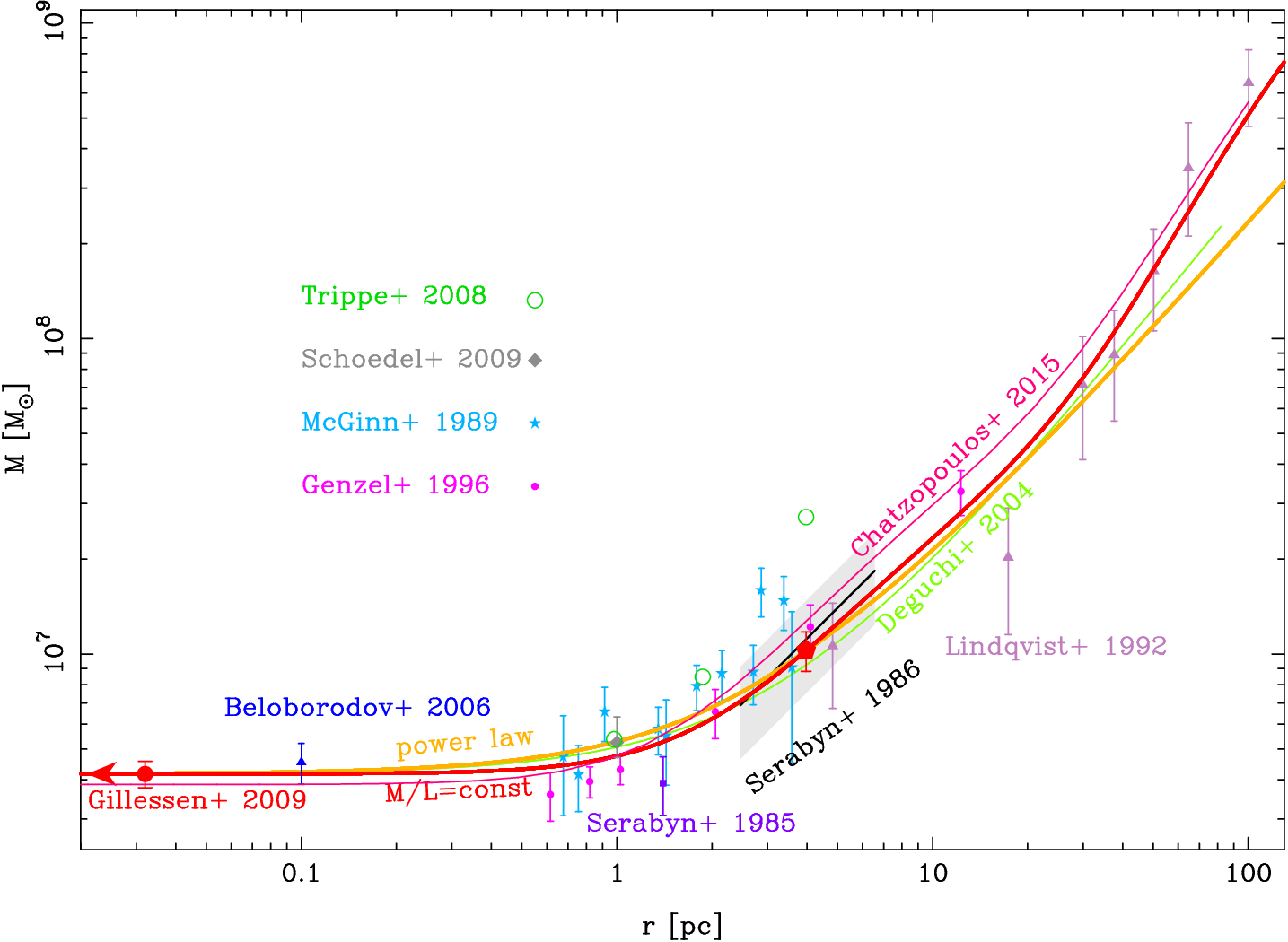}
\caption{Cumulative mass profile of the GC. The main measurement of this work is the red pentagon at 4 pc, through
which the profiles for the power law or constant M/L case pass. The latter is slightly preferred. The stellar orbits based value from \citet{Gillessen_09} is at about 0.002 pc. 
\citet{Beloborodov_06} used an orbit roulette technique to obtain an enclosed mass.
In the work of \citealt{Schoedel_09a} (gray diamond) no formal error is given, we show the largest range mentioned.
For the masses from \citealt{Trippe_08} (green open circles) no errors are given. The value from \citet{Serabyn_85} is the violet square.
For \citealt{Serabyn_86} the thick black line is the value and the light gray area gives the error range.
From \citealt{Lindqvist_92b}  (light violet triangles) and \citealt{McGinn_89} (green stars) we use the Jeans 
modeling values. In case of \citealt{Deguchi_04} (light green line) we show their extended mass model fit using the Boltzmann equation, 
adapting to our assumed SMBH mass.
 We plot Jeans-modeling based values from \citealt{Genzel_96} (pink dots).
We also plot the two-integral fit of  by \citet{Chatzopoulos_14} (pink curve).
We omit the values from \citet{Genzel_96} and \citet{McGinn_89}  inside of 0.55 pc, since more accurate values are available there.
} 
\label{fig:_mass_lit1}
\end{center}
\end{figure*}

\subsubsection{Inner Circumnuclear Disk}

The gas in the in circumnuclear disk (CND) can be used to obtain a mass estimate.
\citet{Serabyn_85} used the emission of [NeII] 12.8 $\mu$m from gas streamers 
 at the inner edge (including the Western arc) of the CND to obtain its rotation velocity. They found   
  $ (3.9 \pm 0.8) \times 10^6 M_{\odot}$ at 1.4 pc. 
  That is 1.7$\,\sigma$ and  2.4$\,\sigma$ less than our constant M/L and  power law estimates. 
Their mass is smaller than the mass of the SMBH.
The projected rotation velocity of the inner CND of $100\,$km/s is well-determined \citep{Genzel_85,Serabyn_85,Guesten_87,Jackson_93,Christopher_05},
and hence the reason for the discrepancy has to be in the model assumed.
 Conceptually, the mass derivation of \citet{Serabyn_85} was simple: they assumed circular motion of all the gas in one ring
 with an inclination of $60-70^{\circ}$. 

In the HCN J-0 data of the inner CND in \citet{Guesten_87} the 
velocities in the southern and western parts follow the model of \citet{Serabyn_85}, but not in the northern and 
eastern parts. The latter can be described by a less inclined ring ($\approx45^{\circ}$) with an intrinsic rotation velocity of 
$137\pm8 $ km/s. This velocity results in a total mass $(6.1\pm 0.7)\times10^6 M_{\odot}$ somewhat larger than our estimates. Using an average inclination 
between \citet{Serabyn_85} and \citet{Guesten_87} would hence yield a mass estimate very similar to ours.

However, 
\citet{Zhao_09} find an inclination consistent with \citet{Serabyn_85}. 
Further \citet{Zhao_09} show that a single 
orbit, even slightly elliptical, cannot fit all gas streamers in the Western arc, and hence the inner CND is likely more complicated than 
assumed by \citet{Serabyn_85}.

Another possibility to solve the discrepancy between our result and the result
of \citet{Serabyn_85} is that the assumption of \citet{Serabyn_85} that the line width is due to nongravitional processes
may be wrong. If the line width is due to asymmetric drift \citep{Binney_08} the true circular velocity $v_{\mathrm{c},\mathrm{true}}$ would be higher
than the measured velocity $v_{\mathrm{c},\mathrm{obs}}$ \citep{Kormendy_13}:  \begin{equation}
v^2_{\mathrm{c},\mathrm{true}}=v^2_{\mathrm{c},\mathrm{obs}}+x \times \sigma^2 
\end{equation}
The factor $x$ depends on the disk profile and other not well known parameters of the 
inner CND. It is between 1 and 3. 
From $\sigma=35\pm5\,$km/s \citep{Serabyn_85} follows for $x=$3
an increase in the mass by 30\% to $ (5.1 \pm 1.04) \times 10^6 M_{\odot}$. This would be consistent with our constant M/L estimate of $5.27 \times 10^6 M_{\odot}$.

 \subsubsection{Outer Circumnuclear Disk}

\citet{Serabyn_86} used CO 1-0 in the outer parts of the CND for 
estimating the mass using the same method as \citet{Serabyn_85}.
This measurement is consistent with our mass. The agreement argues
that other forces besides of gravity are not relevant for at least the outer gas dynamics.

 \subsubsection{\citet{Rieke_88} }

At radii larger than $7\,$pc the gas velocities get unreliable \citep{Serabyn_86,Guesten_87}, but  stellar velocities are available.
\citet{Rieke_88} used a few bright stars and found a dispersion of $75\,$km/s between 6$\arcsec$ and 160$\arcsec$, independent of radius. With our
data we can reject the hypothesis of \citet{McGinn_89}, proposing that a magnitude effect causes the surprisingly flat
dispersion curve. Possibly, \citet{Rieke_88} were limited by the low-number statistics.

\subsubsection{\citet{McGinn_89} }

These authors covered a similar area in size to ours, using integrated velocities and dispersions in large beams.
Broadly their dispersion data is consistent with our data, and they measure a somewhat stronger radial dispersion trend.
They obtain masses similar to our estimates and also find a too small SMBH mass, like most works that apply free Jeans modeling.
At the outer edge the masses of \citet{McGinn_89} are somewhat larger than ours, due to the high rotation velocity in this work
(Section~\ref{sec:rotation}).

\subsubsection{\citet{Lindqvist_92b}}

Using maser velocities and Jeans modeling out to much larger radii \citet{Lindqvist_92b} obtained 
masses consistent with our model. This is not surprising, since we include their velocities in our data set.
The main discrepancy occurs around 17 pc, where their estimate is only half of our extrapolation. 
On the one hand, our modeling has not much flexibility at these large radii. On other hand, their mass profile had possibly too much freedom since the light profile shows no dip there.
 In both our and the analysis of \citet{Lindqvist_92b} the deviation from spherical symmetry is not considered,
which might lead to significant changes at large radii.

\subsubsection{\citet{Deguchi_04} }

\citet{Deguchi_04} used also maser velocities, but in a somewhat smaller area than \citet{Lindqvist_92b}. 
They prefer a result based on a new method developed from the Boltzmann equation. This method includes rotation, but assumes
spherical symmetry and a fixed extended mass slope of $\delta_\mathrm{M}=$1.25. 
To compare their extended mass result with
our work we need to adapt their result to our SMBH mass, which is  within their 1.5$\,\sigma$ range.
In the inner parsec their mass of $(0.90\pm 0.07)\times 10^6 M_{\odot}$ is consistent with our range, but
at 100$\arcsec$ their estimate of $(5.06 \pm 0.39) \times 10^6 M_{\odot}$ deviates by about 1.5$\,\sigma$ from our estimate. 
Due to their fixed slope their mass deviates also at larger radii from the results in \citet{Lindqvist_92b}.
A fixed slope does not describe the stars density in the GC well out to 80 pc.

\subsubsection{\citet{Genzel_96}}

This work obtained the mass distribution out to 20 pc from Jeans modeling of radial velocities,
using  literature radial velocities in combination with newly measured velocities in the center. Also these authors
under predict the SMBH mass. Beyond that, their results are consistent with ours.

\subsubsection{\citet{Chatzopoulos_14}}

As discussed also in Section~\ref{sec:jeans_modelling}, the modeling of \citet{Chatzopoulos_14} obtain a larger mass than we do due to the flattening. That mass is also larger than most other estimates between 2 and 50 pc. The reason is usually the same - the other models are not flattened. For example, that is the reason why \citet{Deguchi_04} obtain a smaller mass.

\subsubsection{Works Using the Light Distribution}

It is possible to obtain mass estimates of the extended mass by using the flux and its distribution of the GC.
It is however difficult due to  
following systematic uncertainties in M/L:  the star formation history, the extinction toward the GC and the IMF of the stars. Thus, in the past \citet{McGinn_89,Lindqvist_92b,Launhardt_02} used a M/L obtained at a selected radius with dynamics
together with the flux profile to estimate the mass of the nuclear cluster. Thus the broad agreement of them with our work shows that the mass of \citet{McGinn_89,Lindqvist_92b,Genzel_96,Genzel_97} agrees with our mass.
With the recent agreeing measurements of extinction \citep{Nishiyama_06b,Fritz_11,Schoedel_09b}
 and star formation history  \citep{Blum_02,Pfuhl_11} two uncertainties in a purely light derived mass are now reduced. 
Still the IMF essentially cannot be constrained using light information alone \citep{Pfuhl_11}.
 Since the IMF of the GC is an interesting subject,  we reverse the argument, 
and use the mass to constrain the IMF (Section~\ref{sec:mass_to_light}).

Recently, \citet{Schoedel_14} used their light decomposition to infer the total mass of the nuclear cluster
of $2.1\pm0.4\times10^7\, M_\odot$. 
That is only half of our estimate but within the error for the total mass.  The $M/L=0.5\pm0.1$ of \citet{Schoedel_14}
is nearly identical with our dynamic value (Section~\ref{sec:mass_to_light}). 
While \citet{Schoedel_14}
used IRAC2 instead of Ks the impact on M/L is rather small (at most 38\%) for the GC star formation history (Section~\ref{sec:mass_to_light}).

\subsection{Mass Cusp or Core?}
\label{sec:mass_cusp_core}

The distribution of old stars does not show the expected central cusp \citep{Bahcall_76} with the inner $\delta_\mathrm{L}=1.25$ slope
 \citep{Buchholz_09,Do_09,Bartko_10}.
The stars follow a shallower slope in the center, 
which can either be a true core ($\delta_\mathrm{L}=3$) or a shallow cusp ($\delta_\mathrm{L}>2$, \citealt{Do_09}).
Also the second case we call core here, since it has much less light in the center than a cusp.
By comparing 
our power law mass models (with a central cusp) and the constant M/L models (with a core)
we can constrain the central mass distribution.

\begin{itemize}  
\item Given that a core can reduce the low dispersion bias, the Jeans modeling of our data 
 favors a core over a cusp. The same is found by \citet{Schoedel_09a},
who used a unrealistically large, fixed break radii for light and especially mass. 
Our results show that the preference of a smaller slope in the center is 
not only due to the large break radius in \citet{Schoedel_09a}.
\item The presence of a warped disk of young stars  \citep{Loeckmann_09a,Bartko_09}
yields an additional constraint.
 \citet{Ayse_13} found that with
an isothermal cusp of 
M($r<0.5$pc)$=10^6\,$M$_{\odot}$ and a flattening of q$=0.9$ an initial warped disk is too quickly destroyed 
 and would not be observable today. For constant M/L we obtain a smaller mass of M($r<0.5\,$pc)$\approx0.15 \times 10^6\,$M$_{\odot}$, which reduces the torque by a factor five, allowing the disk to survive longer. This conclusion
is not firm, since the warping need not be primordial, and a reduced central flattening might also yield less
warping \citep{Kocsis_11}. 
\item The most reliable mass measurement around 1.4 pc apart from 
our analysis, that is
\citet{Serabyn_85}, is more consistent with a core than with a cusp.
\end{itemize}

Overall, it appears likely that the mass profile shows a central core, but better modeling including a solution of the mass bias 
is necessary. 
Another route for constraining the mass profile worth follow up is using direct accelerations of stars at radii between 1$\arcsec$ and 
7$\arcsec$ with
GRAVITY \citep{Eisenhauer_08}.

Our preference for a core in the mass profile is interesting for theories which aim at explaining the missing light cusp.
In some theories the resolvable stars (giants) form a core while dark components (remnants or main sequence stars) 
form a mass cusp. Thus, our result means that mass segregation is less likely to solve the riddle of missing light cusp. 
\citet{Keshet_09} derive a similar conclusion from different arguments. Another mechanism that destroys a light cusp, but not a 
mass cusp, is the destruction of giants by collisions \citep{Dale_09}, and hence our results also disfavor that model.
One model that could work is presented in \citet{Merritt_10} who propose that a 
relatively recent binary black hole (merger) ejected stars. This process would yield a core both in the
light and the mass profile. Still, one would need to fine-tune the timing of such a model, since the early-type stars
are concentrated toward the center, but the giants and even the younger red (super-) giants with ages down to 20 
Myrs \citep{Blum_02,Pfuhl_11} are not.

\subsection{Cumulative Mass Profile}
\label{sec:cum_mass}

\begin{figure}
\begin{center}
\includegraphics[width=0.70 \columnwidth,angle=-90]{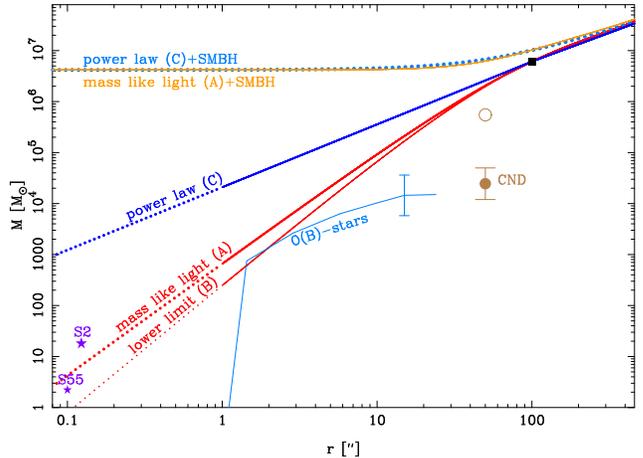}
\caption{Cumulative mass profiles, for a power law model or assuming M/L$_{\mathrm{Ks}}=$const. The normalization
is done for $M_{100\arcsec} = 6.09 \times 10^6 M_\odot$ (black square).
Other masses shown are the ZAMS cumulative mass distribution of the O(B)-star population \citep{Bartko_10}
and with a filled disk the CND from
 \citet{Etxaluze_11,Genzel_85, Mezger_89},  
and
 \citet{Requena_12}. The open circle is the CND mass estimate from \citet{Christopher_05}.
}
\label{fig:_nc2}
\end{center}
\end{figure}

Another way of expressing our results is the cumulative mass profile. In Figure~\ref{fig:_nc2} we show three
cases:
\begin{itemize}
\item We use a constant M/L model with the best fitting inner slope of -0.81 (Row 5 in Table~\ref{tab:_mass_iso1}, model A).
\item We use a constant M/L model with the shallowest possible inner slope of -0.5 (and otherwise row 5 in Table~\ref{tab:_mass_iso1}, model B).
\item We use a power law model with a slope of $\delta_\mathrm{M}=1.232$ (model C), the average of row 3 and 4 in Table~\ref{tab:_mass_iso1}\footnote{Outside of 100$\arcsec$, a range not relevant here, model C uses
 $\delta_\mathrm{M}=1.126$, the average  of rows 11 and 12 in Table~\ref{tab:_mass_iso1}.}. 
\end{itemize}
We normalize these three models to have the overall best-fitting value of $M_{100\arcsec} =6.09 \times 10^6 M_\odot$. Tabulated values of the profiles can
be found in Appendix~\ref{sec_val_cum_mass}. The largest mass difference between the M/L$=$const and the power law model is reached
 at about 35$\arcsec$ with $0.59\times 10^6 M _{\odot}$. 
Since this number exceeds our mass error, it is possible to detect dynamically the cusp, provided the low dispersion bias can be solved.

Extrapolating the constant M/L models down to the regime of the S-stars S2 and S55/S0-102 \citep{Meyer_12} 
yields a mass there, which is smaller than that of the individual stars. In this regime, the S-stars dominate, and
taking into account their Salpeter-like IMF \citep{Bartko_10}
or a potential mass-segregated cusp of stellar remnants 
\citep{Freitag_06,Hopman_06} 
would make the dominance even stronger. 
For power law models it is not clear whether the old stars dominate the mass in the central 
arcsecond. 

Outside the central arcsecond, the mass of the disks of early-type stars 
\citep{Bartko_09,Lu_09} may be important. We derive the spatial distribution from Figure~2 in 
\citet{Bartko_10} and globally deproject to space distances with a factor 1.2 the given projected distances.
We estimate the total ZAMS mass of that population to $1.5\times 10^4\,M_\odot$ \citep{Bartko_10}, 
dominated by the O-stars given the 
 top-heavy shape of the IMF. Across the literature \citep{Paumard_06,Bartko_10,Lu_13} the disk mass is uncertain by a factor 2.5. 
At $r=2\arcsec$ the O-stars might be
comparable in mass to the old stars, if the constant M/L model is correct.

Further out, the only other component is the circumnuclear disk. Most 
publications 
\citep{Genzel_85,Etxaluze_11,Mezger_89,Requena_12} agree that its
 mass is a few $10^4\,M _{\odot}$ and thus  irrelevant compared to the old 
stars. However, the mass of $\approx 10^6 M_{\odot}$ found by \citet{Christopher_05}
would be about a third of our mass estimate at that radius.

Our results also yield a new estimate for the sphere of influence of the SMBH \citep{Alexander_05}, the radial range within which
the extended mass is equal to the mass of the SMBH.
It is completely consistent with previous values \citep{Alexander_05,Genzel_10}. Using model A and B, 
we get r$_{\mathrm{infl}}=76.3\pm5.5\arcsec=3.03\pm0.25\,$pc.

\subsection{Mass to Light Ratio}
\label{sec:mass_to_light}

We now obtain the mass to light ratio from the fits to flux and dynamics with constant M/L.
The mass to light ratio is directly calculated from the output of these models. 
We obtain $M/L=0.51 \pm 0.12 \, M_{\odot}/L_{\odot,\mathrm{Ks}}$ using $M_{\mathrm{Ks}\,\odot}=3.28$ \citep{Binney_98}.
The error consists of 19~\% for the light, 10~\% for the mass, and 8~\% for the distance uncertainty.
The latter number 
follows from the distance uncertainty of 4.1\% \citep{Gillessen_13} multiplied by the exponent 1.83, which is the scaling M/L with
distance given that M scales like 3.83 with distance (Section~\ref{sec:jeans_modelling}).
Our M/L is consistent with the values in \citet{Pfuhl_11} and \citet{Launhardt_02}.
Our error, however, is smaller thanks to the smaller mass error. With the improved mass to light ratio we can constrain the IMF of the old stars 
further.

Firstly, we use the M/L to determine the IMF slope $\alpha$, we assume the respective star formation histories for the old stars (older than 10 Myrs) from \citet{Pfuhl_11}. Their results 
appear to hold also at the larger radii of interest here \citep{Blum_02}.
Already with the accuracy of M/L of \citet{Pfuhl_11} it was possible to discard extreme top-heavy IMFs ($\alpha>-0.6$)
like measured for the O-stars in the GC \citep{Bartko_10}. With our accuracy also the IMF slope
of $\alpha=-0.85$ \citep{Paumard_06} is excluded.

Secondly, we concentrate on variants of the usual -2.3 slope below one solar mass, these are the IMFs of \citet{Salpeter_55,Kroupa_01,Chabrier_03}. 
To check the result on dependence of the used stellar population model we use the models of \citet{Bruzual_03} and \citet{Maraston_05}.
In case of the \citet{Maraston_05} models we obtain 
 $M/L=0.75  M_{\odot}/L_{\odot,\mathrm{Ks}}$ and  $M/L=1.28  M_{\odot}/L_{\odot,\mathrm{Ks}}$ for Kroupa and Salpeter, respectively. 
For the \citet{Bruzual_03} models we get $M/L=1.04  M_{\odot}/L_{\odot,\mathrm{Ks}}$
and $M/L=0.76  M_{\odot}/L_{\odot,\mathrm{Ks}}$ for  Salpeter and Chabrier, respectively. 
The differences in stellar population models cause the differences of up to 20\%. (The stellar population model dependences increase with wavelength: at 4.5 $\mu$m (IRAC2), used by \citet{Schoedel_14} and \citet{Feldmeier_14}, the model of \citet{Bruzual_03} obtains $M/L=0.71  M_{\odot}/L_{\odot,\mathrm{IRAC2}}$  for Chabrier and the model of \citet{Bruzual_03} $M/L=1.01 M_{\odot}/L_{\odot,\mathrm{IRAC2}}$ for Kroupa.) 
Using the star formation history of \citet{Blum_02} instead of the of \citet{Pfuhl_11}
causes 18\% smaller M/L in Ks-band.

Recently, deviations from the Chabrier IMF  were found for different stellar systems:
on the one hand elliptical galaxies seem 
to have  a Salpeter or more bottom heavy IMF 
as \citet{Vandokkum_10,Conroy_12} concluded using  population modeling of integrated spectra and as \citet{Cappellari_06,Cappellari_12}
concluded  from advanced dynamic modeling.
Already a Salpeter IMF is in strong tension with our ratio. A more bottom heavy IMF  of e.g. a slope of -2.8 has an even larger M/L (A factor 1.57 more in M/L compared to Salpeter for the star formation history assumed in \citet{Cappellari_12,Conroy_12}
and is thus firmly excluded.

It seems that (metal rich) globular clusters have M/L smaller than Kroupa \citep{Kruijssen_09a,Bastian_10,Sollima_12}. This was found in the case of M31 by \citet{Conroy_12}
 using  population modeling of integrated spectra. In addition, \citet{Strader_11} found the same by using mass and light measurements.
Precisely, \citet{Strader_11} measured for solar metallicity ($\mathrm{[Z]}=0\pm0.3$) $M/L=0.39  M_{\odot}/L_{\odot,\mathrm{Ks}}$. 
This is already lower than our measurement, although the GC contains more medium old stars than these globulars. 
In an attempt to include these younger stars we reduce in our Kroupa and Chabrier models the mass of the stars $>5$ Gyrs to achieve 
$M/L=0.39  M_{\odot}/L_{\odot,\mathrm{Ks}}$
for these ages. We do not change the populations for the younger stars.
This model has the motivation that it describes a possible formation scenario of the nuclear cluster in which the majority
the old stars originate in big metal rich globular clusters \citep{Capuzzo_08} and later stars formed locally are added to it.
This model results in $M/L=0.31  M_{\odot}/L_{\odot,\mathrm{Ks}}$ deviating by 2$\,\sigma$ from our measurement.  
This argument against a globular cluster origin of the nuclear cluster is weakened by the fact that
when globulars arrive in the GC, they likely have a different M/L due to mass segregation. In case of M31 it is argued
that the low M/L is mostly due to a different IMF and not only due evolutionary effects \citep{Strader_11}. 
However, the preferred loss of low mass objects due to internal mass segregation and the outer tidal field
exists certainly \citep{Bastian_10,Sollima_12} and should be further enhanced close to the GC. 
The problem is that, while mass segregation first reduces the M/L, M/L is enhanced in late stages during the late remnant dominated state of
old populations \citep{Sollima_12,Marks_12}. Thus, without detailed modeling of mass segregation 
the expected M/L of a globular cluster dominated nuclear cluster is uncertain.

The GC shows a normal ratio of diffuse light to the total light \citep{Pfuhl_11}.  
This ratio measures the ratio of main sequence stars to giants and decreases monotonically from a bottom-heavy
to a top-heavy IMF.  Thus it excludes best a cluster dominated by remnants and giants - a possible state of globular
clusters inspiraling to the GC.

The Chabrier IMF  common in Galactic disk and bulge \citep{Bastian_10,Zoccali_00} is close to the M/L of the medium old and 
old stellar populations of the GC. However, the deviation from it is
1.7 $\sigma$ when using the most recent star formation history determination \citep{Pfuhl_11}. 
That deviation vanishes when the bias to small masses due to our spherical modeling is corrected, see \citet{Chatzopoulos_14}. In that case the ratio is  $M/L=0.76 \pm 0.18 \, M_{\odot}/L_{\odot,\mathrm{Ks}}$ matching well the Kroupa/Chabrier IMF.
Therefore a globular cluster origin of the nuclear cluster \citep{Antonini_12,Gnedin_14} is somewhat disfavored by its M/L and 
L$_{\mathrm{diffuse}}$/L$_{\mathrm{total}}$.

\subsection{The Nuclear Cluster of the Milky Way in Comparison}
\label{sec:mass_nucl}

\begin{figure}
\begin{center}
\includegraphics[width=0.70 \columnwidth,angle=-90]{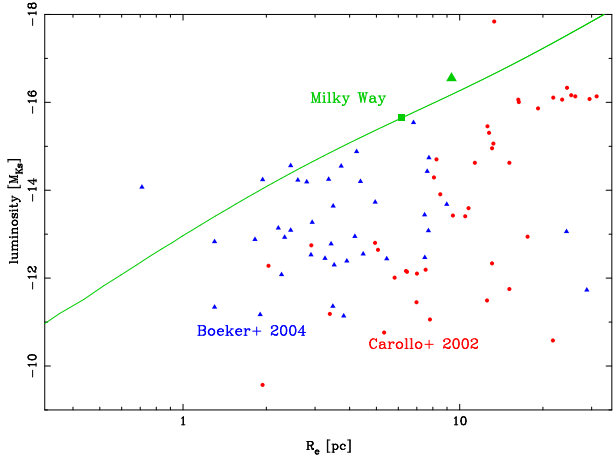} 
\caption{
Sizes and luminosities of nuclear clusters. We show the half light radii for clusters in late-type \citep{Boeker_04} and early-type
spiral galaxies \citep{Carollo_02}.
For the Milky Way we present the inner component of our GALFIT decomposition (green box), of the $\gamma$ model fitting (green triangle) and the total cumulative flux as function of the radius.
} 
\label{fig:_comp_nc}
\end{center}
\end{figure}

We now compare the nuclear cluster of the Milky Way with nuclear clusters in other galaxies.
The literature about the mass of other nuclear clusters \citep{Walcher_05,Barth_09} is sparse and biased toward brighter nuclear clusters. 
We therefore use mainly the light for comparison, for which ample data from HST imaging are available \citep{Carollo_02,Boeker_04}.
\citet{Boeker_04} studied late-type spiral (Scd to Sm)
while \citet{Carollo_02} concentrated on early-type spirals (Sa to Sbc).
 For color-correction we use $\mathrm{'H'}-\mathrm{Ks}=0.15$ and $\mathrm{I}-\mathrm{Ks}=1.1$ for \citet{Carollo_02} and \citet{Boeker_04}, respectively. 
 We compare the obtained magnitudes 
with our GALFIT decomposition for the nuclear cluster (Section~\ref{sec:flattening}), of the double $\gamma$ model fitting (Section~\ref{sec:fitting_profile_gam}).
In addition we use the cumulative flux as function of radius. The latter is not affected by decomposition uncertainties,  (Figure~\ref{fig:_comp_nc}).

The size of the nuclear cluster of the Milky Way is typical. 
It is smaller than many clusters in \citet{Carollo_02} and larger than most clusters in \citet{Boeker_04}. 
This simply might be a consequence from the fact that the galaxy type of the Milky Way is broadly
between the two samples.

It is visible in Figure~\ref{fig:_comp_nc} that the nuclear cluster of the Milky Way has an unusually high surface brightness.
Not only its characteristic brightness is high, but 
also the cumulative brightness profile lies above nearly all other clusters. 
What is the reason for that?
\begin{itemize}
\item
The young ($\approx 6$ Myrs) and medium old ($\approx 200$ Myrs) stars in GC are likely not the reason for this offset. They are subdominant 
compared to the old stars in the Ks-band \citep{Blum_02,Pfuhl_11} and Section~\ref{sec:total_luminosity}.
Further many nuclear clusters \citep{Rossa_06,Walcher_06} contain significant fractions of young or medium old stars.
 Thus, the M/L of the GC is probably typical.
\item
Extinction is probably not an issue although we correct for it. 
The spirals in the sample of \citet{Carollo_02} and \citet{Boeker_04} are seen more or less face-on. 
Further, the mean IR color in \citet{Carollo_02} yields a 
small extinction of A$_{160W}\approx0.2$. Also the spectroscopic study of \citet{Rossa_06} obtained small extinctions for the
same samples.
\item
The higher resolution of our data is also not the reason, since we compare the characteristic brightness within 4 pc, 
a size which can be resolved in most of the galaxies of 
\citet{Carollo_02} and \citet{Boeker_04}.  
\item
The nuclear cluster of the Milky Way has a projected mass density of $\log{\Sigma_e}=5.38 \pm 0.17$ [M$_{\odot}$/pc$^2$]. This is larger than for all 11 nuclear
clusters in the sample of \citet{Walcher_05}. 
\end{itemize}
We conclude that the main reason for high surface light density is the high stellar mass density in the nuclear cluster, 
compared to other galaxies.

For some near-by galaxies a more detailed comparison is possible.
For example M33 shows a central star density which is possibly larger than in the nuclear cluster of the Milky Way  \citep{Lauer_98}.
M33 does not contain a SMBH \citep{Merritt_01,Gebhardt_01}, and the old stars - if present at all - are outshined by young stars.
Some nuclear clusters show flattening and rotation, in case of metallicity the flattening is 16\% \citep{Lauer_98} combined with 
some rotation \citep{Kormendy_93}.  The large nuclear cluster in NGC4244  (R$_e\approx10\,$pc) is visually flattened and has 
$v_{\mathrm{rot}}/\sigma\approx1$ \citep{Seth_08}. NGC404 is a similar case \citep{Seth_10}.
Thus the flattening of the Milky Way nuclear cluster 
 is in the large range of possible shapes. 

 \citet{Carollo_99} finds that nuclear cluster with M$_V\lesssim-12$
are typically associated with signs of circumnuclear star formation, like dust lanes and an HII spectrum.
In this respect the cluster of the Milky Way with M$_V\lesssim-13$ seems typical for its nuclear disk.
Preferentially, bright nuclear clusters are observed together with  nuclear disks, 
 which might be explained 
by a physical coupling of the two systems.   
Since the GC is dominated by old stars this connection is likely not only valid for  the recent star formation event but also for the older
population. 

\section{Conclusions and Summary}
\label{sec:conclusions}

The nuclear cluster in the Milky Way is by far the closest and thus can be studied in more detail
than other nuclear clusters.
In this paper we investigate not only its center, but its full size to compare
it with such clusters in other galaxies.  
To that aim, we obtain its light profile out to 1000$\arcsec$ and measure motions in all three dimensions out to 100$\arcsec$,
expanding on the works of \citet{Trippe_08} and \citet{Schoedel_09a}.\\

\begin{itemize}
 \item 
We construct a stellar density map with sufficient resolution for two dimensional structural analysis out to $r=1000\arcsec$.
This map shows in the central 68$\arcsec$ an axis ratio of
 $q=0.80\pm0.04$. 
 Further out the flattening increases. 
We fit this map by two components, the inner is a Sersic, the outer a Nuker profile.
The more spherical and smaller component is the nuclear cluster, while the more extended component corresponds to the nuclear disk (stellar analog of the central molecular zone).
The decomposition is uncertain because the outer component is relatively bright.
With our preferred decomposition the nuclear cluster has a half light radius 
of 127$\arcsec\approx5.0\,$pc, a flattening of 1/0.80, a Sersic index of n$=1.5\pm0.1$, 
and a total luminosity of M$_{\mathrm{Ks}}=-15.5$. The outer component has a flattening of about 1/0.264 and contributes about 13\% to the projected central flux.
The size and luminosity of the nuclear cluster depend on the functional form of the assumed outer slope, which
is not well constrained by our data. A flatter profile like a $\gamma$ model yields r$_e\approx9\,$pc and 
a total luminosity of M$_{\mathrm{Ks}}\approx-16.5$.
\item Our dynamical analysis shows that
the proper motions are radially 
isotropic out to at least 40$\arcsec$. In the center the velocity distribution shows strong wings
as expected due to the presence of the SMBH. 
Assuming that the fastest stars in the center are bound, we set a lower bound to the SMBH mass of 
$3.6\times 10^6\,$M$_{\odot}$.
Assuming in addition that the black hole mass is known to 1.5\% \citep{Gillessen_09} for a given distance, we can estimate
 $R_0=8.53^{+0.21}_{-0.15}$ kpc. Our new radial velocities show that the projected rotation velocity increases only weakly outside of 30$\arcsec$,
and thus the nuclear cluster rotates less than assumed in \citet{Trippe_08}.

\item  We use the motions of more than 10000 stars for isotropic, spherically symmetric Jeans modeling. As a tracer profile we use either
 stellar number counts or the light profile, and the mass model is either a power law model or assumes constant M/L.
Forcing the mass of the SMBH to its known value
 we measure for the extended mass
a power law slope of $\delta_\mathrm{M}=1.18\pm 0.06$, consistent with the light profile. Our best mass estimate
is obtained at $r_{\mathrm{3D}}=100\arcsec$ as an average over the power-law and constant M/L models. We find
$M_{100\arcsec} = (6.09 \pm 0.53|_{\mathrm{fix} R_0}\pm 0.97|_{R_0} ) \times 10^6 M_{\odot}$ 
The error contains
contributions from the uncertain surface density data and from the uncertainty in $R_0$. 
Deviations from isotropy and spherical symmetry are not included in our error calculation. The most important
deviation is the observed flattening, a model which includes it increases the mass by about 47\%, see \citet{Chatzopoulos_14}.
Our $\gamma$-modeling yields a total cluster mass of
M$_\mathrm{NC}=(4.22 \pm 0.50|_{\mathrm{fix} R_0}\pm 0.67|_{R_0}) \times 10^7$  M$_{\odot} $. As in case of the light, the total mass is model dependent. A model with faster outer decay like an exponential, has a smaller total mass than our fit.

 \item The preference for a too small SMBH mass in the Jeans modeling can be interpreted as an argument for a central, core-like structure, unlike
 to the expected cusp \citep{Bahcall_76}. Hence, the deficit in the center would be present not only in the light, but also in the mass profile.
The missing mass cusp makes other explanations for the missing cusp in the light less likely, such as
 mass segregation and giant destruction 
\citep{Dale_09} less likely.
 However, we think that it is premature to draw this conclusion firmly. One would need to develop a model 
that fits the surface density data and achieves the right SMBH mass. That needs probably at least one of the 
following elements:  (i) a relatively large core-like structure in the tracer distribution,
which possibly can be found by non-parametric fitting of the density profile, 
or (ii) the inclusion of an outer background which contributes a significant number of stars 
also in the center. 

 \item  We obtain a mass to light ratio of $M/L=0.51 \pm 0.12 \, M_{\odot}/L_{\odot,\mathrm{Ks}}$. This
is 1.7$\,\sigma$ smaller than the value for a Chabrier IMF. However, since the mass error does not include all contributions, e.g. the contribution from the flattening is not included, it is consistent with a Chabrier IMF.

 \item The obtained half light radius of the nuclear cluster of 4 to 9 pc is typical compared to extragalactic nuclear clusters. However, it is brighter 
and has a higher light and mass density than most other nuclear clusters. 
Possibly, this large density is connected to the nuclear disk further out whose
surface brightness also seems high. 

 \item The abundance of young stars and molecular clouds in the nuclear disk and the nuclear cluster supports
 the idea that at least the young and medium old stars in the GC formed in-situ. 
 Both the nuclear cluster and the nuclear disk are flattened, although the strength of the flattening varies. The different amount of flattening can be interpreted as an argument for a different origin of the two components, a local origin for the disk and a globular cluster origin for the cluster.  The ratio of diffuse light to total light
appears to be normal, which contradicts a globular cluster origin, since they are likely dominated by stellar remnants when they arrive in the GC. 
Further, the close association of the bright nuclear cluster with the bright nuclear disk is suggestive of a common origin. Finally, the metallicity of most nuclear cluster stars is close to solar \citep{Cunha_07,Ryde_15,Do_15} agreeing with an in-situ origin.  Only 5 of the 83 stars of \citet{Do_15} are compatible with the metallicity [Fe/H]$\approx-0.75$ (\citet{Harris_96}, private communication with Christian Johnson) of bulge globular clusters. Because metallicities are an important discriminator, it would be good to confirm the medium resolution metallicities of \citet{Do_15} with higher resolution spectroscopy.
In conclusion, it seems likely that the majority of the nuclear cluster stars originated not in globular clusters but more locally.

\end{itemize}

To improve further the constraints on mass, shape, and origin of the nuclear cluster we  use the data presented here
for axisymmetric modeling in \citet{Chatzopoulos_14}. This lifts the assumptions of spherical symmetry and add rotation.
\newline

\bibliography{mspap}

\begin{thebibliography}{174}
\expandafter\ifx\csname natexlab\endcsname\relax\def\natexlab#1{#1}\fi

\bibitem[{{Abuter} {et~al.}(2006){Abuter}, {Schreiber}, {Eisenhauer}, {Ott},
  {Horrobin}, \& {Gillesen}}]{Abuter_06}
{Abuter}, R., {Schreiber}, J., {Eisenhauer}, F., {Ott}, T., {Horrobin}, M., \&
  {Gillesen}, S. 2006, 50, 398

\bibitem[{{Alard}(2001)}]{Alard_01}
{Alard}, C. 2001, \aap, 379, L44

\bibitem[{{Alexander}(2005)}]{Alexander_05}
{Alexander}, T. 2005, \physrep, 419, 65

\bibitem[{{Allen} {et~al.}(1983){Allen}, {Hyland}, \& {Jones}}]{Allen_83}
{Allen}, D.~A., {Hyland}, A.~R., \& {Jones}, T.~J. 1983, \mnras, 204, 1145

\bibitem[{{Amorim} {et~al.}(2006){Amorim}, {Lima}, {Alves}, {Rebord{\~a}o},
  {Pinh{\~a}o}, {Gurriana}, {Cabral}, {Marchetti}, {Kolb}, {Tordo}, {Finger},
  {Lizon}, {Santos}, {Marques}, {Alves}, \& {Barros}}]{Amorim_06}
{Amorim}, A., {Lima}, J., {Alves}, J., {Rebord{\~a}o}, J., {Pinh{\~a}o}, J.,
  {Gurriana}, L., {Cabral}, A., {Marchetti}, E., {Kolb}, J., {Tordo}, S.,
  {Finger}, G., {Lizon}, J.-L., {Santos}, F.~D., {Marques}, R.~F., {Alves}, R.,
  \& {Barros}, R. 2006, in Society of Photo-Optical Instrumentation Engineers
  (SPIE) Conference Series, Vol. 6269, Society of Photo-Optical Instrumentation
  Engineers (SPIE) Conference Series

\bibitem[{{Andersen} {et~al.}(2008){Andersen}, {Walcher}, {B{\"o}ker}, {Ho},
  {van der Marel}, {Rix}, \& {Shields}}]{Andersen_08}
{Andersen}, D.~R., {Walcher}, C.~J., {B{\"o}ker}, T., {Ho}, L.~C., {van der
  Marel}, R.~P., {Rix}, H.-W., \& {Shields}, J.~C. 2008, \apj, 688, 990

\bibitem[{{Antonini} {et~al.}(2012){Antonini}, {Capuzzo-Dolcetta},
  {Mastrobuono-Battisti}, \& {Merritt}}]{Antonini_12}
{Antonini}, F., {Capuzzo-Dolcetta}, R., {Mastrobuono-Battisti}, A., \&
  {Merritt}, D. 2012, \apj, 750, 111

\bibitem[{{Bahcall} \& {Wolf}(1976)}]{Bahcall_76}
{Bahcall}, J.~N. \& {Wolf}, R.~A. 1976, \apj, 209, 214

\bibitem[{{Barth} {et~al.}(2009){Barth}, {Strigari}, {Bentz}, {Greene}, \&
  {Ho}}]{Barth_09}
{Barth}, A.~J., {Strigari}, L.~E., {Bentz}, M.~C., {Greene}, J.~E., \& {Ho},
  L.~C. 2009, \apj, 690, 1031

\bibitem[{{Bartko} {et~al.}(2009){Bartko}, {Martins}, {Fritz}, {Genzel},
  {Levin}, {Perets}, {Paumard}, {Nayakshin}, {Gerhard}, {Alexander},
  {Dodds-Eden}, {Eisenhauer}, {Gillessen}, {Mascetti}, {Ott}, {Perrin},
  {Pfuhl}, {Reid}, {Rouan}, {Sternberg}, \& {Trippe}}]{Bartko_09}
{Bartko}, H., {Martins}, F., {Fritz}, T.~K., {Genzel}, R., {Levin}, Y.,
  {Perets}, H.~B., {Paumard}, T., {Nayakshin}, S., {Gerhard}, O., {Alexander},
  T., {Dodds-Eden}, K., {Eisenhauer}, F., {Gillessen}, S., {Mascetti}, L.,
  {Ott}, T., {Perrin}, G., {Pfuhl}, O., {Reid}, M.~J., {Rouan}, D.,
  {Sternberg}, A., \& {Trippe}, S. 2009, \apj, 697, 1741

\bibitem[{{Bartko} {et~al.}(2010){Bartko}, {Martins}, {Trippe}, {Fritz},
  {Genzel}, {Ott}, {Eisenhauer}, {Gillessen}, {Paumard}, {Alexander},
  {Dodds-Eden}, {Gerhard}, {Levin}, {Mascetti}, {Nayakshin}, {Perets},
  {Perrin}, {Pfuhl}, {Reid}, {Rouan}, {Zilka}, \& {Sternberg}}]{Bartko_10}
{Bartko}, H., {Martins}, F., {Trippe}, S., {Fritz}, T.~K., {Genzel}, R., {Ott},
  T., {Eisenhauer}, F., {Gillessen}, S., {Paumard}, T., {Alexander}, T.,
  {Dodds-Eden}, K., {Gerhard}, O., {Levin}, Y., {Mascetti}, L., {Nayakshin},
  S., {Perets}, H.~B., {Perrin}, G., {Pfuhl}, O., {Reid}, M.~J., {Rouan}, D.,
  {Zilka}, M., \& {Sternberg}, A. 2010, \apj, 708, 834

\bibitem[{{Bastian} {et~al.}(2010){Bastian}, {Covey}, \& {Meyer}}]{Bastian_10}
{Bastian}, N., {Covey}, K.~R., \& {Meyer}, M.~R. 2010, \araa, 48, 339

\bibitem[{{Becklin} \& {Neugebauer}(1968)}]{Becklin_68}
{Becklin}, E.~E. \& {Neugebauer}, G. 1968, \apj, 151, 145

\bibitem[{{Beloborodov} {et~al.}(2006){Beloborodov}, {Levin}, {Eisenhauer},
  {Genzel}, {Paumard}, {Gillessen}, \& {Ott}}]{Beloborodov_06}
{Beloborodov}, A.~M., {Levin}, Y., {Eisenhauer}, F., {Genzel}, R., {Paumard},
  T., {Gillessen}, S., \& {Ott}, T. 2006, \apj, 648, 405

\bibitem[{{Bica} {et~al.}(2006){Bica}, {Bonatto}, {Barbuy}, \&
  {Ortolani}}]{Bica_06}
{Bica}, E., {Bonatto}, C., {Barbuy}, B., \& {Ortolani}, S. 2006, \aap, 450, 105

\bibitem[{{Binney} \& {Merrifield}(1998)}]{Binney_98}
{Binney}, J. \& {Merrifield}, M. 1998, {Galactic Astronomy}

\bibitem[{{Binney} \& {Tremaine}(2008)}]{Binney_08}
{Binney}, J. \& {Tremaine}, S. 2008, {Galactic Dynamics: Second Edition}
  (Princeton University Press)

\bibitem[{{Blum} {et~al.}(2003){Blum}, {Ram{\'{\i}}rez}, {Sellgren}, \&
  {Olsen}}]{Blum_02}
{Blum}, R.~D., {Ram{\'{\i}}rez}, S.~V., {Sellgren}, K., \& {Olsen}, K. 2003,
  \apj, 597, 323

\bibitem[{{B{\"o}ker}(2010)}]{Boeker_10}
{B{\"o}ker}, T. 2010, in IAU Symposium, Vol. 266, IAU Symposium, ed. {R.~de
  Grijs \& J.~R.~D.~L{\'e}pine}, 58--63

\bibitem[{{B{\"o}ker} {et~al.}(2002){B{\"o}ker}, {Laine}, {van der Marel},
  {Sarzi}, {Rix}, {Ho}, \& {Shields}}]{Boeker_02}
{B{\"o}ker}, T., {Laine}, S., {van der Marel}, R.~P., {Sarzi}, M., {Rix},
  H.-W., {Ho}, L.~C., \& {Shields}, J.~C. 2002, \aj, 123, 1389

\bibitem[{{B{\"o}ker} {et~al.}(2004){B{\"o}ker}, {Sarzi}, {McLaughlin}, {van
  der Marel}, {Rix}, {Ho}, \& {Shields}}]{Boeker_04}
{B{\"o}ker}, T., {Sarzi}, M., {McLaughlin}, D.~E., {van der Marel}, R.~P.,
  {Rix}, H.-W., {Ho}, L.~C., \& {Shields}, J.~C. 2004, \aj, 127, 105

\bibitem[{{Bonnet} {et~al.}(2003){Bonnet}, {Stroebele}, {Biancat-Marchet},
  {Brynnel}, {Conzelmann}, {Delabre}, {Donaldson}, {Farinato}, {Fedrigo},
  {Hubin}, {Kasper}, \& {Kissler-Patig}}]{Bonnet_03}
{Bonnet}, H., {Stroebele}, S., {Biancat-Marchet}, F., {Brynnel}, J.,
  {Conzelmann}, R.~D., {Delabre}, B., {Donaldson}, R., {Farinato}, J.,
  {Fedrigo}, E., {Hubin}, N.~N., {Kasper}, M.~E., \& {Kissler-Patig}, M. 2003,
  in Society of Photo-Optical Instrumentation Engineers (SPIE) Conference
  Series, Vol. 4839, Society of Photo-Optical Instrumentation Engineers (SPIE)
  Conference Series, ed. {P.~L.~Wizinowich \& D.~Bonaccini}, 329--343

\bibitem[{{Bovy} {et~al.}(2012){Bovy}, {Allende Prieto}, {Beers}, {Bizyaev},
  {da Costa}, {Cunha}, {Ebelke}, {Eisenstein}, {Frinchaboy}, {Garc{\'{\i}}a
  P{\'e}rez}, {Girardi}, {Hearty}, {Hogg}, {Holtzman}, {Maia}, {Majewski},
  {Malanushenko}, {Malanushenko}, {M{\'e}sz{\'a}ros}, {Nidever}, {O'Connell},
  {O'Donnell}, {Oravetz}, {Pan}, {Rocha-Pinto}, {Schiavon}, {Schneider},
  {Schultheis}, {Skrutskie}, {Smith}, {Weinberg}, {Wilson}, \&
  {Zasowski}}]{Bovy_12b}
{Bovy}, J., {Allende Prieto}, C., {Beers}, T.~C., {Bizyaev}, D., {da Costa},
  L.~N., {Cunha}, K., {Ebelke}, G.~L., {Eisenstein}, D.~J., {Frinchaboy},
  P.~M., {Garc{\'{\i}}a P{\'e}rez}, A.~E., {Girardi}, L., {Hearty}, F.~R.,
  {Hogg}, D.~W., {Holtzman}, J., {Maia}, M.~A.~G., {Majewski}, S.~R.,
  {Malanushenko}, E., {Malanushenko}, V., {M{\'e}sz{\'a}ros}, S., {Nidever},
  D.~L., {O'Connell}, R.~W., {O'Donnell}, C., {Oravetz}, A., {Pan}, K.,
  {Rocha-Pinto}, H.~J., {Schiavon}, R.~P., {Schneider}, D.~P., {Schultheis},
  M., {Skrutskie}, M., {Smith}, V.~V., {Weinberg}, D.~H., {Wilson}, J.~C., \&
  {Zasowski}, G. 2012, \apj, 759, 131

\bibitem[{{Bruzual} \& {Charlot}(2003)}]{Bruzual_03}
{Bruzual}, G. \& {Charlot}, S. 2003, \mnras, 344, 1000

\bibitem[{{Buchholz} {et~al.}(2009){Buchholz}, {Sch{\"o}del}, \&
  {Eckart}}]{Buchholz_09}
{Buchholz}, R.~M., {Sch{\"o}del}, R., \& {Eckart}, A. 2009, \aap, 499, 483

\bibitem[{{Cappellari} {et~al.}(2006){Cappellari}, {Bacon}, {Bureau}, {Damen},
  {Davies}, {de Zeeuw}, {Emsellem}, {Falc{\'o}n-Barroso}, {Krajnovi{\'c}},
  {Kuntschner}, {McDermid}, {Peletier}, {Sarzi}, {van den Bosch}, \& {van de
  Ven}}]{Cappellari_06}
{Cappellari}, M., {Bacon}, R., {Bureau}, M., {Damen}, M.~C., {Davies}, R.~L.,
  {de Zeeuw}, P.~T., {Emsellem}, E., {Falc{\'o}n-Barroso}, J., {Krajnovi{\'c}},
  D., {Kuntschner}, H., {McDermid}, R.~M., {Peletier}, R.~F., {Sarzi}, M., {van
  den Bosch}, R.~C.~E., \& {van de Ven}, G. 2006, \mnras, 366, 1126

\bibitem[{{Cappellari} {et~al.}(2012){Cappellari}, {McDermid}, {Alatalo},
  {Blitz}, {Bois}, {Bournaud}, {Bureau}, {Crocker}, {Davies}, {Davis}, {de
  Zeeuw}, {Duc}, {Emsellem}, {Khochfar}, {Krajnovi{\'c}}, {Kuntschner},
  {Lablanche}, {Morganti}, {Naab}, {Oosterloo}, {Sarzi}, {Scott}, {Serra},
  {Weijmans}, \& {Young}}]{Cappellari_12}
{Cappellari}, M., {McDermid}, R.~M., {Alatalo}, K., {Blitz}, L., {Bois}, M.,
  {Bournaud}, F., {Bureau}, M., {Crocker}, A.~F., {Davies}, R.~L., {Davis},
  T.~A., {de Zeeuw}, P.~T., {Duc}, P.-A., {Emsellem}, E., {Khochfar}, S.,
  {Krajnovi{\'c}}, D., {Kuntschner}, H., {Lablanche}, P.-Y., {Morganti}, R.,
  {Naab}, T., {Oosterloo}, T., {Sarzi}, M., {Scott}, N., {Serra}, P.,
  {Weijmans}, A.-M., \& {Young}, L.~M. 2012, \nat, 484, 485

\bibitem[{{Capuzzo-Dolcetta} \& {Miocchi}(2008)}]{Capuzzo_08}
{Capuzzo-Dolcetta}, R. \& {Miocchi}, P. 2008, \apj, 681, 1136

\bibitem[{{Carollo}(1999)}]{Carollo_99}
{Carollo}, C.~M. 1999, \apj, 523, 566

\bibitem[{{Carollo} {et~al.}(1998){Carollo}, {Stiavelli}, \&
  {Mack}}]{Carollo_98}
{Carollo}, C.~M., {Stiavelli}, M., \& {Mack}, J. 1998, \aj, 116, 68

\bibitem[{{Carollo} {et~al.}(2002){Carollo}, {Stiavelli}, {Seigar}, {de Zeeuw},
  \& {Dejonghe}}]{Carollo_02}
{Carollo}, C.~M., {Stiavelli}, M., {Seigar}, M., {de Zeeuw}, P.~T., \&
  {Dejonghe}, H. 2002, \aj, 123, 159

\bibitem[{{Catchpole} {et~al.}(1990){Catchpole}, {Whitelock}, \&
  {Glass}}]{Catchpole_90}
{Catchpole}, R.~M., {Whitelock}, P.~A., \& {Glass}, I.~S. 1990, \mnras, 247,
  479

\bibitem[{{Chabrier}(2003)}]{Chabrier_03}
{Chabrier}, G. 2003, \pasp, 115, 763

\bibitem[{{Chatzopoulos} {et~al.}(2015){Chatzopoulos}, {Fritz}, {Gerhard},
  {Gillessen}, {Wegg}, {Genzel}, \& {Pfuhl}}]{Chatzopoulos_14}
{Chatzopoulos}, S., {Fritz}, T.~K., {Gerhard}, O., {Gillessen}, S., {Wegg}, C.,
  {Genzel}, R., \& {Pfuhl}, O. 2015, \mnras, 447, 948

\bibitem[{{Christopher} {et~al.}(2005){Christopher}, {Scoville}, {Stolovy}, \&
  {Yun}}]{Christopher_05}
{Christopher}, M.~H., {Scoville}, N.~Z., {Stolovy}, S.~R., \& {Yun}, M.~S.
  2005, \apj, 622, 346

\bibitem[{{Clarkson} {et~al.}(2012){Clarkson}, {Ghez}, {Morris}, {Lu},
  {Stolte}, {McCrady}, {Do}, \& {Yelda}}]{Clarkson_12}
{Clarkson}, W.~I., {Ghez}, A.~M., {Morris}, M.~R., {Lu}, J.~R., {Stolte}, A.,
  {McCrady}, N., {Do}, T., \& {Yelda}, S. 2012, \apj, 751, 132

\bibitem[{{Conroy} \& {van Dokkum}(2012)}]{Conroy_12}
{Conroy}, C. \& {van Dokkum}, P.~G. 2012, \apj, 760, 71

\bibitem[{{Cox}(2000)}]{Cox_00}
{Cox}, A.~N. {Introduction}, ed. {Cox, A.~N.}, 1

\bibitem[{{Cunha} {et~al.}(2007){Cunha}, {Sellgren}, {Smith}, {Ramirez},
  {Blum}, \& {Terndrup}}]{Cunha_07}
{Cunha}, K., {Sellgren}, K., {Smith}, V.~V., {Ramirez}, S.~V., {Blum}, R.~D.,
  \& {Terndrup}, D.~M. 2007, \apj, 669, 1011

\bibitem[{{Dale} {et~al.}(2009){Dale}, {Davies}, {Church}, \&
  {Freitag}}]{Dale_09}
{Dale}, J.~E., {Davies}, M.~B., {Church}, R.~P., \& {Freitag}, M. 2009, \mnras,
  393, 1016

\bibitem[{{Das} {et~al.}(2011){Das}, {Gerhard}, {Mendez}, {Teodorescu}, \& {de
  Lorenzi}}]{Das_11}
{Das}, P., {Gerhard}, O., {Mendez}, R.~H., {Teodorescu}, A.~M., \& {de
  Lorenzi}, F. 2011, \mnras, 415, 1244

\bibitem[{{Deguchi} {et~al.}(2004){Deguchi}, {Imai}, {Fujii}, {Glass}, {Ita},
  {Izumiura}, {Kameya}, {Miyazaki}, {Nakada}, \& {Nakashima}}]{Deguchi_04}
{Deguchi}, S., {Imai}, H., {Fujii}, T., {Glass}, I.~S., {Ita}, Y., {Izumiura},
  H., {Kameya}, O., {Miyazaki}, A., {Nakada}, Y., \& {Nakashima}, J.-I. 2004,
  \pasj, 56, 261

\bibitem[{{Dehnen}(1993)}]{Dehnen_93}
{Dehnen}, W. 1993, \mnras, 265, 250

\bibitem[{{Dehnen} \& {Binney}(1998)}]{Dehnen_98}
{Dehnen}, W. \& {Binney}, J.~J. 1998, \mnras, 298, 387

\bibitem[{{Diolaiti} {et~al.}(2000){Diolaiti}, {Bendinelli}, {Bonaccini},
  {Close}, {Currie}, \& {Parmeggiani}}]{Diolaiti_00}
{Diolaiti}, E., {Bendinelli}, O., {Bonaccini}, D., {Close}, L., {Currie}, D.,
  \& {Parmeggiani}, G. 2000, \aaps, 147, 335

\bibitem[{{Do} {et~al.}(2012){Do}, {Ghez}, {Lu}, {Morris}, {Yelda}, {Martinez},
  {Peter}, {Wright}, {Bullock}, {Kaplinghat}, \& {Matthews}}]{Do_12}
{Do}, T., {Ghez}, A., {Lu}, J.~R., {Morris}, M.~R., {Yelda}, S., {Martinez},
  G.~D., {Peter}, A.~H.~G., {Wright}, S., {Bullock}, J., {Kaplinghat}, M., \&
  {Matthews}, K. 2012, Journal of Physics Conference Series, 372, 012016

\bibitem[{{Do} {et~al.}(2009){Do}, {Ghez}, {Morris}, {Lu}, {Matthews}, {Yelda},
  \& {Larkin}}]{Do_09}
{Do}, T., {Ghez}, A.~M., {Morris}, M.~R., {Lu}, J.~R., {Matthews}, K., {Yelda},
  S., \& {Larkin}, J. 2009, \apj, 703, 1323

\bibitem[{{Do} {et~al.}(2015){Do}, {Kerzendorf}, {Winsor}, {St{\o}stad},
  {Morris}, {Lu}, \& {Ghez}}]{Do_15}
{Do}, T., {Kerzendorf}, W., {Winsor}, N., {St{\o}stad}, M., {Morris}, M.~R.,
  {Lu}, J.~R., \& {Ghez}, A.~M. 2015, \apj, 809, 143

\bibitem[{{Do} {et~al.}(2013{\natexlab{a}}){Do}, {Lu}, {Ghez}, {Morris},
  {Yelda}, {Martinez}, {Wright}, \& {Matthews}}]{Do_13}
{Do}, T., {Lu}, J.~R., {Ghez}, A.~M., {Morris}, M.~R., {Yelda}, S., {Martinez},
  G.~D., {Wright}, S.~A., \& {Matthews}, K. 2013{\natexlab{a}}, \apj, 764, 154

\bibitem[{{Do} {et~al.}(2013{\natexlab{b}}){Do}, {Martinez}, {Yelda}, {Ghez},
  {Bullock}, {Kaplinghat}, {Lu}, {Peter}, \& {Phifer}}]{Do_13b}
{Do}, T., {Martinez}, G.~D., {Yelda}, S., {Ghez}, A., {Bullock}, J.,
  {Kaplinghat}, M., {Lu}, J.~R., {Peter}, A.~H.~G., \& {Phifer}, K.
  2013{\natexlab{b}}, \apjl, 779, L6

\bibitem[{{D'Souza} \& {Rix}(2013)}]{Dsouza_13}
{D'Souza}, R. \& {Rix}, H.-W. 2013, \mnras, 439

\bibitem[{{Eisenhauer} {et~al.}(2003{\natexlab{a}}){Eisenhauer}, {Abuter},
  {Bickert}, {Bianchet-Marchet}, {Bonnet}, {Brynnel}, {Conzelmann}, {Delabre},
  {Donaldson}, {Farinato}, {Fedrigo}, {Genzel}, {Hubin}, {Iserlohe}, {Kasper},
  {Kissler-Patig}, {Monnet}, {Roehrle}, {Scheiber}, {Stroebele}, {Tecza},
  {Thatte}, \& {Weisz}}]{Eisenhauer_etal2003}
{Eisenhauer}, F., {Abuter}, R., {Bickert}, K., {Bianchet-Marchet}, F.,
  {Bonnet}, H., {Brynnel}, J., {Conzelmann}, R.~D., {Delabre}, B., {Donaldson},
  R., {Farinato}, J., {Fedrigo}, E., {Genzel}, R., {Hubin}, N.~N., {Iserlohe},
  C., {Kasper}, M.~E., {Kissler-Patig}, M., {Monnet}, G.~J., {Roehrle}, C.,
  {Scheiber}, J., {Stroebele}, S., {Tecza}, M., {Thatte}, N.~A., \& {Weisz}, H.
  2003{\natexlab{a}}, SPIE, 4841, 1548

\bibitem[{{Eisenhauer} {et~al.}(2005){Eisenhauer}, {Genzel}, {Alexander},
  {Abuter}, {Paumard}, {Ott}, {Gilbert}, {Gillessen}, {Horrobin}, {Trippe},
  {Bonnet}, {Dumas}, {Hubin}, {Kaufer}, {Kissler-Patig}, {Monnet},
  {Str{\"o}bele}, {Szeifert}, {Eckart}, {Sch{\"o}del}, \&
  {Zucker}}]{Eisenhauer_05}
{Eisenhauer}, F., {Genzel}, R., {Alexander}, T., {Abuter}, R., {Paumard}, T.,
  {Ott}, T., {Gilbert}, A., {Gillessen}, S., {Horrobin}, M., {Trippe}, S.,
  {Bonnet}, H., {Dumas}, C., {Hubin}, N., {Kaufer}, A., {Kissler-Patig}, M.,
  {Monnet}, G., {Str{\"o}bele}, S., {Szeifert}, T., {Eckart}, A.,
  {Sch{\"o}del}, R., \& {Zucker}, S. 2005, \apj, 628, 246

\bibitem[{{Eisenhauer} {et~al.}(2008){Eisenhauer}, {Perrin}, {Brandner},
  {Straubmeier}, {Richichi}, {Gillessen}, {Berger}, {Hippler}, {Eckart},
  {Sch{\"o}ller}, {Rabien}, {Cassaing}, {Lenzen}, {Thiel}, {Cl{\'e}net},
  {Ramos}, {Kellner}, {F{\'e}dou}, {Baumeister}, {Hofmann}, {Gendron}, {Boehm},
  {Bartko}, {Haubois}, {Klein}, {Dodds-Eden}, {Houairi}, {Hormuth},
  {Gr{\"a}ter}, {Jocou}, {Naranjo}, {Genzel}, {Kervella}, {Henning}, {Hamaus},
  {Lacour}, {Neumann}, {Haug}, {Malbet}, {Laun}, {Kolmeder}, {Paumard},
  {Rohloff}, {Pfuhl}, {Perraut}, {Ziegleder}, {Rouan}, \&
  {Rousset}}]{Eisenhauer_08}
{Eisenhauer}, F., {Perrin}, G., {Brandner}, W., {Straubmeier}, C., {Richichi},
  A., {Gillessen}, S., {Berger}, J.~P., {Hippler}, S., {Eckart}, A.,
  {Sch{\"o}ller}, M., {Rabien}, S., {Cassaing}, F., {Lenzen}, R., {Thiel}, M.,
  {Cl{\'e}net}, Y., {Ramos}, J.~R., {Kellner}, S., {F{\'e}dou}, P.,
  {Baumeister}, H., {Hofmann}, R., {Gendron}, E., {Boehm}, A., {Bartko}, H.,
  {Haubois}, X., {Klein}, R., {Dodds-Eden}, K., {Houairi}, K., {Hormuth}, F.,
  {Gr{\"a}ter}, A., {Jocou}, L., {Naranjo}, V., {Genzel}, R., {Kervella}, P.,
  {Henning}, T., {Hamaus}, N., {Lacour}, S., {Neumann}, U., {Haug}, M.,
  {Malbet}, F., {Laun}, W., {Kolmeder}, J., {Paumard}, T., {Rohloff}, R.-R.,
  {Pfuhl}, O., {Perraut}, K., {Ziegleder}, J., {Rouan}, D., \& {Rousset}, G.
  2008, in Society of Photo-Optical Instrumentation Engineers (SPIE) Conference
  Series, Vol. 7013, Society of Photo-Optical Instrumentation Engineers (SPIE)
  Conference Series

\bibitem[{{Eisenhauer} {et~al.}(2003{\natexlab{b}}){Eisenhauer}, {Sch{\"o}del},
  {Genzel}, {Ott}, {Tecza}, {Abuter}, {Eckart}, \& {Alexander}}]{Eisenhauer_03}
{Eisenhauer}, F., {Sch{\"o}del}, R., {Genzel}, R., {Ott}, T., {Tecza}, M.,
  {Abuter}, R., {Eckart}, A., \& {Alexander}, T. 2003{\natexlab{b}}, \apjl,
  597, L121

\bibitem[{{Emsellem} \& {van de Ven}(2008)}]{Emsellem_08}
{Emsellem}, E. \& {van de Ven}, G. 2008, \apj, 674, 653

\bibitem[{{Etxaluze} {et~al.}(2011){Etxaluze}, {Smith}, {Tolls}, {Stark}, \&
  {Gonz{\'a}lez-Alfonso}}]{Etxaluze_11}
{Etxaluze}, M., {Smith}, H.~A., {Tolls}, V., {Stark}, A.~A., \&
  {Gonz{\'a}lez-Alfonso}, E. 2011, \aj, 142, 134

\bibitem[{{Feigelson} \& {Jogesh Babu}(2012)}]{Feigelson_12}
{Feigelson}, E.~D. \& {Jogesh Babu}, G. 2012, {Modern Statistical Methods for
  Astronomy}

\bibitem[{{Feldmeier} {et~al.}(2014){Feldmeier}, {Neumayer}, {Seth},
  {Sch{\"o}del}, {L{\"u}tzgendorf}, {de Zeeuw}, {Kissler-Patig}, {Nishiyama},
  \& {Walcher}}]{Feldmeier_14}
{Feldmeier}, A., {Neumayer}, N., {Seth}, A., {Sch{\"o}del}, R.,
  {L{\"u}tzgendorf}, N., {de Zeeuw}, P.~T., {Kissler-Patig}, M., {Nishiyama},
  S., \& {Walcher}, C.~J. 2014, \aap, 570, A2

\bibitem[{{Figer} {et~al.}(2003){Figer}, {Gilmore}, {Kim}, {Morris}, {Becklin},
  {McLean}, {Gilbert}, {Graham}, {Larkin}, {Levenson}, \& {Teplitz}}]{Figer_03}
{Figer}, D.~F., {Gilmore}, D., {Kim}, S.~S., {Morris}, M., {Becklin}, E.~E.,
  {McLean}, I.~S., {Gilbert}, A.~M., {Graham}, J.~R., {Larkin}, J.~E.,
  {Levenson}, N.~A., \& {Teplitz}, H.~I. 2003, \apj, 599, 1139

\bibitem[{{Forrest} {et~al.}(1987){Forrest}, {Shure}, {Pipher}, \&
  {Woodward}}]{Forrest_87}
{Forrest}, W.~J., {Shure}, M.~A., {Pipher}, J.~L., \& {Woodward}, C.~E. 1987,
  in American Institute of Physics Conference Series, Vol. 155, The Galactic
  Center, ed. {D.~C.~Backer}, 153--156

\bibitem[{{Freitag} {et~al.}(2006){Freitag}, {Amaro-Seoane}, \&
  {Kalogera}}]{Freitag_06}
{Freitag}, M., {Amaro-Seoane}, P., \& {Kalogera}, V. 2006, \apj, 649, 91

\bibitem[{{Fritz} {et~al.}(2010{\natexlab{a}}){Fritz}, {Gillessen}, {Trippe},
  {Ott}, {Bartko}, {Pfuhl}, {Dodds-Eden}, {Davies}, {Eisenhauer}, \&
  {Genzel}}]{Fritz_09}
{Fritz}, T., {Gillessen}, S., {Trippe}, S., {Ott}, T., {Bartko}, H., {Pfuhl},
  O., {Dodds-Eden}, K., {Davies}, R., {Eisenhauer}, F., \& {Genzel}, R.
  2010{\natexlab{a}}, \mnras, 401, 1177

\bibitem[{{Fritz} {et~al.}(2011){Fritz}, {Gillessen}, {Dodds-Eden}, {Lutz},
  {Genzel}, {Raab}, {Ott}, {Pfuhl}, {Eisenhauer}, \& {Yusef-Zadeh}}]{Fritz_11}
{Fritz}, T.~K., {Gillessen}, S., {Dodds-Eden}, K., {Lutz}, D., {Genzel}, R.,
  {Raab}, W., {Ott}, T., {Pfuhl}, O., {Eisenhauer}, F., \& {Yusef-Zadeh}, F.
  2011, \apj, 737, 73

\bibitem[{{Fritz} {et~al.}(2010{\natexlab{b}}){Fritz}, {Gillessen},
  {Dodds-Eden}, {Martins}, {Bartko}, {Genzel}, {Paumard}, {Ott}, {Pfuhl},
  {Trippe}, {Eisenhauer}, \& {Gratadour}}]{Fritz_10b}
{Fritz}, T.~K., {Gillessen}, S., {Dodds-Eden}, K., {Martins}, F., {Bartko}, H.,
  {Genzel}, R., {Paumard}, T., {Ott}, T., {Pfuhl}, O., {Trippe}, S.,
  {Eisenhauer}, F., \& {Gratadour}, D. 2010{\natexlab{b}}, \apj, 721, 395

\bibitem[{{Gebhardt} {et~al.}(2001){Gebhardt}, {Lauer}, {Kormendy}, {Pinkney},
  {Bower}, {Green}, {Gull}, {Hutchings}, {Kaiser}, {Nelson}, {Richstone}, \&
  {Weistrop}}]{Gebhardt_01}
{Gebhardt}, K., {Lauer}, T.~R., {Kormendy}, J., {Pinkney}, J., {Bower}, G.~A.,
  {Green}, R., {Gull}, T., {Hutchings}, J.~B., {Kaiser}, M.~E., {Nelson},
  C.~H., {Richstone}, D., \& {Weistrop}, D. 2001, \aj, 122, 2469

\bibitem[{{Genzel} {et~al.}(1985){Genzel}, {Crawford}, {Townes}, \&
  {Watson}}]{Genzel_85}
{Genzel}, R., {Crawford}, M.~K., {Townes}, C.~H., \& {Watson}, D.~M. 1985,
  \apj, 297, 766

\bibitem[{{Genzel} {et~al.}(1997){Genzel}, {Eckart}, {Ott}, \&
  {Eisenhauer}}]{Genzel_97}
{Genzel}, R., {Eckart}, A., {Ott}, T., \& {Eisenhauer}, F. 1997, \mnras, 291,
  219

\bibitem[{{Genzel} {et~al.}(2010){Genzel}, {Eisenhauer}, \&
  {Gillessen}}]{Genzel_10}
{Genzel}, R., {Eisenhauer}, F., \& {Gillessen}, S. 2010, Reviews of Modern
  Physics, 82, 3121

\bibitem[{{Genzel} {et~al.}(2000){Genzel}, {Pichon}, {Eckart}, {Gerhard}, \&
  {Ott}}]{Genzel_00}
{Genzel}, R., {Pichon}, C., {Eckart}, A., {Gerhard}, O.~E., \& {Ott}, T. 2000,
  \mnras, 317, 348

\bibitem[{{Genzel} {et~al.}(2003){Genzel}, {Sch{\"o}del}, {Ott}, {Eisenhauer},
  {Hofmann}, {Lehnert}, {Eckart}, {Alexander}, {Sternberg}, {Lenzen},
  {Cl{\'e}net}, {Lacombe}, {Rouan}, {Renzini}, \&
  {Tacconi-Garman}}]{Genzel_03b}
{Genzel}, R., {Sch{\"o}del}, R., {Ott}, T., {Eisenhauer}, F., {Hofmann}, R.,
  {Lehnert}, M., {Eckart}, A., {Alexander}, T., {Sternberg}, A., {Lenzen}, R.,
  {Cl{\'e}net}, Y., {Lacombe}, F., {Rouan}, D., {Renzini}, A., \&
  {Tacconi-Garman}, L.~E. 2003, \apj, 594, 812

\bibitem[{{Genzel} {et~al.}(1996){Genzel}, {Thatte}, {Krabbe}, {Kroker}, \&
  {Tacconi-Garman}}]{Genzel_96}
{Genzel}, R., {Thatte}, N., {Krabbe}, A., {Kroker}, H., \& {Tacconi-Garman},
  L.~E. 1996, \apj, 472, 153

\bibitem[{{Ghez} {et~al.}(2008){Ghez}, {Salim}, {Weinberg}, {Lu}, {Do}, {Dunn},
  {Matthews}, {Morris}, {Yelda}, {Becklin}, {Kremenek}, {Milosavljevic}, \&
  {Naiman}}]{Ghez_09}
{Ghez}, A.~M., {Salim}, S., {Weinberg}, N.~N., {Lu}, J.~R., {Do}, T., {Dunn},
  J.~K., {Matthews}, K., {Morris}, M.~R., {Yelda}, S., {Becklin}, E.~E.,
  {Kremenek}, T., {Milosavljevic}, M., \& {Naiman}, J. 2008, \apj, 689, 1044

\bibitem[{{Gillessen} {et~al.}(2013){Gillessen}, {Eisenhauer}, {Fritz},
  {Pfuhl}, {Ott}, \& {Genzel}}]{Gillessen_13}
{Gillessen}, S., {Eisenhauer}, F., {Fritz}, T.~K., {Pfuhl}, O., {Ott}, T., \&
  {Genzel}, R. 2013, in IAU Symposium, Vol. 289, IAU Symposium, ed. R.~{de
  Grijs}, 29--35

\bibitem[{{Gillessen} {et~al.}(2009){Gillessen}, {Eisenhauer}, {Trippe},
  {Alexander}, {Genzel}, {Martins}, \& {Ott}}]{Gillessen_09}
{Gillessen}, S., {Eisenhauer}, F., {Trippe}, S., {Alexander}, T., {Genzel}, R.,
  {Martins}, F., \& {Ott}, T. 2009, \apj, 692, 1075

\bibitem[{{Gnedin} {et~al.}(2014){Gnedin}, {Ostriker}, \&
  {Tremaine}}]{Gnedin_14}
{Gnedin}, O.~Y., {Ostriker}, J.~P., \& {Tremaine}, S. 2014, \apj, 785, 71

\bibitem[{{Graham} \& {Spitler}(2009)}]{Graham_09}
{Graham}, A.~W. \& {Spitler}, L.~R. 2009, \mnras, 397, 2148

\bibitem[{{Graves} {et~al.}(1998){Graves}, {Northcott}, {Roddier}, {Roddier},
  \& {Close}}]{Graves_98}
{Graves}, J.~E., {Northcott}, M.~J., {Roddier}, F.~J., {Roddier}, C.~A., \&
  {Close}, L.~M. 1998, in Society of Photo-Optical Instrumentation Engineers
  (SPIE) Conference Series, Vol. 3353, Society of Photo-Optical Instrumentation
  Engineers (SPIE) Conference Series, ed. {D.~Bonaccini \& R.~K.~Tyson}, 34--43

\bibitem[{{Guesten} {et~al.}(1987){Guesten}, {Genzel}, {Wright}, {Jaffe},
  {Stutzki}, \& {Harris}}]{Guesten_87}
{Guesten}, R., {Genzel}, R., {Wright}, M.~C.~H., {Jaffe}, D.~T., {Stutzki}, J.,
  \& {Harris}, A.~I. 1987, \apj, 318, 124

\bibitem[{{Haller} {et~al.}(1996){Haller}, {Rieke}, {Rieke}, {Tamblyn},
  {Close}, \& {Melia}}]{Haller_96}
{Haller}, J.~W., {Rieke}, M.~J., {Rieke}, G.~H., {Tamblyn}, P., {Close}, L., \&
  {Melia}, F. 1996, \apj, 456, 194

\bibitem[{{Harris}(1996)}]{Harris_96}
{Harris}, W.~E. 1996, \aj, 112, 1487

\bibitem[{{Hills}(1988)}]{Hills_88}
{Hills}, J.~G. 1988, \nat, 331, 687

\bibitem[{{Hodapp} {et~al.}(1996){Hodapp}, {Hora}, {Hall}, {Cowie}, {Metzger},
  {Irwin}, {Vural}, {Kozlowski}, {Cabelli}, {Chen}, {Cooper}, {Bostrup},
  {Bailey}, \& {Kleinhans}}]{Hodapp_96}
{Hodapp}, K.-W., {Hora}, J.~L., {Hall}, D.~N.~B., {Cowie}, L.~L., {Metzger},
  M., {Irwin}, E., {Vural}, K., {Kozlowski}, L.~J., {Cabelli}, S.~A., {Chen},
  C.~Y., {Cooper}, D.~E., {Bostrup}, G.~L., {Bailey}, R.~B., \& {Kleinhans},
  W.~E. 1996, New Astronomy, 1, 177

\bibitem[{{Hopman} \& {Alexander}(2006)}]{Hopman_06}
{Hopman}, C. \& {Alexander}, T. 2006, \apjl, 645, L133

\bibitem[{{Hunter} \& {Qian}(1993)}]{Hunter_93}
{Hunter}, C. \& {Qian}, E. 1993, \mnras, 262, 401

\bibitem[{{Jackson} {et~al.}(1993){Jackson}, {Geis}, {Genzel}, {Harris},
  {Madden}, {Poglitsch}, {Stacey}, \& {Townes}}]{Jackson_93}
{Jackson}, J.~M., {Geis}, N., {Genzel}, R., {Harris}, A.~I., {Madden}, S.,
  {Poglitsch}, A., {Stacey}, G.~J., \& {Townes}, C.~H. 1993, \apj, 402, 173

\bibitem[{{Keshet} {et~al.}(2009){Keshet}, {Hopman}, \&
  {Alexander}}]{Keshet_09}
{Keshet}, U., {Hopman}, C., \& {Alexander}, T. 2009, \apjl, 698, L64

\bibitem[{{Kocsis} \& {Tremaine}(2011)}]{Kocsis_11}
{Kocsis}, B. \& {Tremaine}, S. 2011, \mnras, 412, 187

\bibitem[{{Kormendy} \& {Ho}(2013)}]{Kormendy_13}
{Kormendy}, J. \& {Ho}, L.~C. 2013, \araa, 51, 511

\bibitem[{{Kormendy} \& {McClure}(1993)}]{Kormendy_93}
{Kormendy}, J. \& {McClure}, R.~D. 1993, \aj, 105, 1793

\bibitem[{{Krabbe} {et~al.}(1991){Krabbe}, {Genzel}, {Drapatz}, \&
  {Rotaciuc}}]{Krabbe_91}
{Krabbe}, A., {Genzel}, R., {Drapatz}, S., \& {Rotaciuc}, V. 1991, \apjl, 382,
  L19

\bibitem[{{Kroupa}(2001)}]{Kroupa_01}
{Kroupa}, P. 2001, \mnras, 322, 231

\bibitem[{{Kruijssen} \& {Mieske}(2009)}]{Kruijssen_09a}
{Kruijssen}, J.~M.~D. \& {Mieske}, S. 2009, \aap, 500, 785

\bibitem[{{Lauer} {et~al.}(1995){Lauer}, {Ajhar}, {Byun}, {Dressler}, {Faber},
  {Grillmair}, {Kormendy}, {Richstone}, \& {Tremaine}}]{Lauer_95}
{Lauer}, T.~R., {Ajhar}, E.~A., {Byun}, Y.-I., {Dressler}, A., {Faber}, S.~M.,
  {Grillmair}, C., {Kormendy}, J., {Richstone}, D., \& {Tremaine}, S. 1995,
  \aj, 110, 2622

\bibitem[{{Lauer} {et~al.}(1998){Lauer}, {Faber}, {Ajhar}, {Grillmair}, \&
  {Scowen}}]{Lauer_98}
{Lauer}, T.~R., {Faber}, S.~M., {Ajhar}, E.~A., {Grillmair}, C.~J., \&
  {Scowen}, P.~A. 1998, \aj, 116, 2263

\bibitem[{{Launhardt} {et~al.}(2002){Launhardt}, {Zylka}, \&
  {Mezger}}]{Launhardt_02}
{Launhardt}, R., {Zylka}, R., \& {Mezger}, P.~G. 2002, \aap, 384, 112

\bibitem[{{Lenzen} {et~al.}(2003){Lenzen}, {Hartung}, {Brandner}, {Finger},
  {Hubin}, {Lacombe}, {Lagrange}, {Lehnert}, {Moorwood}, \&
  {Mouillet}}]{Lenzen_03}
{Lenzen}, R., {Hartung}, M., {Brandner}, W., {Finger}, G., {Hubin}, N.~N.,
  {Lacombe}, F., {Lagrange}, A.-M., {Lehnert}, M.~D., {Moorwood}, A.~F.~M., \&
  {Mouillet}, D. 2003, in Society of Photo-Optical Instrumentation Engineers
  (SPIE) Conference Series, Vol. 4841, Society of Photo-Optical Instrumentation
  Engineers (SPIE) Conference Series, ed. {M.~Iye \& A.~F.~M.~Moorwood},
  944--952

\bibitem[{{Leonard} \& {Merritt}(1989)}]{Leonard_89}
{Leonard}, P.~J.~T. \& {Merritt}, D. 1989, \apj, 339, 195

\bibitem[{{Lindqvist} {et~al.}(1992{\natexlab{a}}){Lindqvist}, {Habing}, \&
  {Winnberg}}]{Lindqvist_92b}
{Lindqvist}, M., {Habing}, H.~J., \& {Winnberg}, A. 1992{\natexlab{a}}, \aap,
  259, 118

\bibitem[{{Lindqvist} {et~al.}(1992{\natexlab{b}}){Lindqvist}, {Winnberg},
  {Habing}, \& {Matthews}}]{Lindqvist_92a}
{Lindqvist}, M., {Winnberg}, A., {Habing}, H.~J., \& {Matthews}, H.~E.
  1992{\natexlab{b}}, \aaps, 92, 43

\bibitem[{{L{\"o}ckmann} \& {Baumgardt}(2009)}]{Loeckmann_09a}
{L{\"o}ckmann}, U. \& {Baumgardt}, H. 2009, \mnras, 394, 1841

\bibitem[{{Lu} {et~al.}(2013){Lu}, {Do}, {Ghez}, {Morris}, {Yelda}, \&
  {Matthews}}]{Lu_13}
{Lu}, J.~R., {Do}, T., {Ghez}, A.~M., {Morris}, M.~R., {Yelda}, S., \&
  {Matthews}, K. 2013, \apj, 764, 155

\bibitem[{{Lu} {et~al.}(2009){Lu}, {Ghez}, {Hornstein}, {Morris}, {Becklin}, \&
  {Matthews}}]{Lu_09}
{Lu}, J.~R., {Ghez}, A.~M., {Hornstein}, S.~D., {Morris}, M.~R., {Becklin},
  E.~E., \& {Matthews}, K. 2009, \apj, 690, 1463

\bibitem[{{L{\"u}tzgendorf} {et~al.}(2012){L{\"u}tzgendorf}, {Kissler-Patig},
  {Gebhardt}, {Baumgardt}, {Noyola}, {Jalali}, {de Zeeuw}, \&
  {Neumayer}}]{Luetzgendorf_12}
{L{\"u}tzgendorf}, N., {Kissler-Patig}, M., {Gebhardt}, K., {Baumgardt}, H.,
  {Noyola}, E., {Jalali}, B., {de Zeeuw}, P.~T., \& {Neumayer}, N. 2012, \aap,
  542, A129

\bibitem[{{Lynden-Bell}(1960)}]{Lynden-Bell_60}
{Lynden-Bell}, D. 1960, \mnras, 120, 204

\bibitem[{{Madigan} {et~al.}(2014){Madigan}, {Pfuhl}, {Levin}, {Gillessen},
  {Genzel}, \& {Perets}}]{Madigan_13}
{Madigan}, A.-M., {Pfuhl}, O., {Levin}, Y., {Gillessen}, S., {Genzel}, R., \&
  {Perets}, H.~B. 2014, \apj, 784, 23

\bibitem[{{Maraston}(2005)}]{Maraston_05}
{Maraston}, C. 2005, \mnras, 362, 799

\bibitem[{{Marchetti} {et~al.}(2004){Marchetti}, {Brast}, {Delabre},
  {Donaldson}, {Fedrigo}, {Frank}, {Hubin}, {Kolb}, {Le Louarn}, {Lizon},
  {Oberti}, {Reiss}, {Santos}, {Tordo}, {Ragazzoni}, {Arcidiacono},
  {Baruffolo}, {Diolaiti}, {Farinato}, \& {Vernet-Viard}}]{Marchetti_04}
{Marchetti}, E., {Brast}, R., {Delabre}, B., {Donaldson}, R., {Fedrigo}, E.,
  {Frank}, C., {Hubin}, N.~N., {Kolb}, J., {Le Louarn}, M., {Lizon}, J.-L.,
  {Oberti}, S., {Reiss}, R., {Santos}, J., {Tordo}, S., {Ragazzoni}, R.,
  {Arcidiacono}, C., {Baruffolo}, A., {Diolaiti}, E., {Farinato}, J., \&
  {Vernet-Viard}, E. 2004, in Society of Photo-Optical Instrumentation
  Engineers (SPIE) Conference Series, Vol. 5490, Society of Photo-Optical
  Instrumentation Engineers (SPIE) Conference Series, ed. {D.~Bonaccini Calia,
  B.~L.~Ellerbroek, \& R.~Ragazzoni}, 236--247

\bibitem[{{Marks} {et~al.}(2012){Marks}, {Kroupa}, {Dabringhausen}, \&
  {Pawlowski}}]{Marks_12}
{Marks}, M., {Kroupa}, P., {Dabringhausen}, J., \& {Pawlowski}, M.~S. 2012,
  \mnras, 422, 2246

\bibitem[{{Markwardt}(2009)}]{Markwardt_09}
{Markwardt}, C.~B. 2009, in Astronomical Society of the Pacific Conference
  Series, Vol. 411, Astronomical Data Analysis Software and Systems XVIII, ed.
  D.~A. {Bohlender}, D.~{Durand}, \& P.~{Dowler}, 251

\bibitem[{{Matthews} \& {Gallagher}(1997)}]{Matthews_97}
{Matthews}, L.~D. \& {Gallagher}, III, J.~S. 1997, \aj, 114, 1899

\bibitem[{{McGinn} {et~al.}(1989){McGinn}, {Sellgren}, {Becklin}, \&
  {Hall}}]{McGinn_89}
{McGinn}, M.~T., {Sellgren}, K., {Becklin}, E.~E., \& {Hall}, D.~N.~B. 1989,
  \apj, 338, 824

\bibitem[{{Merritt}(2010)}]{Merritt_10}
{Merritt}, D. 2010, \apj, 718, 739

\bibitem[{{Merritt} {et~al.}(2001){Merritt}, {Ferrarese}, \&
  {Joseph}}]{Merritt_01}
{Merritt}, D., {Ferrarese}, L., \& {Joseph}, C.~L. 2001, Science, 293, 1116

\bibitem[{{Merritt} \& {Tremblay}(1994)}]{Merritt_94}
{Merritt}, D. \& {Tremblay}, B. 1994, \aj, 108, 514

\bibitem[{{Meyer} {et~al.}(2012){Meyer}, {Ghez}, {Sch{\"o}del}, {Yelda},
  {Boehle}, {Lu}, {Do}, {Morris}, {Becklin}, \& {Matthews}}]{Meyer_12}
{Meyer}, L., {Ghez}, A.~M., {Sch{\"o}del}, R., {Yelda}, S., {Boehle}, A., {Lu},
  J.~R., {Do}, T., {Morris}, M.~R., {Becklin}, E.~E., \& {Matthews}, K. 2012,
  Science, 338, 84

\bibitem[{{Mezger} {et~al.}(1996){Mezger}, {Duschl}, \& {Zylka}}]{Mezger_96}
{Mezger}, P.~G., {Duschl}, W.~J., \& {Zylka}, R. 1996, \aapr, 7, 289

\bibitem[{{Mezger} {et~al.}(1989){Mezger}, {Zylka}, {Salter}, {Wink}, {Chini},
  {Kreysa}, \& {Tuffs}}]{Mezger_89}
{Mezger}, P.~G., {Zylka}, R., {Salter}, C.~J., {Wink}, J.~E., {Chini}, R.,
  {Kreysa}, E., \& {Tuffs}, R. 1989, \aap, 209, 337

\bibitem[{{Milosavljevi{\'c}}(2004)}]{Milsavljevic_04}
{Milosavljevi{\'c}}, M. 2004, \apjl, 605, L13

\bibitem[{{Nishiyama} {et~al.}(2006){Nishiyama}, {Nagata}, {Kusakabe},
  {Matsunaga}, {Naoi}, {Kato}, {Nagashima}, {Sugitani}, {Tamura}, {Tanab{\'e}},
  \& {Sato}}]{Nishiyama_06b}
{Nishiyama}, S., {Nagata}, T., {Kusakabe}, N., {Matsunaga}, N., {Naoi}, T.,
  {Kato}, D., {Nagashima}, C., {Sugitani}, K., {Tamura}, M., {Tanab{\'e}}, T.,
  \& {Sato}, S. 2006, \apj, 638, 839

\bibitem[{{Nishiyama} \& {Sch{\"o}del}(2013)}]{Nishiyama_12}
{Nishiyama}, S. \& {Sch{\"o}del}, R. 2013, \aap, 549, A57

\bibitem[{{Paumard} {et~al.}(2006){Paumard}, {Genzel}, {Martins}, {Nayakshin},
  {Beloborodov}, {Levin}, {Trippe}, {Eisenhauer}, {Ott}, {Gillessen}, {Abuter},
  {Cuadra}, {Alexander}, \& {Sternberg}}]{Paumard_06}
{Paumard}, T., {Genzel}, R., {Martins}, F., {Nayakshin}, S., {Beloborodov},
  A.~M., {Levin}, Y., {Trippe}, S., {Eisenhauer}, F., {Ott}, T., {Gillessen},
  S., {Abuter}, R., {Cuadra}, J., {Alexander}, T., \& {Sternberg}, A. 2006,
  \apj, 643, 1011

\bibitem[{{Peng} {et~al.}(2002){Peng}, {Ho}, {Impey}, \& {Rix}}]{Peng_02}
{Peng}, C.~Y., {Ho}, L.~C., {Impey}, C.~D., \& {Rix}, H.-W. 2002, \aj, 124, 266

\bibitem[{{Perets} {et~al.}(2007){Perets}, {Hopman}, \&
  {Alexander}}]{Perets_07}
{Perets}, H.~B., {Hopman}, C., \& {Alexander}, T. 2007, \apj, 656, 709

\bibitem[{{Pfuhl} {et~al.}(2011){Pfuhl}, {Fritz}, {Zilka}, {Maness},
  {Eisenhauer}, {Genzel}, {Gillessen}, {Ott}, {Dodds-Eden}, \&
  {Sternberg}}]{Pfuhl_11}
{Pfuhl}, O., {Fritz}, T.~K., {Zilka}, M., {Maness}, H., {Eisenhauer}, F.,
  {Genzel}, R., {Gillessen}, S., {Ott}, T., {Dodds-Eden}, K., \& {Sternberg},
  A. 2011, \apj, 741, 108

\bibitem[{{Philipp} {et~al.}(1999){Philipp}, {Zylka}, {Mezger}, {Duschl},
  {Herbst}, \& {Tuffs}}]{Philipp_99}
{Philipp}, S., {Zylka}, R., {Mezger}, P.~G., {Duschl}, W.~J., {Herbst}, T., \&
  {Tuffs}, R.~J. 1999, \aap, 348, 768

\bibitem[{{Phillips} {et~al.}(1996){Phillips}, {Illingworth}, {MacKenty}, \&
  {Franx}}]{Phillips_96}
{Phillips}, A.~C., {Illingworth}, G.~D., {MacKenty}, J.~W., \& {Franx}, M.
  1996, \aj, 111, 1566

\bibitem[{{Press} {et~al.}(1986){Press}, {Flannery}, \& {Teukolsky}}]{Press_86}
{Press}, W.~H., {Flannery}, B.~P., \& {Teukolsky}, S.~A. 1986, {Numerical
  recipes. The art of scientific computing}

\bibitem[{{Rayner} {et~al.}(2009){Rayner}, {Cushing}, \& {Vacca}}]{Rayner_09}
{Rayner}, J.~T., {Cushing}, M.~C., \& {Vacca}, W.~D. 2009, \apjs, 185, 289

\bibitem[{{Reid}(1993)}]{Reid_93}
{Reid}, M.~J. 1993, \araa, 31, 345

\bibitem[{{Reid} \& {Brunthaler}(2004)}]{Reid_04}
{Reid}, M.~J. \& {Brunthaler}, A. 2004, \apj, 616, 872

\bibitem[{{Reid} {et~al.}(2014){Reid}, {Menten}, {Brunthaler}, {Zheng}, {Dame},
  {Xu}, {Wu}, {Zhang}, {Sanna}, {Sato}, {Hachisuka}, {Choi}, {Immer},
  {Moscadelli}, {Rygl}, \& {Bartkiewicz}}]{Reid_14}
{Reid}, M.~J., {Menten}, K.~M., {Brunthaler}, A., {Zheng}, X.~W., {Dame},
  T.~M., {Xu}, Y., {Wu}, Y., {Zhang}, B., {Sanna}, A., {Sato}, M., {Hachisuka},
  K., {Choi}, Y.~K., {Immer}, K., {Moscadelli}, L., {Rygl}, K.~L.~J., \&
  {Bartkiewicz}, A. 2014, \apj, 783, 130

\bibitem[{{Reid} {et~al.}(2007){Reid}, {Menten}, {Trippe}, {Ott}, \&
  {Genzel}}]{Reid_07}
{Reid}, M.~J., {Menten}, K.~M., {Trippe}, S., {Ott}, T., \& {Genzel}, R. 2007,
  \apj, 659, 378

\bibitem[{{Requena-Torres} {et~al.}(2012){Requena-Torres}, {G{\"u}sten},
  {Wei{\ss}}, {Harris}, {Mart{\'{\i}}n-Pintado}, {Stutzki}, {Klein},
  {Heyminck}, \& {Risacher}}]{Requena_12}
{Requena-Torres}, M.~A., {G{\"u}sten}, R., {Wei{\ss}}, A., {Harris}, A.~I.,
  {Mart{\'{\i}}n-Pintado}, J., {Stutzki}, J., {Klein}, B., {Heyminck}, S., \&
  {Risacher}, C. 2012, \aap, 542, L21

\bibitem[{{Rieke} \& {Rieke}(1988)}]{Rieke_88}
{Rieke}, G.~H. \& {Rieke}, M.~J. 1988, \apjl, 330, L33

\bibitem[{{Rossa} {et~al.}(2006){Rossa}, {van der Marel}, {B{\"o}ker},
  {Gerssen}, {Ho}, {Rix}, {Shields}, \& {Walcher}}]{Rossa_06}
{Rossa}, J., {van der Marel}, R.~P., {B{\"o}ker}, T., {Gerssen}, J., {Ho},
  L.~C., {Rix}, H.-W., {Shields}, J.~C., \& {Walcher}, C.-J. 2006, \aj, 132,
  1074

\bibitem[{{Rousset} {et~al.}(2003){Rousset}, {Lacombe}, {Puget}, {Hubin},
  {Gendron}, {Fusco}, {Arsenault}, {Charton}, {Feautrier}, {Gigan}, {Kern},
  {Lagrange}, {Madec}, {Mouillet}, {Rabaud}, {Rabou}, {Stadler}, \&
  {Zins}}]{Rousset_03}
{Rousset}, G., {Lacombe}, F., {Puget}, P., {Hubin}, N.~N., {Gendron}, E.,
  {Fusco}, T., {Arsenault}, R., {Charton}, J., {Feautrier}, P., {Gigan}, P.,
  {Kern}, P.~Y., {Lagrange}, A.-M., {Madec}, P.-Y., {Mouillet}, D., {Rabaud},
  D., {Rabou}, P., {Stadler}, E., \& {Zins}, G. 2003, in Society of
  Photo-Optical Instrumentation Engineers (SPIE) Conference Series, Vol. 4839,
  Society of Photo-Optical Instrumentation Engineers (SPIE) Conference Series,
  ed. {P.~L.~Wizinowich \& D.~Bonaccini}, 140--149

\bibitem[{{Ryde} \& {Schultheis}(2015)}]{Ryde_15}
{Ryde}, N. \& {Schultheis}, M. 2015, \aap, 573, A14

\bibitem[{{Saito} {et~al.}(2012){Saito}, {Hempel}, {Minniti}, {Lucas},
  {Rejkuba}, {Toledo}, {Gonzalez}, {Alonso-Garc{\'{\i}}a}, {Irwin},
  {Gonzalez-Solares}, {Hodgkin}, {Lewis}, {Cross}, {Ivanov}, {Kerins},
  {Emerson}, {Soto}, {Am{\^o}res}, {Gurovich}, {D{\'e}k{\'a}ny}, {Angeloni},
  {Beamin}, {Catelan}, {Padilla}, {Zoccali}, {Pietrukowicz}, {Moni Bidin},
  {Mauro}, {Geisler}, {Folkes}, {Sale}, {Borissova}, {Kurtev}, {Ahumada},
  {Alonso}, {Adamson}, {Arias}, {Bandyopadhyay}, {Barb{\'a}}, {Barbuy},
  {Baume}, {Bedin}, {Bellini}, {Benjamin}, {Bica}, {Bonatto}, {Bronfman},
  {Carraro}, {Chen{\`e}}, {Clari{\'a}}, {Clarke}, {Contreras}, {Corvill{\'o}n},
  {de Grijs}, {Dias}, {Drew}, {Fari{\~n}a}, {Feinstein},
  {Fern{\'a}ndez-Laj{\'u}s}, {Gamen}, {Gieren}, {Goldman},
  {Gonz{\'a}lez-Fern{\'a}ndez}, {Grand}, {Gunthardt}, {Hambly}, {Hanson},
  {He{\l}miniak}, {Hoare}, {Huckvale}, {Jord{\'a}n}, {Kinemuchi}, {Longmore},
  {L{\'o}pez-Corredoira}, {Maccarone}, {Majaess}, {Mart{\'{\i}}n}, {Masetti},
  {Mennickent}, {Mirabel}, {Monaco}, {Morelli}, {Motta}, {Palma}, {Parisi},
  {Parker}, {Pe{\~n}aloza}, {Pietrzy{\'n}ski}, {Pignata}, {Popescu}, {Read},
  {Rojas}, {Roman-Lopes}, {Ruiz}, {Saviane}, {Schreiber}, {Schr{\"o}der},
  {Sharma}, {Smith}, {Sodr{\'e}}, {Stead}, {Stephens}, {Tamura}, {Tappert},
  {Thompson}, {Valenti}, {Vanzi}, {Walton}, {Weidmann}, \&
  {Zijlstra}}]{Saito_12}
{Saito}, R.~K., {Hempel}, M., {Minniti}, D., {Lucas}, P.~W., {Rejkuba}, M.,
  {Toledo}, I., {Gonzalez}, O.~A., {Alonso-Garc{\'{\i}}a}, J., {Irwin}, M.~J.,
  {Gonzalez-Solares}, E., {Hodgkin}, S.~T., {Lewis}, J.~R., {Cross}, N.,
  {Ivanov}, V.~D., {Kerins}, E., {Emerson}, J.~P., {Soto}, M., {Am{\^o}res},
  E.~B., {Gurovich}, S., {D{\'e}k{\'a}ny}, I., {Angeloni}, R., {Beamin}, J.~C.,
  {Catelan}, M., {Padilla}, N., {Zoccali}, M., {Pietrukowicz}, P., {Moni
  Bidin}, C., {Mauro}, F., {Geisler}, D., {Folkes}, S.~L., {Sale}, S.~E.,
  {Borissova}, J., {Kurtev}, R., {Ahumada}, A.~V., {Alonso}, M.~V., {Adamson},
  A., {Arias}, J.~I., {Bandyopadhyay}, R.~M., {Barb{\'a}}, R.~H., {Barbuy}, B.,
  {Baume}, G.~L., {Bedin}, L.~R., {Bellini}, A., {Benjamin}, R., {Bica}, E.,
  {Bonatto}, C., {Bronfman}, L., {Carraro}, G., {Chen{\`e}}, A.~N.,
  {Clari{\'a}}, J.~J., {Clarke}, J.~R.~A., {Contreras}, C., {Corvill{\'o}n},
  A., {de Grijs}, R., {Dias}, B., {Drew}, J.~E., {Fari{\~n}a}, C., {Feinstein},
  C., {Fern{\'a}ndez-Laj{\'u}s}, E., {Gamen}, R.~C., {Gieren}, W., {Goldman},
  B., {Gonz{\'a}lez-Fern{\'a}ndez}, C., {Grand}, R.~J.~J., {Gunthardt}, G.,
  {Hambly}, N.~C., {Hanson}, M.~M., {He{\l}miniak}, K.~G., {Hoare}, M.~G.,
  {Huckvale}, L., {Jord{\'a}n}, A., {Kinemuchi}, K., {Longmore}, A.,
  {L{\'o}pez-Corredoira}, M., {Maccarone}, T., {Majaess}, D., {Mart{\'{\i}}n},
  E.~L., {Masetti}, N., {Mennickent}, R.~E., {Mirabel}, I.~F., {Monaco}, L.,
  {Morelli}, L., {Motta}, V., {Palma}, T., {Parisi}, M.~C., {Parker}, Q.,
  {Pe{\~n}aloza}, F., {Pietrzy{\'n}ski}, G., {Pignata}, G., {Popescu}, B.,
  {Read}, M.~A., {Rojas}, A., {Roman-Lopes}, A., {Ruiz}, M.~T., {Saviane}, I.,
  {Schreiber}, M.~R., {Schr{\"o}der}, A.~C., {Sharma}, S., {Smith}, M.~D.,
  {Sodr{\'e}}, L., {Stead}, J., {Stephens}, A.~W., {Tamura}, M., {Tappert}, C.,
  {Thompson}, M.~A., {Valenti}, E., {Vanzi}, L., {Walton}, N.~A., {Weidmann},
  W., \& {Zijlstra}, A. 2012, \aap, 537, A107

\bibitem[{{Salpeter}(1955)}]{Salpeter_55}
{Salpeter}, E.~E. 1955, \apj, 121, 161

\bibitem[{{Sch{\"o}del}(2011)}]{Schoedel_11a}
{Sch{\"o}del}, R. 2011, in Astronomical Society of the Pacific Conference
  Series, Vol. 439, The Galactic Center: a Window to the Nuclear Environment of
  Disk Galaxies, ed. {M.~R.~Morris, Q.~D.~Wang, \& F.~Yuan}, 222

\bibitem[{{Sch{\"o}del} {et~al.}(2007){Sch{\"o}del}, {Eckart}, {Alexander},
  {Merritt}, {Genzel}, {Sternberg}, {Meyer}, {Kul}, {Moultaka}, {Ott}, \&
  {Straubmeier}}]{Schoedel_07}
{Sch{\"o}del}, R., {Eckart}, A., {Alexander}, T., {Merritt}, D., {Genzel}, R.,
  {Sternberg}, A., {Meyer}, L., {Kul}, F., {Moultaka}, J., {Ott}, T., \&
  {Straubmeier}, C. 2007, \aap, 469, 125

\bibitem[{{Sch{\"o}del} {et~al.}(2014){Sch{\"o}del}, {Feldmeier}, {Kunneriath},
  {Stolovy}, {Neumayer}, {Amaro-Seoane}, \& {Nishiyama}}]{Schoedel_14}
{Sch{\"o}del}, R., {Feldmeier}, A., {Kunneriath}, D., {Stolovy}, S.,
  {Neumayer}, N., {Amaro-Seoane}, P., \& {Nishiyama}, S. 2014, \aap, 566, A47

\bibitem[{{Sch{\"o}del} {et~al.}(2008){Sch{\"o}del}, {Merritt}, \&
  {Eckart}}]{Schoedel_08a}
{Sch{\"o}del}, R., {Merritt}, D., \& {Eckart}, A. 2008, Journal of Physics
  Conference Series, 131, 012044

\bibitem[{{Sch{\"o}del} {et~al.}(2009){Sch{\"o}del}, {Merritt}, \&
  {Eckart}}]{Schoedel_09a}
---. 2009, \aap, 502, 91

\bibitem[{{Sch{\"o}del} {et~al.}(2010){Sch{\"o}del}, {Najarro}, {Muzic}, \&
  {Eckart}}]{Schoedel_09b}
{Sch{\"o}del}, R., {Najarro}, F., {Muzic}, K., \& {Eckart}, A. 2010, \aap, 511,
  A18+

\bibitem[{{Sch{\"o}del} {et~al.}(2002){Sch{\"o}del}, {Ott}, {Genzel},
  {Hofmann}, {Lehnert}, {Eckart}, {Mouawad}, {Alexander}, {Reid}, {Lenzen},
  {Hartung}, {Lacombe}, {Rouan}, {Gendron}, {Rousset}, {Lagrange}, {Brandner},
  {Ageorges}, {Lidman}, {Moorwood}, {Spyromilio}, {Hubin}, \&
  {Menten}}]{Schoedel_02}
{Sch{\"o}del}, R., {Ott}, T., {Genzel}, R., {Hofmann}, R., {Lehnert}, M.,
  {Eckart}, A., {Mouawad}, N., {Alexander}, T., {Reid}, M.~J., {Lenzen}, R.,
  {Hartung}, M., {Lacombe}, F., {Rouan}, D., {Gendron}, E., {Rousset}, G.,
  {Lagrange}, A.-M., {Brandner}, W., {Ageorges}, N., {Lidman}, C., {Moorwood},
  A.~F.~M., {Spyromilio}, J., {Hubin}, N., \& {Menten}, K.~M. 2002, \nat, 419,
  694

\bibitem[{{Sch{\"o}nrich} {et~al.}(2010){Sch{\"o}nrich}, {Binney}, \&
  {Dehnen}}]{Schoenrich_10}
{Sch{\"o}nrich}, R., {Binney}, J., \& {Dehnen}, W. 2010, \mnras, 403, 1829

\bibitem[{{Schreiber} {et~al.}(2004){Schreiber}, {Thatte}, {Eisenhauer},
  {Tecza}, {Abuter}, \& {Horrobin}}]{Schreiber_04}
{Schreiber}, J., {Thatte}, N., {Eisenhauer}, F., {Tecza}, M., {Abuter}, R., \&
  {Horrobin}, M. 2004, in Astronomical Society of the Pacific Conference
  Series, Vol. 314, Astronomical Data Analysis Software and Systems (ADASS)
  XIII, ed. {F.~Ochsenbein, M.~G.~Allen, \& D.~Egret}, 380--+

\bibitem[{{Scott}(1992)}]{Scott_92}
{Scott}, D.~W. 1992, {Multivariate Density Estimation}

\bibitem[{{Sellgren} {et~al.}(1987){Sellgren}, {Hall}, {Kleinmann}, \&
  {Scoville}}]{Sellgren_87}
{Sellgren}, K., {Hall}, D.~N.~B., {Kleinmann}, S.~G., \& {Scoville}, N.~Z.
  1987, \apj, 317, 881

\bibitem[{{Serabyn} {et~al.}(1986){Serabyn}, {Guesten}, {Walmsley}, {Wink}, \&
  {Zylka}}]{Serabyn_86}
{Serabyn}, E., {Guesten}, R., {Walmsley}, J.~E., {Wink}, J.~E., \& {Zylka}, R.
  1986, \aap, 169, 85

\bibitem[{{Serabyn} \& {Lacy}(1985)}]{Serabyn_85}
{Serabyn}, E. \& {Lacy}, J.~H. 1985, \apj, 293, 445

\bibitem[{{Serabyn} \& {Morris}(1996)}]{Serabyn_96}
{Serabyn}, E. \& {Morris}, M. 1996, \nat, 382, 602

\bibitem[{{Sersic}(1968)}]{Sersic_68}
{Sersic}, J.~L. 1968, {Atlas de galaxias australes}

\bibitem[{{Seth} {et~al.}(2008){Seth}, {Blum}, {Bastian}, {Caldwell}, \&
  {Debattista}}]{Seth_08}
{Seth}, A.~C., {Blum}, R.~D., {Bastian}, N., {Caldwell}, N., \& {Debattista},
  V.~P. 2008, \apj, 687, 997

\bibitem[{{Seth} {et~al.}(2010){Seth}, {Cappellari}, {Neumayer}, {Caldwell},
  {Bastian}, {Olsen}, {Blum}, {Debattista}, {McDermid}, {Puzia}, \&
  {Stephens}}]{Seth_10}
{Seth}, A.~C., {Cappellari}, M., {Neumayer}, N., {Caldwell}, N., {Bastian}, N.,
  {Olsen}, K., {Blum}, R.~D., {Debattista}, V.~P., {McDermid}, R., {Puzia}, T.,
  \& {Stephens}, A. 2010, \apj, 714, 713

\bibitem[{{Sollima} {et~al.}(2012){Sollima}, {Bellazzini}, \&
  {Lee}}]{Sollima_12}
{Sollima}, A., {Bellazzini}, M., \& {Lee}, J.-W. 2012, \apj, 755, 156

\bibitem[{{Strader} {et~al.}(2011){Strader}, {Caldwell}, \&
  {Seth}}]{Strader_11}
{Strader}, J., {Caldwell}, N., \& {Seth}, A.~C. 2011, \aj, 142, 8

\bibitem[{{Tegmark} {et~al.}(2004){Tegmark}, {Strauss}, {Blanton}, {Abazajian},
  {Dodelson}, {Sandvik}, {Wang}, {Weinberg}, {Zehavi}, {Bahcall}, {Hoyle},
  {Schlegel}, {Scoccimarro}, {Vogeley}, {Berlind}, {Budavari}, {Connolly},
  {Eisenstein}, {Finkbeiner}, {Frieman}, {Gunn}, {Hui}, {Jain}, {Johnston},
  {Kent}, {Lin}, {Nakajima}, {Nichol}, {Ostriker}, {Pope}, {Scranton},
  {Seljak}, {Sheth}, {Stebbins}, {Szalay}, {Szapudi}, {Xu}, {Annis},
  {Brinkmann}, {Burles}, {Castander}, {Csabai}, {Loveday}, {Doi}, {Fukugita},
  {Gillespie}, {Hennessy}, {Hogg}, {Ivezi{\'c}}, {Knapp}, {Lamb}, {Lee},
  {Lupton}, {McKay}, {Kunszt}, {Munn}, {O'Connell}, {Peoples}, {Pier},
  {Richmond}, {Rockosi}, {Schneider}, {Stoughton}, {Tucker}, {vanden Berk},
  {Yanny}, \& {York}}]{Tegmark_04}
{Tegmark}, M., {Strauss}, M.~A., {Blanton}, M.~R., {Abazajian}, K., {Dodelson},
  S., {Sandvik}, H., {Wang}, X., {Weinberg}, D.~H., {Zehavi}, I., {Bahcall},
  N.~A., {Hoyle}, F., {Schlegel}, D., {Scoccimarro}, R., {Vogeley}, M.~S.,
  {Berlind}, A., {Budavari}, T., {Connolly}, A., {Eisenstein}, D.~J.,
  {Finkbeiner}, D., {Frieman}, J.~A., {Gunn}, J.~E., {Hui}, L., {Jain}, B.,
  {Johnston}, D., {Kent}, S., {Lin}, H., {Nakajima}, R., {Nichol}, R.~C.,
  {Ostriker}, J.~P., {Pope}, A., {Scranton}, R., {Seljak}, U., {Sheth}, R.~K.,
  {Stebbins}, A., {Szalay}, A.~S., {Szapudi}, I., {Xu}, Y., {Annis}, J.,
  {Brinkmann}, J., {Burles}, S., {Castander}, F.~J., {Csabai}, I., {Loveday},
  J., {Doi}, M., {Fukugita}, M., {Gillespie}, B., {Hennessy}, G., {Hogg},
  D.~W., {Ivezi{\'c}}, {\v Z}., {Knapp}, G.~R., {Lamb}, D.~Q., {Lee}, B.~C.,
  {Lupton}, R.~H., {McKay}, T.~A., {Kunszt}, P., {Munn}, J.~A., {O'Connell},
  L., {Peoples}, J., {Pier}, J.~R., {Richmond}, M., {Rockosi}, C., {Schneider},
  D.~P., {Stoughton}, C., {Tucker}, D.~L., {vanden Berk}, D.~E., {Yanny}, B.,
  \& {York}, D.~G. 2004, \prd, 69, 103501

\bibitem[{{Tremaine} {et~al.}(2002){Tremaine}, {Gebhardt}, {Bender}, {Bower},
  {Dressler}, {Faber}, {Filippenko}, {Green}, {Grillmair}, {Ho}, {Kormendy},
  {Lauer}, {Magorrian}, {Pinkney}, \& {Richstone}}]{Tremaine_02}
{Tremaine}, S., {Gebhardt}, K., {Bender}, R., {Bower}, G., {Dressler}, A.,
  {Faber}, S.~M., {Filippenko}, A.~V., {Green}, R., {Grillmair}, C., {Ho},
  L.~C., {Kormendy}, J., {Lauer}, T.~R., {Magorrian}, J., {Pinkney}, J., \&
  {Richstone}, D. 2002, \apj, 574, 740

\bibitem[{{Tremaine} {et~al.}(1994){Tremaine}, {Richstone}, {Byun}, {Dressler},
  {Faber}, {Grillmair}, {Kormendy}, \& {Lauer}}]{Tremaine_94}
{Tremaine}, S., {Richstone}, D.~O., {Byun}, Y.-I., {Dressler}, A., {Faber},
  S.~M., {Grillmair}, C., {Kormendy}, J., \& {Lauer}, T.~R. 1994, \aj, 107, 634

\bibitem[{{Tremaine} {et~al.}(1975){Tremaine}, {Ostriker}, \&
  {Spitzer}}]{Tremaine_75}
{Tremaine}, S.~D., {Ostriker}, J.~P., \& {Spitzer}, Jr., L. 1975, \apj, 196,
  407

\bibitem[{{Trippe} {et~al.}(2008){Trippe}, {Gillessen}, {Gerhard}, {Bartko},
  {Fritz}, {Maness}, {Eisenhauer}, {Martins}, {Ott}, {Dodds-Eden}, \&
  {Genzel}}]{Trippe_08}
{Trippe}, S., {Gillessen}, S., {Gerhard}, O.~E., {Bartko}, H., {Fritz}, T.~K.,
  {Maness}, H.~L., {Eisenhauer}, F., {Martins}, F., {Ott}, T., {Dodds-Eden},
  K., \& {Genzel}, R. 2008, \aap, 492, 419

\bibitem[{{Ulubay-Siddiki} {et~al.}(2013){Ulubay-Siddiki}, {Bartko}, \&
  {Gerhard}}]{Ayse_13}
{Ulubay-Siddiki}, A., {Bartko}, H., \& {Gerhard}, O. 2013, \mnras, 428, 1986

\bibitem[{{van de Ven} {et~al.}(2006){van de Ven}, {van den Bosch}, {Verolme},
  \& {de Zeeuw}}]{Vandeven_06}
{van de Ven}, G., {van den Bosch}, R.~C.~E., {Verolme}, E.~K., \& {de Zeeuw},
  P.~T. 2006, \aap, 445, 513

\bibitem[{{van der Marel} \& {Anderson}(2010)}]{Marel_10}
{van der Marel}, R.~P. \& {Anderson}, J. 2010, \apj, 710, 1063

\bibitem[{{van Dokkum} \& {Conroy}(2010)}]{Vandokkum_10}
{van Dokkum}, P.~G. \& {Conroy}, C. 2010, \nat, 468, 940

\bibitem[{{Vollmer} {et~al.}(2003){Vollmer}, {Zylka}, \& {Duschl}}]{Vollmer_03}
{Vollmer}, B., {Zylka}, R., \& {Duschl}, W.~J. 2003, \aap, 407, 515

\bibitem[{{Walcher} {et~al.}(2006){Walcher}, {B{\"o}ker}, {Charlot}, {Ho},
  {Rix}, {Rossa}, {Shields}, \& {van der Marel}}]{Walcher_06}
{Walcher}, C.~J., {B{\"o}ker}, T., {Charlot}, S., {Ho}, L.~C., {Rix}, H.-W.,
  {Rossa}, J., {Shields}, J.~C., \& {van der Marel}, R.~P. 2006, \apj, 649, 692

\bibitem[{{Walcher} {et~al.}(2005){Walcher}, {van der Marel}, {McLaughlin},
  {Rix}, {B{\"o}ker}, {H{\"a}ring}, {Ho}, {Sarzi}, \& {Shields}}]{Walcher_05}
{Walcher}, C.~J., {van der Marel}, R.~P., {McLaughlin}, D., {Rix}, H.-W.,
  {B{\"o}ker}, T., {H{\"a}ring}, N., {Ho}, L.~C., {Sarzi}, M., \& {Shields},
  J.~C. 2005, \apj, 618, 237

\bibitem[{{Yu} \& {Tremaine}(2003)}]{Yu_03}
{Yu}, Q. \& {Tremaine}, S. 2003, \apj, 599, 1129

\bibitem[{{Zhao} {et~al.}(2009){Zhao}, {Morris}, {Goss}, \& {An}}]{Zhao_09}
{Zhao}, J.-H., {Morris}, M.~R., {Goss}, W.~M., \& {An}, T. 2009, \apj, 699, 186

\bibitem[{{Zoccali} {et~al.}(2000){Zoccali}, {Cassisi}, {Frogel}, {Gould},
  {Ortolani}, {Renzini}, {Rich}, \& {Stephens}}]{Zoccali_00}
{Zoccali}, M., {Cassisi}, S., {Frogel}, J.~A., {Gould}, A., {Ortolani}, S.,
  {Renzini}, A., {Rich}, R.~M., \& {Stephens}, A.~W. 2000, \apj, 530, 418

\end{thebibliography}

\appendix

 \section{A: Derivation of Proper Motions}
\label{sec:ap_prom_mot}

In this section we explain how we derived the proper motions. We progress from the center outwards.
The star numbers for each field are obtained after exclusion of the stars described in Appendix~\ref{sec:app_clean}.

\begin{deluxetable}{llllll} 
\tabletypesize{\scriptsize}
\tablecolumns{6}
\tablewidth{30pc}
\tablecaption{Images used for the extended field \label{tab:trip_27}}
\tablehead{ time [mjd] & time & Band & stars on image &  fraction of good stars & median position error [mas]}
\startdata
52397.5 & 2002.334 &  Ks  & 4151 & 0.932 & 0.98 \\
52769.5 & 2003.352 &  Ks  & 4914 & 0.974 & 0.65 \\
53168.5 & 2004.445 &  IB2.06  & 5265 & 0.981 & 0.47 \\
53168.5 & 2004.445 &  IB2.24  & 5125 & 0.961 & 0.66 \\
53169.5 & 2004.448 &  IB2.33  & 5194 & 0.961 & 0.51 \\
53169.5 & 2004.448 &  NB2.17  & 5246 & 0.905 & 1.86 \\
53191.5 & 2004.508 &  Ks  & 5271 & 0.948 & 1.10 \\
53502.5 & 2005.359 &  Ks  & 6033 & 0.984 & 0.71 \\
53540.5 & 2005.463 &  Ks  & 5991 & 0.807 & 3.09 \\
53854.5 & 2006.323 &  H  & 5338 & 0.935 & 1.16 \\
53854.5 & 2006.323 &  Ks  & 5340 & 0.964 & 1.00 \\
53975.5 & 2006.654 &  H  & 5673 & 0.522 & 3.28 \\
53975.5 & 2006.654 &  Ks  & 5687 & 0.698 & 2.70 \\
54175.5 & 2007.202 &  Ks  & 5768 & 0.985 & 0.41 \\
54190.5 & 2007.243 &  Ks  & 5995 & 0.949 & 0.73 \\
54561.5 & 2008.259 &  Ks  & 5921 & 0.974 & 0.44 \\
54561.5 & 2008.259 &  Ks  & 5746 & 0.987 & 0.75 \\
54683.5 & 2008.593 &  Ks  & 5742 & 0.986 & 0.36 \\
54725.5 & 2008.708 &  Ks  & 5958 & 0.977 & 0.5 \\
54919.5 & 2009.239 &  Ks  & 5982 & 0.984 & 0.73 \\
55094.5 & 2009.718 &  Ks  & 5979 & 0.980 & 0.45 \\
55325.5 & 2010.350 &  H  & 5971 & 0.843 & 1.78 \\
55325.5 & 2010.350 &  Ks  & 5978 & 0.967 & 0.99 \\
55467.5 & 2010.739 &  Ks  & 5981 & 0.948 & 0.97 \\
55652.5 & 2011.246 &  Ks  & 5975 & 0.867 & 1.45 \\
55697.5 & 2011.369 &  Ks  & 6037 & 0.966 & 0.61 \\
\enddata
\tablecomments{
All images are obtained in the 27 mas scales with NACO/VLT. At maximum all 6037 stars can be detected on an image.
The calculation of the part of good stars excludes the stars not
on the image.
}
\end{deluxetable}

\begin{deluxetable}{llll} 
\tabletypesize{\scriptsize}
\tablecolumns{4}
\tablewidth{18pc}
\tablecaption{Images used for derivation of proper motions in the large and outer field \label{tab:big_im}}
\tablehead{ Telescope/Instrument & epoch & R.A.  [$\arcsec$] &  Dec.  [$\arcsec$]
}
\startdata

Gemini/Hokupa'a & 2000-07-02 & -13 to 31 & -6 to 75 \\
VLT/NACO & 2004-05-06 & -37 to 36 & -38 to 35 \\
VLT/NACO & 2006-04-29 & -20 to 22 & -16 to 25 \\
VLT/MAD & 2008-08-21 & -37 to 35 & -48 to 27 \\
VLT/NACO & 2011-05-16 & -44 to 44 & -44 to 44 \\
VLT/NACO & 2011-05-29 & 0 to 28 & 45 to 73 \\
\enddata
 \tablecomments{The field of view of the first two datasets is not approximately rectangular. 
 }
\end{deluxetable}

\subsection{The Central Field}
\label{sec:pm_center}

In the center (R$\leq 2\arcsec$) the crowding is strong. Thus it is difficult to fit the stars on the 
images with simple Gaussians. Instead, we extract the point spread function (PSF) from the images and 
deconvolve the image with the Lucy-Richardson algorithm, as in \citet{Gillessen_09}. Compared to that
work we expand the time baseline and enlarge the field of view from about 1$\arcsec$ to 2$\arcsec$.

We fit the astrometric data points and their errors for all stars by linear fits. We rescale the $\chi^2/d.o.f.$ to 1 as in \citet{Gillessen_09}.
 Some of the stars have significant accelerations \citep{Gillessen_09}. However, to avoid complicating our analysis by including acceleration for a 
very small fraction of the stars, we calculate for all stars only the linear motion. The velocity errors are small,
 with an average 0.038 mas/yr compared to the velocity
dispersion of 5.23 mas/yr. Furthermore, due to the large number of data points (more than 100 for many stars) the errors can be determined well from rescaling.
For the stars with significant acceleration the errors are not strictly right, since the error calculation assumes random errors. 
However, since the cubic deviation from linear motion is nearly never significant \citep{Gillessen_09}
 the derived velocity is close to the velocity at the mean time of the observing interval. The derived error is
slightly too big for this definition, since the scatter is mainly caused by non-random accelerations.

All the fit errors, even the largest of 0.71 mas/yr for the velocity of -19.09 mas/yr for S38, are too small
 to have any influence compared to the large 
dispersion of $\sigma_{\mathrm{1D}}=$5.25 mas/yr and its associated Poisson error 
of 0.30 mas/yr. ($\sigma_{\mathrm{1D}}$ is as in the following the average dispersion using all 1D velocities in x and y.) Therefore, we neglect the error uncertainties for proper motions in the central two arcseconds in our analysis.

\subsection{The Extended Field}
\label{sec:tri_fie}

This data set is an update of \citet{Trippe_08}. This field extends out to 20$\arcsec$.
We exclude stars, which are within in the central field (Appendix~\ref{sec:pm_center}). 

Compared to \citet{Trippe_08} we add new epochs (see Table~\ref{tab:trip_27} for a list of all images) and 
a new conversion to absolute coordinates aligned with \citet{Gillessen_09}. The positions of the stars in these images are obtained
 by fitting Gaussians to distortion corrected images. The distortion correction 
 is applied in the
same way as in \citet{Trippe_08}.
Like them we use in addition following linear transformation:

\begin{equation}
\begin {split}
x'=a_0+a_{1} \times x+a_{2} \times y 
\\
y'=b_0+b_{1} \times x+b_{2} \times y 
\end{split}
\label{eq:lin_traf}
\end{equation}

 This transformation contains also crossterms. Thus, it corrects also automatically for linear effects like differential atmospheric distortion.
Only sources with no close neighbor on the images are included in the dataset. When a source
 consists of two very close neighbors it can be included in our data set. In such cases the velocity is a flux weighted average, which
 reduces the absolute velocities and therefore in average also the dispersion. In the center we have also
deconvolved images available over 10 years. Checking them shows
that source confusion affects less than 1/10 of the sources in the extended field. Further out the source density is smaller and thus
source confusion introduces no relevant dispersion bias in the extended field data set.

\begin{figure}
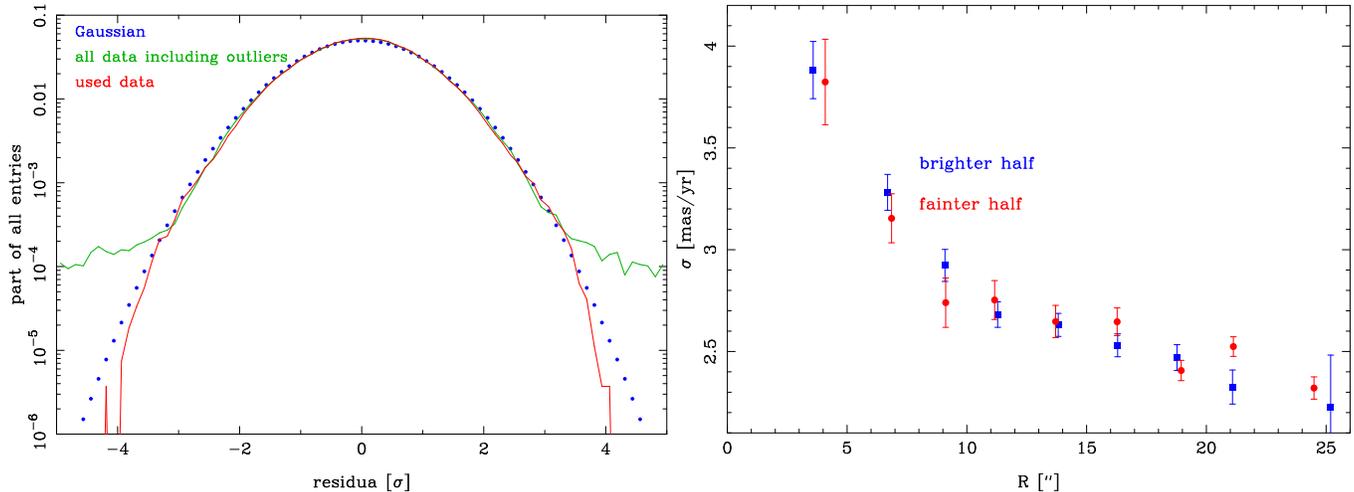

\begin{center}
\includegraphics[width=0.36 \columnwidth,angle=-90]{fa1a.eps}
\includegraphics[width=0.36 \columnwidth,angle=-90]{fa1b.eps}
\caption{Residua and dispersion in the extended field.
Left: Histogram of the position residuals compared to the velocity fits.  The red line shows the histogram of the 
position residua
which we use for our final $2\times6037$ velocity fits.  
The green line also includes the outliers which are not fit.
 It can be seen
that the residuals to our fit are close to a Gaussian (blue dots). 
Right: Influence of magnitude on the dispersion. We divide
the stars in two groups by their flux and bin the stars then. The same flux threshold is used in all bins.
} 
\label{fig:_residua0}
\end{center}
\end{figure}

We change the procedure of the outlier rejection and error calculation compared to \citet{Trippe_08}.
For the first fit we use for each stars the position uncertainty of the Gaussian on the image and rescale the $\chi^2/d.o.f.$ of the fit to 1.
We then calculate for each image the residua 
distribution of all stars compared to the fit.
The width including 68.3~\%  ($\equiv$W1S) 
of all entries is often quite different from 1$\,\sigma$. To change this we iterate the fitting.
In the iteration we rescale the errors on each image
by the W1S in $\sigma$. We repeat this progress a second time, then the variation of the W1S between different images is less than 0.1$\,\sigma$.
Other factors are then more important like magnitude with up to 0.2$\,\sigma$ deviation. However, we ignore these effects because the sign of the 
deviations is randomly distributed over
our 26 images. For the final fit we exclude all 5 $\sigma$ outliers. Outliers between 2.5 and 5 $\sigma$ we reject by drawing random numbers to 
achieve that the distribution of all residual 
approximates a Gaussian distribution. 
 We give in Table~\ref{tab:trip_27} the median error for all images used.
The final residual distribution is approximately Gaussian up to $|3.7|\,\sigma$, see Figure~\ref{fig:_residua0}.

Assuming only Poisson errors the derived dispersion is $\sigma_{\mathrm{1D}}=2.677 \pm 0.018$.
The error on the dispersion in this field is so small that other error sources could be important. 
The median and average velocity errors are 0.150 and 0.208 mas/yr, respectively.
They are so small, that it does not matter which of them is used. However, this assumes that they are correctly determined.
One possible check is to measure the dependence of the dispersion on magnitude \citep{Clarkson_12}.
For this test we divide our data into radial bins, to avoid that the radial trend dilutes a signal, and measure the dispersion in these bins
for the brighter and fainter stars, using the same threshold magnitude in all bins, see Figure~\ref{fig:_residua0}.
Taking the average of these bins, the difference between fainter and brighter stars is $\delta\sigma_{1D}=0.007\pm0.041$mas/yr. The error
is the scatter divided by $\sqrt{N_{\mathrm{bins}}}$ which is consistent with Poisson errors.
The consistent dispersions imply that our stars are bright and isolated enough that magnitude dependent errors like photon or halo noise 
\citep{Fritz_09} do not contribute notably. Thus, Poisson errors are likely the most important error source for the
dispersion and we use only them.

\subsection{The Large Field}
\label{sec:big_fie}

For deriving proper motions between 20 and 45$\arcsec$ we use the images listed in Table~\ref{tab:big_im}. 
(The last image of this table is used only for deriving proper motions in the outer field, see Appendix~\ref{sec:out_fie}.) 

We choose as the master image the image from the 16$^{\mathrm{th}}$ May 2011. This image is the largest in most directions and 
has, thanks to 0.4$\arcsec$ DIMM seeing,
the best resolution. The overlap between its 16 different pointings is too small to obtain a reliable distortion correction in the way
 of \citet{Trippe_08}. The NACO distortion is, however, stable within instruments interventions, see \citet{Fritz_09}. We therefore use the distortion solution for the distortion correction
that is valid  during this epoch. Apart from this higher-than-linear distortion it is also necessary 
to align the scales and pointings of the images by linear transformations. For this we follow also the procedure of \citet{Trippe_08}.

We again use the full linear transformation (Formula~\ref{eq:lin_traf}).
 On this final distortion corrected 
and aligned image we search for stars and fit them with two-dimensional Gaussians. 
We thereby exclude stars with close neighbors to reduce the influence of neighboring seeing halos \citep{Fritz_09} on the position of the target
 stars. We then translate the pixel positions of the stars to arcseconds by using the known positions and motions
 of bright stars in the central 20$\arcsec$ using our extended field sample.

For the NACO images in the other epochs we removed the distortion from all single pointing images with the distortion solution valid during this time. 
For the  image from the 6$^{\mathrm{th}}$ May 2004, the overlap of the different pointings is too small for a reliable alignment of the different pointings. 
In contrast, the pointing of the images from the  29$^{\mathrm{th}}$ April 2006 overlap enough to apply a reliable alignment.
In consequence,  we have in most epochs many, mostly single pointing 
 images of different parts of the GC. 
The images overlap partly.  We identify on all images ten bright stars from the master image for preliminary alignment of the images. 
We then search on these images around the expected star positions from 2011 within a radius of 3 pixels for the local maxima. 
When a maximum is fittable by a Gaussian similar to the PSF core of the image we treat the star as identified on this single pointing 
image. We then use all stars identified on the single pointing image to obtain the cubic transformation to the master image. 
This transformation is defined in the following way:

\begin{equation*}
\begin {split}
x'=a_0+a_{1} \times x+a_{2} \times y + a_{3} \times x^2+ a_4 \times xy+
a_5 \times y^2 \\+ a_6 \times x^3 
+a_7 \times x^2y+a_8 \times xy^2+a_9 \times y^3
\end{split}
\end{equation*}

\begin{equation}
\begin {split}
y'=b_0+b_{1} \times x+b_{2} \times y + b_{3} \times x^2+ b_4 \times xy +b_5 \times
y^2\\+ b_6 \times x^3+b_7 \times x^2y +b_8 \times xy^2+b_9 \times y^3
\end{split}
\label{eq:cub_traf}
\end{equation}

The primary purpose of applying this cubic transformation is to correct unknown distortions together with the other mostly linear effects 
\citep{Fritz_09}.
The number of stars used for the transformation depends on the field of view of the detector and the density of well detected stars. 
For most epochs we have on average about 4500, and at least 1000 stars. For the single NACO pointings of the 6$^{\mathrm{th}}$ May 2004 
there are in average about 836 stars, and at least 660 stars.
Only for the smaller Gemini field of view we need to use fewer stars. On these image, there are on average 644, and on one only 58. 
Thus, all images have enough stars to average out the influence of the intrinsic motion of the stars on the cubic transformation.
We use the median of all detections in each epoch as the position of the star in this epoch.

\begin{figure}
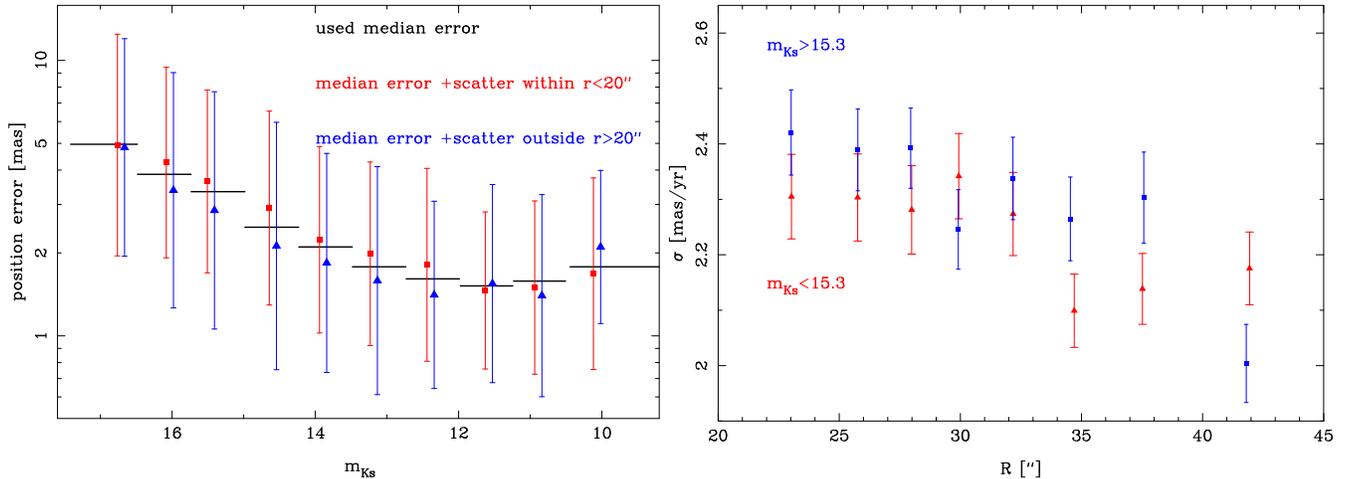

\begin{center}
  \includegraphics[width=0.35 \columnwidth,angle=-90]{fa2a.eps}
 \includegraphics[width=0.35 \columnwidth,angle=-90]{fa2b.eps} 
\caption{
Position errors and dispersion in the large field.
Left: dependence of the error on magnitude.
As the scatter we plot $1.483\,\times$ the median deviation (This measure is identical to 1$\,\sigma$ if the distribution is Gaussian.) of the stars in this
 bin. The errors in magnitude bins for stars within and outside the central r$=20\arcsec$ box are similar.
The black line shows the error as function of magnitude which we use for velocity error calculation. 
Right: dependence of the dispersion on magnitude. We present in each distance bin the dispersion
of brighter and fainter stars.
} 
\label{fig:_poser1}
\end{center}
\end{figure}

To obtain the velocity errors we start with  the same position error for all epochs and one that is three times smaller for 
the master image, because it is the best image and is mostly 
corrected for distortion. We rescale the errors such that the reduced $\chi^2$ has the 
expected value. The errors obtained this way vary a lot at each magnitude. 
In the magnitude bins the error of the stars at the 75~\% quantile of the error distribution is 3.4 times 
larger than the error at the 25~\% quantile of the error distribution. 
 A broad fit error distribution is expected also for perfectly Gaussian distribution of position
errors in our case  of fitting only three to five data points. Thus, obtaining the errors by rescaling is not possible. 
Since a clear correlation of the errors with magnitude is
 visible, see Figure~\ref{fig:_poser1}, we calculate the median error in magnitude bins and use it as the error for most stars. 
We use the rescaled error only if the rescaled error is more than four times larger the median error. We exclude from our sample stars those that have in at least one dimension errors larger than 10 mas/yr. When the error is so large, the 
velocity cannot be determined reliably.
For the other stars we use the same error for both dimensions.
The faintest stars have an error three times larger than bright stars (Figure~\ref{fig:_poser1}). This behavior is expected, since many errors are more 
important for fainter stars,
 see e.g. \citet{Fritz_09}. Due to saturation effects the errors increase somewhat again for very bright stars. 
Another possible error source is that  the remaining distortion could be larger in the outer parts of the field than in the center. To test 
this we compare the errors inside and outside the central r$=20\arcsec$ box, see Figure~\ref{fig:_poser1}. We find no major difference between the errors
 inside and outside this box: the errors inside have a median 23~\% bigger than the errors outside, possibly due to the higher source density there.

In principle it is possible to calculate velocities from only two epochs. In this case it is however, not possible to calculate the errors
 from the fit to the data points. Thus, great care is necessary.
 To test the reliability of velocity measurements using only two images 
we compare, in the central part of the field of view, 
 robust dispersion measurements (the median deviation) using all epochs and using only the master epoch (2011) and a single other epoch. 
We find that the dispersions obtained from two epochs are compatible with the dispersion 
 from most data pairings with the master
epoch. The only exception is the 2008 MAD image. The dispersion from this
 image is about 16 \% greater than the dispersion when using all images.
 The reason for the greater dispersion is probably the small time baseline between 2008 and 2011 together with the fact that
 two different instruments (with partly unknown distortion) are used for this velocity measurement. 
By comparison to the dispersion using all data, it follows that the velocity error is 40 \% of the dispersion for the 2008-2011 pair. 
If we could be sure that the error has this value, we could use the dispersion after correcting for the biasing velocity error. 
However, the errors due to distortion, and maybe others, are probably not constant over the field of view. 
Even if the errors would increase only to 50 \% this would enlarge the dispersion values by 8\%. 
This error induced uncertainty is much larger for this data pair than for the other data pairs.
We therefore exclude stars only detected in 2011 and 2008 from the analysis. The stars detected only in two other datasets 
(Due to the covered fields they are mostly in the 2000 and 2004 data set.) show no higher dispersion. We include them in our analysis.

As in the extended field we further test if the dispersion depends on the magnitude, see Figure~\ref{fig:_poser1}. 
For this test we exclude  stars with only two measurements.
For this comparison, subtracting the errors is more important than closer to Sgr~A*, since the median error is 0.464 mas/yr.
For the subtraction we use  $1.2\times$ the  median error, since this is between the median error (which is too low because 
 larger errors contribute more to the overall error) and the average (which is affected by few unrealistically large errors).

The half with m$_{\mathrm{Ks}}>15.3$ has, after velocity error subtraction by
 $0.076 \pm 0.037$ mas/yr, larger errors than the brighter half.  
Thus, the dispersion possibly increases by 2$\,\sigma$ with magnitude. 
This is barely significant. Therefore, we do not include a systematic dispersion error uncertainty for this data set.
We then compare this data set with the extended field data set and exclude all sources within this inner field.
In total we have 3826 late-types stars in the large field.

\subsection{The Outer Field}
\label{sec:out_fie}

Of the two epochs of this field (Gemini in 2000 and NACO in 2011), 
the NACO data are of higher quality, so we use them 
 as the master images. We remove distortions with the distortion solution. We use STARFINDER \citep{Diolaiti_00} to obtain first estimates for
 the stellar positions with a single PSF extracted from the full
 field. In a second step we fit these stars with two-dimensional Gaussians and keep only the stars which are well fit by a Gaussian. 
We fit the same stars also on three other NACO images in other filters but with the same pointing.
The position in 2011 is the average of the four filters and the error is the scatter.  
The error in this epoch is negligible compared to the error in
 the Gemini epoch. The Gemini data of 2000 consists of 12 reduced images, covering 3 pointings. Thereby one pointing creates the overlap between the 
other two pointings, thus allowing 
estimates for distortion effects. We use ten stars on each Gemini image to establish a first linear transformation to the NACO data. We 
search then in a radius of three pixels around the expected star positions for maxima and fit them with two-dimensional Gaussian functions. All stars that are
found to be offset by less than 3 pixel from the NACO epoch position are used for a full cubic transformation (equation~\ref{eq:cub_traf}) of each Gemini 
image to the NACO image. These transformations use on average 470 and at least 133 stars. 

For the calculation of errors we use stars, which are on more than one pointing. Firstly, we calculate the average position of the different
images in one pointing. Secondly, we obtain the final position and its error by using the average and the scatter of
the different pointings.
Since errors obtained from the scatter of two or three images have a large scatter we construct a magnitude dependent error model. 
 Therefore we bin the stars by magnitude to obtain the median error.
We interpolate these 
data by a curve that has three components:  an error floor for the brightest magnitudes, photon noise  at intermediate magnitudes, and
sky/read-out noise at the faintest level \citep{Fritz_09}. If a star has an error more than three times the expected error this larger error is used. 
The floor of the error model results in a total velocity error of 0.090 mas/yr. This is much less than the dispersion of 2 mas/yr.
Thus magnitude-independent errors like distortion are not important in the outer field.

\begin{figure}
\begin{center}
\includegraphics[width=0.35 \columnwidth,angle=-90]{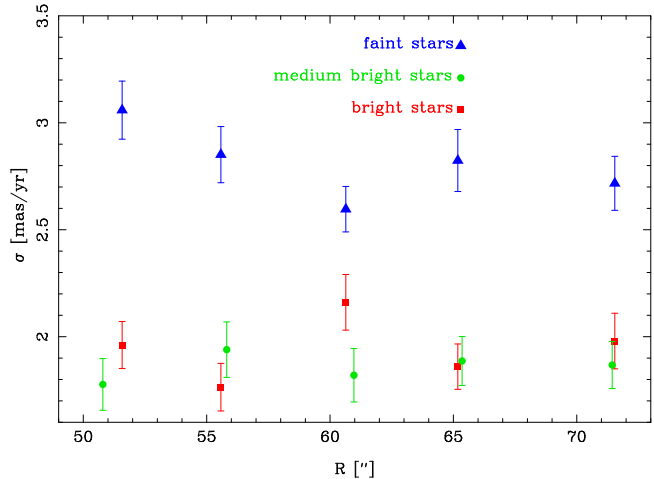}
\caption{
Dependence of the dispersion on the magnitude in the outer field.
The bin borders are m$_K=14.9$ and m$_K=15.6$. 
} 
\label{fig:_poser3}
\end{center}
\end{figure}

\begin{figure}
\begin{center}
\includegraphics[width=0.35 \columnwidth,angle=-90]{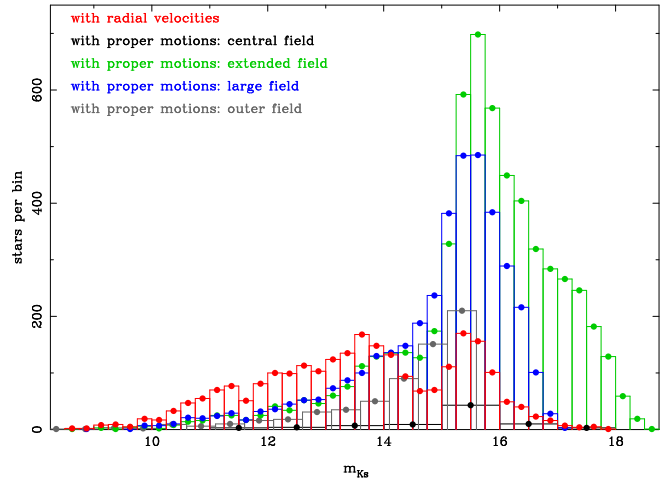}
\caption{
Magnitude histograms of the four proper motion samples and the CO radial velocity sample.  The bin sizes vary between the different samples to improve the visualization.} 
\label{fig:mag_his}
\end{center}
\end{figure}

We now test down to which magnitude 
the dispersions are reliable.
For this test we split the stars in three magnitude bins, see Figure~\ref{fig:_poser3}. In the brightest bins we use bright stars 
with m$_K<14.9$ for which the luminosity does
not limit astrometric accuracy. The upper magnitude edge of the second bin we choose such that the dispersion in this bin is identical to the dispersion in the brightest bin.
 This results in m$_K=15.6$ and a dispersion difference of $-0.087\pm 0.075$ mas/yr between the median and the brightest bin.
The fainter stars have a $0.864\pm 0.078$ mas/yr larger dispersion than the brightest stars. 
Partly the high dispersion in this bin is due to the difficulty to clean it from outliers.
We exclude the faint stars from the analysis and use further on only the 633
stars in two brighter bins. Therefore, the magnitude distribution has a sharp cutoff in that  field in contrast to the other proper motion fields (Figure~\ref{fig:mag_his}). There the selection is very complicated; it depends not only the magnitude of a star but also on the magnitude of the neighboring stars. However, the magnitude difference between the different proper motion fields is smaller than the magnitude difference between the proper motions sample and the radial velocity sample. Also, the magnitude difference between proper motions and radial velocities is not important (Section~\ref{sec:rad_vel}).
The total dispersion of both dimensions together is $\sigma_{\mathrm{1D}}=1.919 \pm 0.038\,$mas/yr.

 \section{B: Radial Velocities}
\label{sec:app_rv}

Here we present how we derive and collect the radial velocities within r$<4\,$pc from our SINFONI data. Further out
we retrieve data from the literature.

 \subsection{Our CO Radial Velocities}
\label{sec:app_our}

We extract the star spectra from the data cubes using on and off spaxels. 
For  measuring the radial velocities from the SINFONI data we use the strong and sharp CO band head features between 2.29 and 
2.37~$\mu$m. We cross-correlate the spectra with a template and correct the radial velocities to the LSR.
The statistical error on the radial velocity is obtained by the uncertainty of the best crossmatch in the correlation.
This error is typically larger than 4 km/s. The median velocity error obtained in this way is 7 km/s.
For a subset of more than 300 stars we have velocities available extracted from two or more entirely independent different data sets, which allows us to 
check the velocity error. The median velocity difference is 8.4 km/s, consistent with our error estimate. 

\begin{figure}
\begin{center}
\includegraphics[width=0.35 \columnwidth,angle=-90]{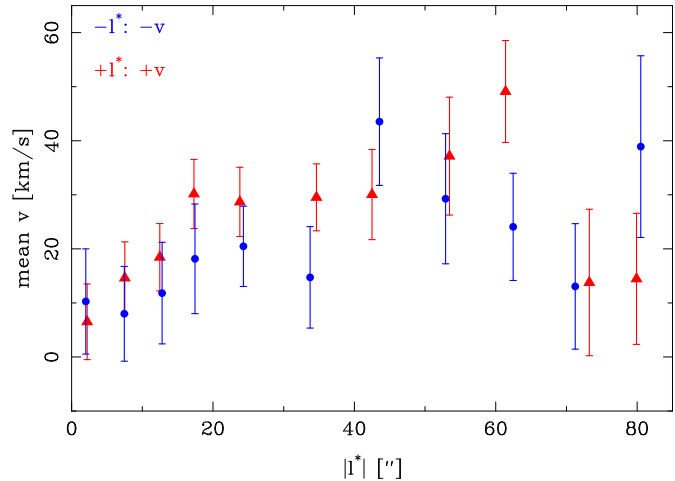}
\caption{ Mean radial velocities along the Galactic plane for our SINFONI sample. 
The signs of the velocities at negative $l^*$-values are reversed for better comparison. 
} 
\label{fig:_lsr}
\end{center}
\end{figure}

Compared to the total dispersion of  $\sigma_z=102.2$ km/s the discussed uncertainty of the velocities is small and negligible. However, this uncertainty does not include systematic problems due to calibration 
errors or template mismatch.
A possibility to check our velocity accuracy is the comparison with the LSR. No relative motion is expected between the average motion of the solar neighborhood and the nuclear cluster in the radial direction. The determination of the LSR  
is difficult in the direction of Galactic rotation
 \citep{Dehnen_98, Schoenrich_10,Bovy_12b}, but here only the radial component U matters. We use the LSR calculation of the ATCA 
array\footnote{http://www.narrabri.atnf.csiro.au/cgi-bin/obstools/velo.cgi?radec=17\%3A45\%3A40+-29\%3A00\%3A28\&velo=0\&frame=bary\&type=radio
\&date= 03\%2F07\%2F11\&freq1=0\&freq2=100\&\-telescope=atca}, which uses U$=$10.25 km/s, consistent with the recent values of 10.5 km/s 
\citep{Bovy_12b} and
 $11.1 \pm 1.2$ km/s \citep{Schoenrich_10}. 
The nuclear cluster is rotating in the Galactic plane \citep{McGinn_89,Lindqvist_92b,Genzel_96,Trippe_08}. Since our coverage is not homogeneous we 
cancel the rotation by using stellar number count weighted bins in $|l^*|$, see Figure~\ref{fig:_lsr}.
The mean radial velocity over all bins is $6.1 \pm 3.8$ km/s, consistent with 0. Reverting the argument, assuming that the 
calibration is correct, we can conclude that the nuclear cluster is moving less than 15 km/s radially relative to the LSR. 

In conclusion it seems likely that Poisson errors are dominating the dispersion uncertainty for our radial velocity sample, and we only include those errors.

 \subsection{Maser Velocities from the Literature}
\label{sec:app_maser}

In the central 100$\arcsec$ we ignore the literature of single star radial velocities \citep{Sellgren_87,Rieke_88,Genzel_96,Figer_03} because our sample is much larger than all previous 
samples. (The exception is \citet{Trippe_08}, but we incorporate all of their velocities corrected by a constant velocity to account 
for the different CO template used.) Although our sample has 2513 velocities it only covers the R$<4\,$pc. Since this is less than 
the size of the nuclear cluster we scan the literature for old star velocities further out. The sample
of \citet{Rieke_88} contains 11 CO based velocities there. However, we do not use them due to the small number and
 also because IR spectroscopy at this time possibly had difficulties with calibration as the rather different dynamics
in  \citet{Rieke_88} and \citet{McGinn_89} indicates (Section~\ref{sec:rotation}).

Instead we use the maser based velocities from \citet{Lindqvist_92a} and \citet{Deguchi_04}.
\citet{Lindqvist_92a} used the VLA for detecting OH masers in a blind Galactic Center survey. This survey covered the GC out to 
$\approx\,$3000$\arcsec$ from Sgr A*.  Stars with $|v|>$217 km/s were not detectable
due to the limited spectral range. Since only very few stars have a velocity close to this value probably only very few stars were missed. 
\citet{Lindqvist_92b} subdivided the masers into two classes according to their age (as estimated from the the expansion velocity 
v$_{\mathrm{exp}}$ of the 
masers), but found that both follow the same projected rotation curve, with the same sign and orientation as the Galactic rotation. We thus use both populations.

\citet{Deguchi_04} targeted large amplitude variables within 950$\arcsec$ from Sgr~A* as SiO masers 
 with the 45 m Nobeyama radio telescope. The beam 
of the observations was 40$\arcsec$. Their sample probably contains multiple identifications, as there are suspiciously many close neighbors (R$<40\arcsec$) that 
have velocity differences of less than 5 km/s. We clean the sample and remove also stars already present in the \citet{Lindqvist_92a} sample. 
As a side product of this matching, we confirm that the typical velocity uncertainty is less than 3 km/s as stated in \citet{Lindqvist_92a} and \citet{Deguchi_04} and therefore irrelevant.
Due to the big position errors for the masers it is difficult to find the corresponding IR stars.
We therefore exclude from the combined list the eleven stars that overlap spatially with the areas in which we obtained spectra, with the aim of 
avoiding using stars twice. Overall we use 274 radial velocities outside the central field. 
 After exclusion of outliers, see Section~\ref{sec:out_rej}, 261 stars remain in our final maser sample.

 \section{C: Sample cleaning}
 \label{sec:app_clean}

We here describe how we exclude stars from the sample which do not belong to the main old, probably relaxed stellar 
population of the GC.

 \subsection{By Stellar Type}
\label{sec:app_early}

Young stars are predicted to be unrelaxed \citep{Alexander_05} in the nuclear cluster.
For the really young stars, the early-type stars, the WR-, O- and B-stars \citep{Paumard_06}
and the red supergiant IRS7, the not relaxed state is confirmed by the observed dynamics \citep{Genzel_03b,Gillessen_09,Bartko_09,Bartko_10,Madigan_13}.
Surprisingly, the intermediate-age stars with ages of about 20 to 200 Myrs that constitute about 10\% of all late-type stars \citep{Pfuhl_11} share in dynamics and radial distribution the properties of the majority of the stars, which are more than 5 Gyrs old. 
Therefore we do not exclude them. 
Hence our selection criterion is a simple spectroscopic one: we exclude the early-type stars and IRS7
and use all late-type stars.  

Since the ratio of young to old stars is a strong function of radius 
\citep{Krabbe_91,Buchholz_09,Bartko_10,Do_13}, our selection criterion for old stars is radius dependent. 
In the central arcsecond spectroscopically we have identified more early-type stars than late-type stars, see also 
Figure~1 of \citet{Gillessen_09}. Since the KLF slope of the young stars in the central 
arcsecond \citep{Bartko_10} is identical to the one of the late-type stars \citep{Buchholz_09,Pfuhl_11}, the majority of the stars without spectral identification is 
probably young there. Outside the central arcsecond the fraction of young stars decreases, but remains high for R$\leq2\arcsec$
 \citep{Buchholz_09,Do_09,Bartko_10}. We therefore include for R$\leq2\arcsec$ only stars in our
sample that are spectroscopically confirmed late-type stars. Therefore, as always for star classifications,
 we use our SINFONI spectra for spectral classifications.

Outside of R$\geq2\arcsec$ there are less early-type than late-type stars \citep{Bartko_10,Do_09,Buchholz_09,Pfuhl_11}. 
In this radial range we include all stars, for which we did not record an early-type spectrum. Since we do not have spectra for all stars,
early-type can be in the sample.
Our selection is not radius independent, because we have a better spectral coverage close  to the center \citep{Bartko_10}:  
at 2$\arcsec$ we have spectral identifications for about 50\% of the stars, for which we have dynamics. This
fraction decreases to 13~\% at 20$\arcsec$. 
Further, our spectroscopic completeness decreases toward fainter stars. In conjunction with the top-heavy IMF for young stars in this radial 
range \citep{Bartko_10,Lu_13}
this means that the pollution of our sample with young stars is reduced compared to what would follow from the overall fractions. 
Quantitatively, we calculate the number of young stars that we still expect in our sample by a simple completeness correction, multiplying in each radius 
and magnitude bin the number of unidentified stars with the locally measured early-type fraction. We obtain that we include about 140 early-type stars
in the extended sample of 5864 stars. The total
pollution fraction of $\approx2.4$\% increases only slightly to $\approx3.7$\% in the inner radial half. 
For $r>20\arcsec$ we even find only 6 early-type stars, less than 1\% of all stars with spectra. 
Thus, we are confident that our dispersions (or higher moments) are not biased in a significant way.

\begin{figure}
\begin{center}
\includegraphics[width=0.50 \columnwidth,angle=-90]{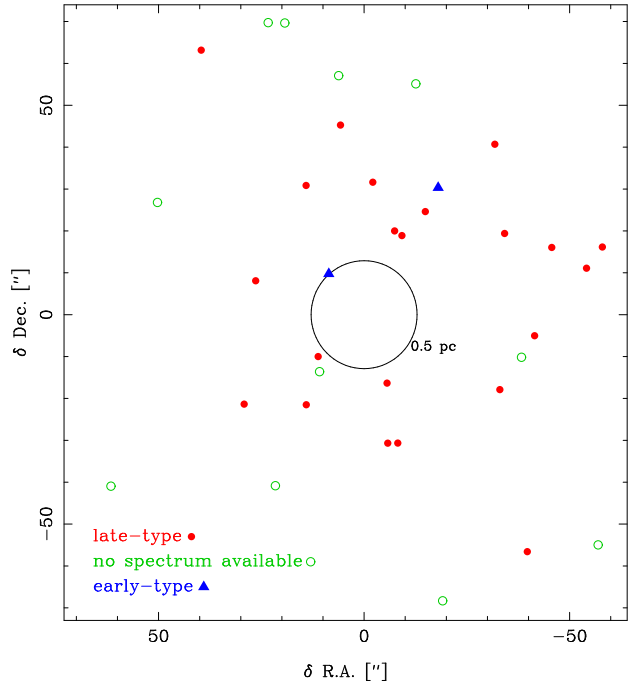}
\caption{Spectral identifications of the early-type candidates of \citet{Nishiyama_12}. We only check the candidates outside of 0.5 pc,
inside this radius most candidates were already crossmatched with spectra in the literature by \citet{Nishiyama_12}.
} 
\label{fig:nish_comp}
\end{center}
\end{figure}

We find in our spectra that there are very few early-type stars outside of 20$\arcsec$.
Surprisingly, \citet{Nishiyama_12} found from narrow band imaging a rather high early-type candidate fraction of $\approx7$~\% for bright
 (m$_{\mathrm{Ks}}<12.25$) stars. At $R>0.5\,$pc they found 35 stars. We have spectra for 24 of them, and see clear late-type signatures in 22 of them, see Figure~\ref{fig:nish_comp}. Only 2 
 of our spectra show early-type signatures. Thus, the contamination fraction is around 90~\%, and not 20~\% as claimed by \citet{Nishiyama_12}. 
The main problem in their analysis is probably that they misidentify stars from the population of  warm giants \citep{Pfuhl_11}.

 \subsection{By Foreground}
\label{sec:app_foreg}

Foreground and background stars should obviously not be used to constrain the GC potential. 
Similar to \citet{Buchholz_09} we use color to exclude foreground stars. 
The selection criterion for foreground
stars is $\mathrm{H-Ks}<1.3$, except in regions with unusual extinction where we use a local value. 
In total we exclude 97 stars as foreground stars.  For many stars no H-band magnitude is available. 
These stars are included in our sample, since from the more than 7000 stars with H-band information 
we can estimate that only $\approx1$\% of all stars belong to the foreground. Since foreground stars are not clustered
this fraction is rather constant over the field of view.
Color selecting does not work for the background. Most of the red stars in the GC are red due to intrinsic redness as a detailed spectroscopic analysis, see e.g. \citet{Fritz_10b}, shows.
Due to faint magnitudes of background stars this is even less
 of a problem than the foreground.

 \subsection{Velocity Outliers}
\label{sec:app_fast}

Here, we describe how we reject some high velocity stars from our sample.
Only 63 of over 10000 stars are excluded.

Mainly due to our mostly automatic procedure for finding and fitting stars it is possibly 
that extreme proper motions are not real.
This problem does not affect the motions in the central  and extended field. In these two fields all target stars selected on a single image have 
a reliable velocity due to the high number of epochs used. 
In contrast outside of 20$\arcsec$ we have at most five epochs for motions and the velocities might be affected by outliers.
The velocity histogram for stars at $R > 20\arcsec$ shows an excess of stars at v$_{\mathrm{2D}}$ above 0.8 v$_{\mathrm{esc}}$
compared to the velocity distribution obtained from stars between 7'' and 20''. 
In order to avoid such a bias to high velocities we look on the images of each star
and exclude stars which are not well identified, also considering v$_{\mathrm{2D}}$.
This procedure excludes 30 stars in the extended field and makes the shape of the velocity 
histogram similar to the inner sample.
Using the same procedure we exclude 20 stars in the outer field. Since only two epochs are available the fraction
of excluded stars is with $\approx3$\% higher.
This procedure reduces the dispersion significantly by $0.147 \pm 0.020 $ mas/yr in the large field and by  $0.172 \pm 0.041 $ mas/yr
in the outer field.
Afterwards the velocity histogram for the outer stars looks more similar to the one in the radial range from 7$\arcsec$ to 20$\arcsec$ (Section~\ref{sec:fast_stars}).
Since it is possible that we exclude somewhat too few or too
many stars in this exclusion procedure the tails of velocity distribution are still unreliable. 
Thus, we do not analyze the wings of the velocity distribution in the large and outer fields.

In the maser sample there are some surprisingly fast stars, see  
\citet{Lindqvist_92a,Deguchi_04}
 and Figure~\ref{fig:_velocity_his}.
After spatial binning of these stars (Section~\ref{sec:binning}) we inspect the binwise velocity histograms and exclude
stars that are apparent outliers. We exclude 13 of 274 stars.
The scatter in mean velocity and dispersion of the bins is notably reduced by this procedure.
Since with single and partly even two epoch radio observations a high fluke velocity is  difficult to produce, these stars
are probably truly fast stars from another population of stars.

 \section{D: Obtaining the Luminosity Properties}
\label{sec:obt_lum}

We describe here in detail how we obtain the luminosity properties.

\subsection{VISTA Star Counts}
\label{sec:ap_vis_star_count}

For measuring stars from the VISTA data we extract point sources from  the central R$_{\mathrm{box}}=1000\arcsec$ in H and Ks using STARFINDER \citep{Diolaiti_00}.
We divide the field of view into nine subimages: a central image and a ring of 8 images around. 
We use standard parameters of STARFINDER and PSF with a size of 15 pixels. Since we do not use the faintest stars, the exact parameters do not
have a relevant influence on the result.
The magnitudes obtained are adjusted to the VISTA catalog by matching bright, but not saturated stars. Essentially all stars outside the 
central 20$\arcsec$ are old 
stars, see Section~\ref{sec:out_rej}. Thus, we do not need to exclude young field stars from the VISTA data, except the
 Arches and Quintuplet clusters.

For source counts completeness is a concern.
To measure the source confusion we insert in the central Ks-image (R$_{\mathrm{box}}=344\arcsec$) artificial stars, using
 the PSF extracted with STARFINDER.  The PSFs are separated 
by 20 pixels to avoid artificial confusion.
 In this way we create six images with artificial stars with 11$\,<\,$Ks$\,<\,$16 in steps of 1 mag., 
to cover all unsaturated 
magnitudes for which significant numbers are detected.
In all these image we use STARFINDER for detecting the inserted stars in the same way as for
 the original image. The resulting completeness maps have a resolution of 20 pixels. Locally this may not be sufficient for a good  map. 
However, since we are mainly interested in the global radial and azimuthal profiles outside the very center, the number of artificial stars used (10201)
is sufficient for our purposes. 
Outside the central field the confusion is smaller. In the ring fields we measure confusion in the same way for Ks$=$14  and Ks$=16$ and then 
extrapolate to the other 
magnitudes using the multi magnitude completeness curve in the center which matches best to the completeness at Ks$=$14  and Ks$=16$.

We correct for extinction using stellar colors. The H-Ks color is nearly independent of stellar type 
\citep{Cox_00}: H-Ks$=0.29$ for M$_\mathrm{Ks}=-6.26$ giants and H-Ks$=0.15$ for M$_\mathrm{Ks}=-3.46$ giants. 
We use a color of H-Ks$=0.2$ for all stars.
For the extinction correction we use the extinction law toward the GC from \citet{Fritz_11} which implies A$_{\mathrm{Ks}}=(\mathrm{A}_{\mathrm{H}}-\mathrm{A}_{\mathrm{Ks}})/0.753$ for VISTA
filters. If possible we use $\mathrm{H-Ks}$ of each star for extinction correction. This way we can better account for extinction variations in the line 
of sight than if we would use the mean extinction. We use this method for Ks-sources with a H counterpart within two pixels and no other source 
within R$_{\mathrm{closest}}+1$ pixels. 
The vast majority of all bright Ks-sources has such a H counterpart.  
For the other sources we use the local extinction distribution of the matched sources in bins R$_{\mathrm{box}}=12.7\arcsec$. 
The primary result of this procedure are stellar density maps with pixels of R$_{\mathrm{box}}=12.7\arcsec$. 
We made these maps for different extinction corrected magnitudes.  
The brightest magnitude which is not significantly affected by saturation is, after extinction correction, m$_{\mathrm{Ks},\mathrm{excor}}=9$. 
For our final density map we use stars fainter than this magnitude and
 brighter than m$_{\mathrm{Ks},\mathrm{excor}}=10.5$ to exclude magnitudes which are severely incomplete in the center.
With this faint magnitude cut the completeness is 52~\% at r$\approx25\arcsec$ for the VISTA data. 
The use of stars fainter than m$_{\mathrm{Ks},\mathrm{excor}}=9$ also has the advantage that we avoid stars brighter than
 the tip of the giant branch which are partly
younger than most old stars in the GC. These younger and brighter stars are more concentrated to Sgr~A* on a large scale \citep{Catchpole_90}.

\begin{figure}
\begin{center}
 \includegraphics[width=0.50 \columnwidth,angle=0]{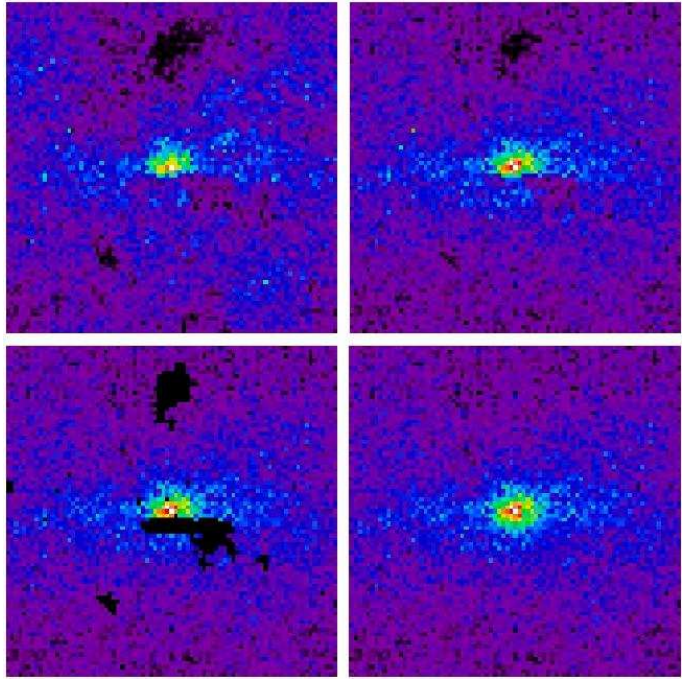} 
\caption{Stellar surface density maps in the GC. 
The  number of stars are obtained from  VISTA
data. The images show R$=1000\arcsec$ boxes around Sgr A*.
The Galactic plane runs horizontally. The upper left image shows the  completeness corrected star density
for stars with $11.5<$m$_{\mathrm{Ks}}<13$. The upper right image shows the completeness and 
extinction corrected
density of stars with $9<$m$_{\mathrm{Ks},\mathrm{excor}}<10.5$.
 Since each star is corrected for its extinction,  the same stars
are not necessarily used in the upper left and upper right image. The magnitude range in the raw magnitude map is chosen with the aim 
to use approximately the same number of stars in both maps.
In the lower left image we mask out the pixels with unusually low (high) counts.
In the lower right image we replace the masked pixels with the average flux of the pixels at the same $|l^*|$ and $|b^*|$ assuming such
symmetry with respect to Sgr A* in these coordinates.
} 
\label{fig:_surf_bri5}
\end{center}
\end{figure}

The result of this procedure is a star density map (Figure~\ref{fig:_surf_bri5}).
Symmetry relative to the Galactic plane and the concentration toward the GC is visible on this map.
A more complicated symmetry is also physically unlikely 
because the Galactic center is so small compared to its distance that any bar like structure would appear symmetric along its axis, independent of the orientation of the bar to the line of sight. 
Some Ks-dark clouds are still visible on the map.   
We mask out map pixels which show too low a density compared to their neighbors and others pixels at the same $|b^*|$ and $|l^*|$. 
In areas where completeness is no issue at fainter magnitudes
 we base the masking also on density maps for fainter stars, to reduce the influence of small number statistics.
At some radii the masked areas are mostly at $l^*\approx0$. At these radii a calculation of the radial profile from the unmasked area would result
in a density biased to $b^*\approx0$. To avoid this we replace in our final map the counts in the masked pixels by the average of the unmasked 
pixels at the same $|b^*|$ and $|l^*|$, see Figure~\ref{fig:_surf_bri5}.
To obtain a radial profile at higher resolution we also binned the stars more finely by a factor 5. Thereby, we use still the larger binning  
for masking and completeness correction since on smaller areas Poisson noise dominates.

To obtain error estimates for the general 
radial profile and the profiles in $l^*$ and $b^*$  we first calculate the lowest possible error from
 Poisson statistics. 
We obtain another error estimate by calculating the difference between the upper and lower halves, and the left and right halves of the maps
respectively.
This second estimate is typically 4.4~\%, is larger than the first and thus used for most radial bins. With these, the reduced $\chi^2$ of log-polynomials 
of fourth degree is smaller than 1. This shows that the errors of neighboring bins are correlated.
To estimate
the flattening we measure the density along the major and minor axes using data with less than  38$\arcsec$ separation from these axes. 
(Within 66$\arcsec$ where this method obviously smooths out the flattening we don't use the VISTA data to estimate it.)
For these profile we obtain the non-Poisson errors by comparing the two sides and obtain an error of 12.5$\,$\% which we scale up by a factor $\sqrt{2}$ 
for points which have only data on one side. At large separations, polynomial fits yield a $\chi^2/d.o.f.$ larger than 1.
 We therefore scale up the errors there to obtain $\chi^2/d.o.f.=1$.

\subsection{WFC3/IR Star Counts}
\label{sec:ap_wf_st_count}

For WFC3/IR data we use the final Multidrizzle product images in M127 and M153. We use STARFINDER \citep{Diolaiti_00} with standard parameters
and a PSF box of 2$\arcsec$ diameter. The fluxes obtained with this PSF are converted to magnitudes using the zeropoints for a 0.4$\arcsec$ PSF and
an infinitely large PSF. We convert the pixel coordinates of both star lists in Galactic center coordinates by using bright stars which are detected in WFC3/IR data and in the NACO data.
We match sources in the two filters and use only sources which are detected in both filters. Precisely, this means that the 
closest neighbor in M127 is within 1 pixel and second closest is at least 2.3 pixel further away.
We use the colors in the following way to apply the extinction correction on a per star basis: 
A$_{\mathrm{M153}}=2.12\,\cdot(m_{\mathrm{M127}}-m_{\mathrm{M153}}-0.374)$.
The factor is obtained from the -2.11 power law extinction in \citet{Fritz_11}. The giant color 0.374 is obtained from observed giant spectra, 
published in \citet{Rayner_09}, approximately K4III.

\begin{figure}
\begin{center}
  \includegraphics[width=0.50 \columnwidth,angle=0]{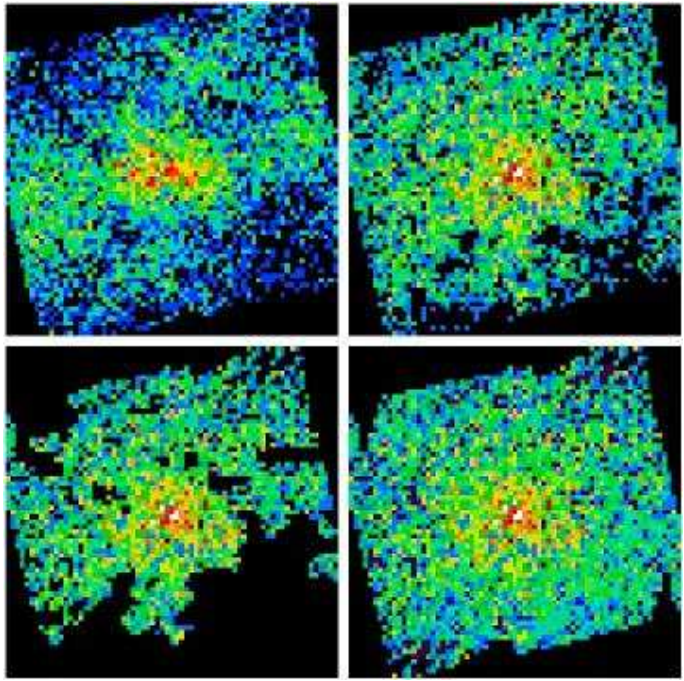} 
\caption{Stellar surface density maps in the inner GC. 
The number of stars is obtained from  WFC3/IR
data. The images show R$=68\arcsec$ boxes around Sgr A*.
The Galactic plane runs horizontal. The upper left image shows the star density of stars with m$_{\mathrm{M153}}<18.3$.
The upper right image shows the 
extinction corrected
density of stars with m$_{\mathrm{M153},\mathrm{excor}}<12.5$.
 Since each star is corrected for its own extinction, not everywhere the same stars
are used in the upper left and upper right image. The magnitude range in the raw magnitude map is chosen with the aim 
to use approximately the same number of stars in both maps.
In the lower left image we mask out the pixels with unusual low (high) counts.
In the lower right image we replace the masked pixels with the average flux of the pixels at the same $|l^*|$ and $|b^*|$ assuming such
symmetry with respect to Sgr A* in these coordinates.
} 
\label{fig:_surf_bri_wfc}
\end{center}
\end{figure}

 From these extinction corrected stars we construct stellar density maps in galactic coordinates using pixels of 2$\arcsec$ size.
We use stars brighter than $m_{\mathrm{M153}\, \mathrm{excor}}=12.5$ which in addition have $ m_{\mathrm{M127}}-m_{\mathrm{M153}}>2$ in order to
exclude foreground stars.
With this brightness cut more than 87$\,$\% of the unextincted stars, which have typically $m_{\mathrm{M153},\mathrm{excor}}<12.5$, have matches in M127. 
In the very center (R$<8\arcsec$) we recover with WFC3 90\% of the $m_{\mathrm{M153}\, \mathrm{excor}}<12.5$ stars found in the higher resolution NACO data. 
This completeness increases to 98\% at 20$\arcsec$. Since we use WFC3 data only outside of 20$\arcsec$, we do not need  to correct for 
completeness. Saturation is no worry, since only very few stars are saturated and many more fainter stars are included in the sample. 
As in the VISTA data NIR dark clouds can locally block all GC light, see Figure~\ref{fig:_surf_bri_wfc}. We use a density map of all stars with 
$m_{\mathrm{M153}\,\mathrm{excor}}<15$
to identify these clouds and mask them out. This second density map saturates in the center due to crowding but is superior in areas of low star
density, such as these dark clouds. 

Early-types are not important in the region of the map we use (R$>20\arcsec$), see Section~\ref{sec:out_rej}.
For visualization, as in case of the VISTA data, we replace the masked out areas with their symmetric partners, see Figure~\ref{fig:_surf_bri_wfc}.
There it can be seen that the cluster is, after extinction correction, much less flattened than at
first sight.
To quantify the weak flattening we measure the stellar density separately in $l^*$ and $b^*$ using stars in halves which have $|l^*|>|b^*|$, 
and $|l^*|<|b^*|$ respectively. As the error we use the larger of the following: either Poisson statistics or the average difference between the density at $-l^*$ and $+l^*$
and $-b^*$ and $+b^*$ which is 7.8~\%. We obtain the full radial profile from the average of the two profiles in $l^*$ and $b^*$.

\subsection{NACO Star Counts}
\label{sec:ap_nac_st_cou}

In the center we use our NACO data. We extract stars from a NACO image with STARFIN\-DER \citep{Diolaiti_00}. We calibrate the magnitudes of
these stars locally on the source list of \citep{Schoedel_09b}.
We correct again for extinction.
In the NACO field the use of an extinction map is superior to a star by star color since the extinction does not vary much in the line of sight. 
Thus, a map that uses the local median reduces the noise.
We use the
extinction map of \citet{Schoedel_09b} scaled by a factor of 0.976 to align the map with the extinction of \citet{Fritz_11}.
We select stars with $m_{\mathrm{Ks},\mathrm{excor}}=12.65$. 
With this magnitude cut we still have some stars in the central arcsecond and are, at the same time, complete. In imaging we are essentially
complete till this magnitude, we thus do not apply a completeness correction. 
Dark clouds are no issue in the central 20$\arcsec$, we therefore do not apply masks.
In the NACO data the exclusion of early-type stars is important, see Appendix~\ref{sec:app_early} for the detailed procedure. For the density profile we
 use from 0 to 20$\arcsec$ the late-type fraction of the spectroscopically typed stars with $m_{\mathrm{Ks},\mathrm{excor}}=12.65$ to convert the density profile of all stars to the late-type density profile.  
To obtain estimates for the flattening in the center we also separately calculate the late-type stellar density in the same two halves of the data
as for the WFC3/IR data. 
The three profiles from different instruments are aligned at their transition radial ranges.

\subsection{Extinction Corrected Flux}
\label{sec:ap_flux}

We now describe how we obtain the extinction corrected flux of the GC. We use VISTA and NACO data.

In case of the VISTA data we subtract the median counts toward some dark clouds in the H- and Ks-band as sky.
We bin the flux in H and Ks, again in pixels of R$_{\mathrm{box}}=12.7\arcsec$. This flux is then used for extinction correction, assuming a somewhat bluer
color of $\mathrm{H-Ks}=0.15$ since  bright cool giants are less important in total flux than in source counts of bright stars, 
especially when as in our case, the brightest stars are saturated. In the 
extinction corrected flux map the dark clouds are better visible than in the stellar density map, since few bright blue foreground stars
can bias the extinction estimate toward values that are too low. We therefore construct a new mask for the flux and apply it to the data. To obtain a profile
 with finer radial sampling we bin the flux finer but still use the same extinction and  mask maps.  
 The final profile is similar but not identical to the stellar density profile.
In the very center of the VISTA image, saturation affects many sources.

 We use there a flux calibrated NACO image, from which the sky is already subtracted, and  
 the extinction map described in Appendix~\ref{sec:ap_nac_st_cou} above to obtain the light profile. 
In the center, a few bright young stars dominate the total flux. We find in the star list used in Appendix~\ref{sec:ap_nac_st_cou}
the spectroscopic young stars (early-type stars and the young red-supergiant IRS7) and subtract their total flux.
Their flux is only relevant in the inner 10$\arcsec$. Due to the top-heavy IMF \citep{Bartko_10,Lu_13}, the fact that we do not have spectra
for all stars has a negligible influence on the total flux.

Finally we use the transition region around 20$\arcsec$ to align the
VISTA flux to the NACO flux. 
We estimate the errors on the resulting profile by fitting step wise low order log-polynomials to the data  and setting $\chi^2/d.o.f.=1$.
As expected for noise, which is dominated by few bright stars, the error decreases with radius.

\subsection{Measuring the flattening}
\label{app:meas_flat}

\begin{figure}
\begin{center}
  \includegraphics[width=0.60 \columnwidth,angle=-90]{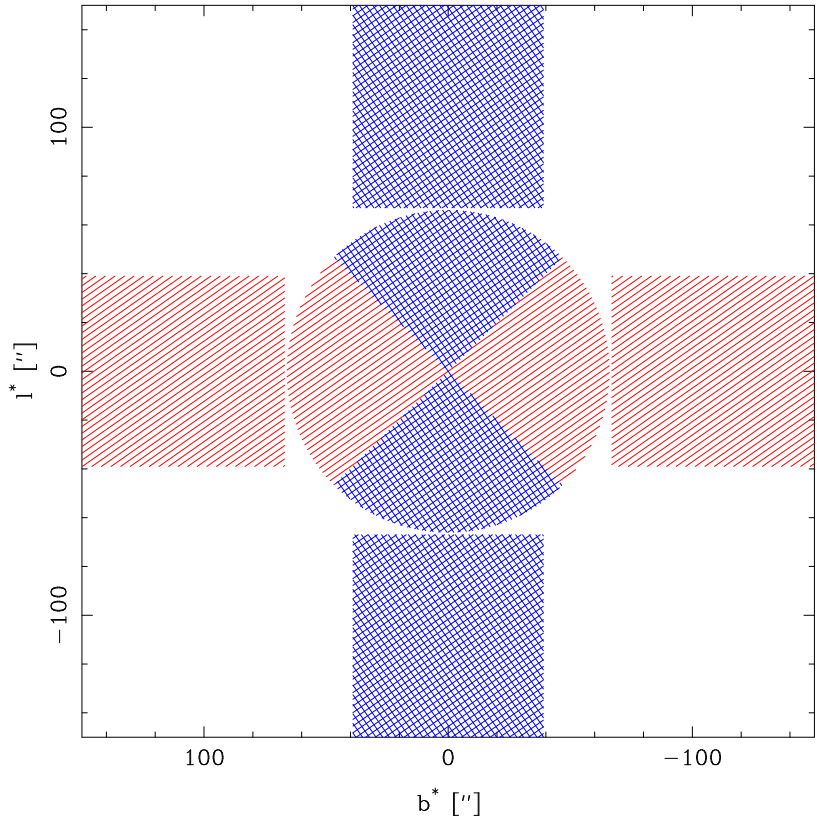} 
\caption{Bins for obtaining the density in $|l^*|$ and $|b^*|$. The red single   areas shows the bins for the density in $|l^*|$,
the blue double hatched area shows the bins for measuring density in $|b^*|$. Outside of 150$\arcsec$ we continue to use the rather small strip around the axes to measure the density along the axes.
} 
\label{fig:_lb_dens_bins}
\end{center}
\end{figure}

For the binning of the density in $|l^*|$ and $|b^*|$ we use two different schemes dependent on the radius, see Figure~\ref{fig:_lb_dens_bins}.

 For the NACO and WFC3 data the number counts in 
$|l^*|$ and $|b^*|$ are measured using the angles which  are closer to $|l^*|$ respectively to $|b^*|$. 
With this method we use all available information and are such able to measure the small flattening at $R<68\arcsec$. 
Further out the signal is high enough that the flattening can be measured directly.
There we use for the density along major and minor axis
only the data which is closer than 39'' to the respective axis.

Since our data is of too low SNR, it is not possible to measure the flattening directly by fitting 
constant density ellipses to it. 
For our measure we use Galfit in bins.
The bins are elliptical, which we adapt iteratively to the flattening in the bins. We fit a single Sersic to the data in each bin. The flattening is not sensitive to the radial (Sersic or other) functional  form.
We use the usual mask and fix the flattening to align with the Galactic plane. The inner most bin covers the 
NACO and WFC3 area: $|l^*|<68\arcsec$. The fit errors, the dependence on the used functional form and 
the dependence on the used $|b^*|$ range together add to the total errors.

\subsection{Reliability of the measured flattening}
\label{app:flat_syst_err}

The axis ratio is $0.80\pm0.04$
within $|l^*|<68\arcsec$. In this radial range the ratio is rather constant with a slightly higher value around 20$\arcsec$ than elsewhere.  The flattening increases outside of 68$\arcsec$. Within 68$\arcsec$ we obtain the flattening from 
from NACO and WFC3/IR data. Outside of this range we use VISTA data and there the flattening  increases
out to the end of our field of view. 
Due to this change of the data source where the flattening increases it is natural to ask the question if the
flattening measured is affected by the data source used. We concentrate thereby on the WFC3/IR data range since that radial range
is the most important range. 
For both transition regions of WFC3/IR,  the inner (to NACO) and the outer (to VISTA) we obtain 
consistent flattenings in the different data sets.
The signal of the VISTA data is too low that this comparison is constraining.

We test now whether one of the steps, which we use for WFC3/IR data inhibits a known systematic uncertainty, which is relevant for flattening compared to the statistical uncertainty.
A priori it may be of worry, that the WFC3/IR data is obtained at a smaller wavelength (1.53~$\mu$m) compared to
the  NACO and VISTA data (2.16~$\mu$m). 
 The stellar color varies more at wavelength shortward of the Ks-band.  However, compared to other effects like extinction the importance of stellar color does not increase and can change q at most by 0.005.

At 1.53~$\mu$m the extinction is stronger than in Ks-band by a factor 2.05. Extinction can have two different effects on stars. Firstly, it can be weak, such that it can be corrected by two-color 
information using a known extinction law.
Secondly, the extinction can be so strong that the stars are invisible in all our bands.
The slope uncertainty of 0.06 in the extinction power law \citep{Fritz_11} propagates to a systematic error of 0.005 in q.
As an independent test of the extinction law, we use the red clump magnitude in our data like \citet{Schoedel_09b}. We obtain a by 0.12 steeper extinction law. A steeper extinction law decreases the flattening for extinction distribution. The change in q is however only 0.01, for the observed in the extinction law.

As another test whether extinction issues are important we analyze H-/Ks-data, namely good seeing (0.4$\arcsec$ FWHM) ISAAC data. 
The Figure~5 of \citet{Schoedel_07} is based on such data.
No clear elongation in the Galactic plane is visible 
after extinction correction, as also stated in their text, 
confirming our WFC3/IR result. To get a more quantitative estimate we also analyze ISAAC images ourselves.
We use therefore mainly newer ISAAC images, which were also used by \citet{Nishiyama_12}. Precisely we use data in the filters IB171 and IB225,
which are intermediate band filters close but not exactly in the middle of H and Ks-band.
We then proceed as in case of the other data to obtain a star based density map corrected from effects like extinction. 
We obtain from these data $q\approx0.83$ consistent with the WFC3/IR data.  Finally, we do not use these data because the ISAAC detector shows at GC fluxes a relevant change in the bias \citep{Nishiyama_12}
which is difficult to correct. 

 The blocking effect of extinction is more difficult to quantize. When we vary the blocked area by a reasonable amount, estimated per eye,  q can be changed at most by 0.02. 
There are two possibilities how blocking dust clouds in the Galactic plane can cause relevant changes in the flattening without being detectable in the star density map.
In the first case, large dark clouds are slightly behind the GC, and block all light from behind, but not from front. In that cause the clouds decrease the flattening slightly, but do not cause obvious dark patches.
In the second case, the clouds are
very small clumps undetectable at the resolution of our star density maps. No special location in the line of sight is required in this case. Both solutions require too much designed structure to be likely. The first solution
would also result in particular dynamics which are not evident in our data. The second solution requires very dense rather small clumps: 
for example if the clouds are at the upper limit of possible
diameter of 2$\arcsec=0.08$ pc and have an extinction of four times the usual one, that are A$_{\mathrm{Ks}}=10$, a density of 10$^6$ cm$^{-2}$ follows from 
 1 A$_{\mathrm{Ks}}\approx 3 \times 10^{22} \mathrm{cm}^{-2}\,$N$_{H}$ \citep{Fritz_11}. Such dust clouds should be obvious also
in the far-IR/sub-mm data.

 \section{E: Binning of the dynamics data}
\label{app_binning}

\subsection{Proper Motion Bins}

\begin{figure}
\begin{center}
  \includegraphics[width=0.60 \columnwidth,angle=-90]{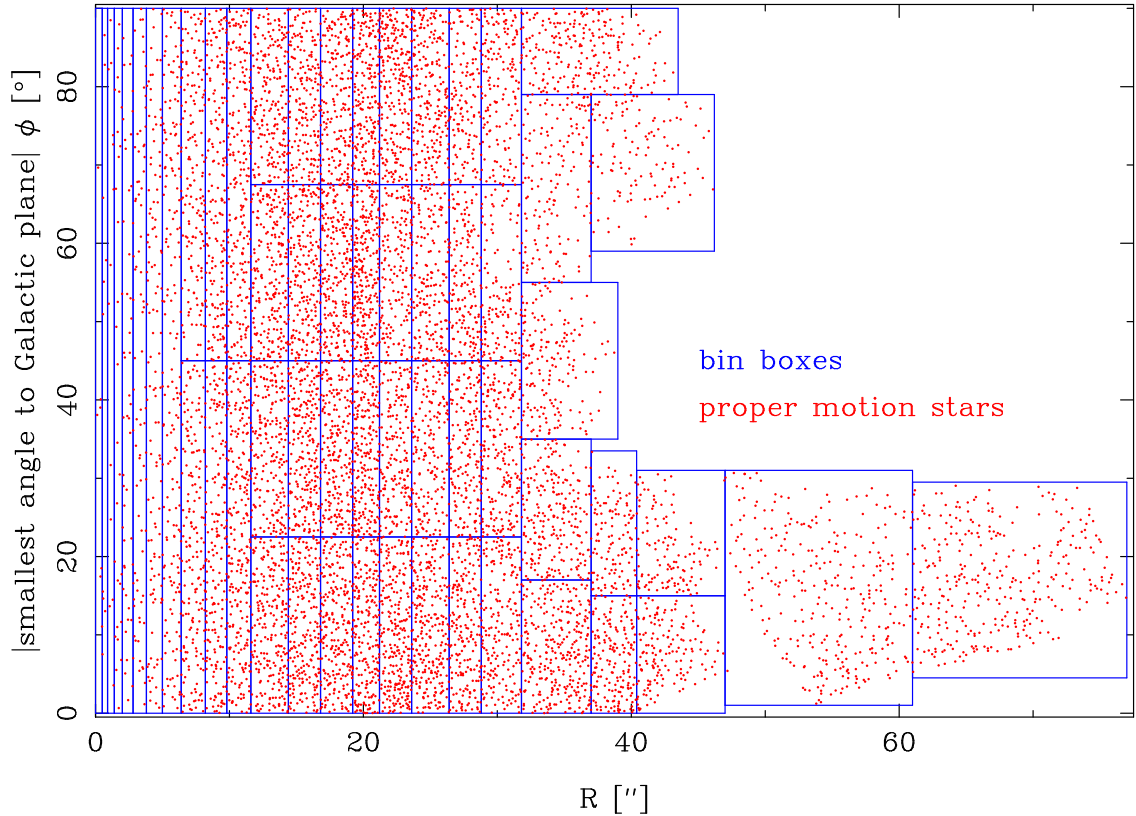}  
\caption{Binning of the proper motion stars. 
The figure shows the parameter space which we use for binning, radial distance from Sgr~A* and the
smallest absolute value of $\mathrm{angular}$ distance from the Galactic plane. 
} 
\label{fig:_bin_prop}
\end{center}
\end{figure}

Rotation and flattening of the GC cluster are expected to occur in direction along the Galactic plane, and hence we choose proper motion bins 
reflecting that. We use two-dimensional circular coordinates $(R,\phi)$, where $R$ is the distance from the center and $\phi$ is the absolute value
of the smallest angle
between the Galactic plane and the respective star, thus $0^{\circ}\leq \phi \leq 90^{\circ}$.
This definition of $\phi$ uses the symmetry of the edge-on system, in which the dispersion can only vary by $R$ and $\phi$. 
Our bins contain a relative similar
 number of stars (Figure~\ref{fig:_bin_prop}), between 123 and 325 with an average of 198 stars outside of 12$\arcsec$. Inside of that
 we have less stars, but 
we need higher resolution in order to resolve the increased velocity dispersion caused by the 
SMBH. The innermost bin with $R<0.5\arcsec$ contains 7 stars. Between 12$\arcsec$ and 32$\arcsec$ we choose four azimuthal bins for being able to measure 
robustly azimuthal variations, for $R<5\arcsec$  we do not bin the stars azimuthally.

\subsubsection{Radial Velocity Bins}

\begin{figure}
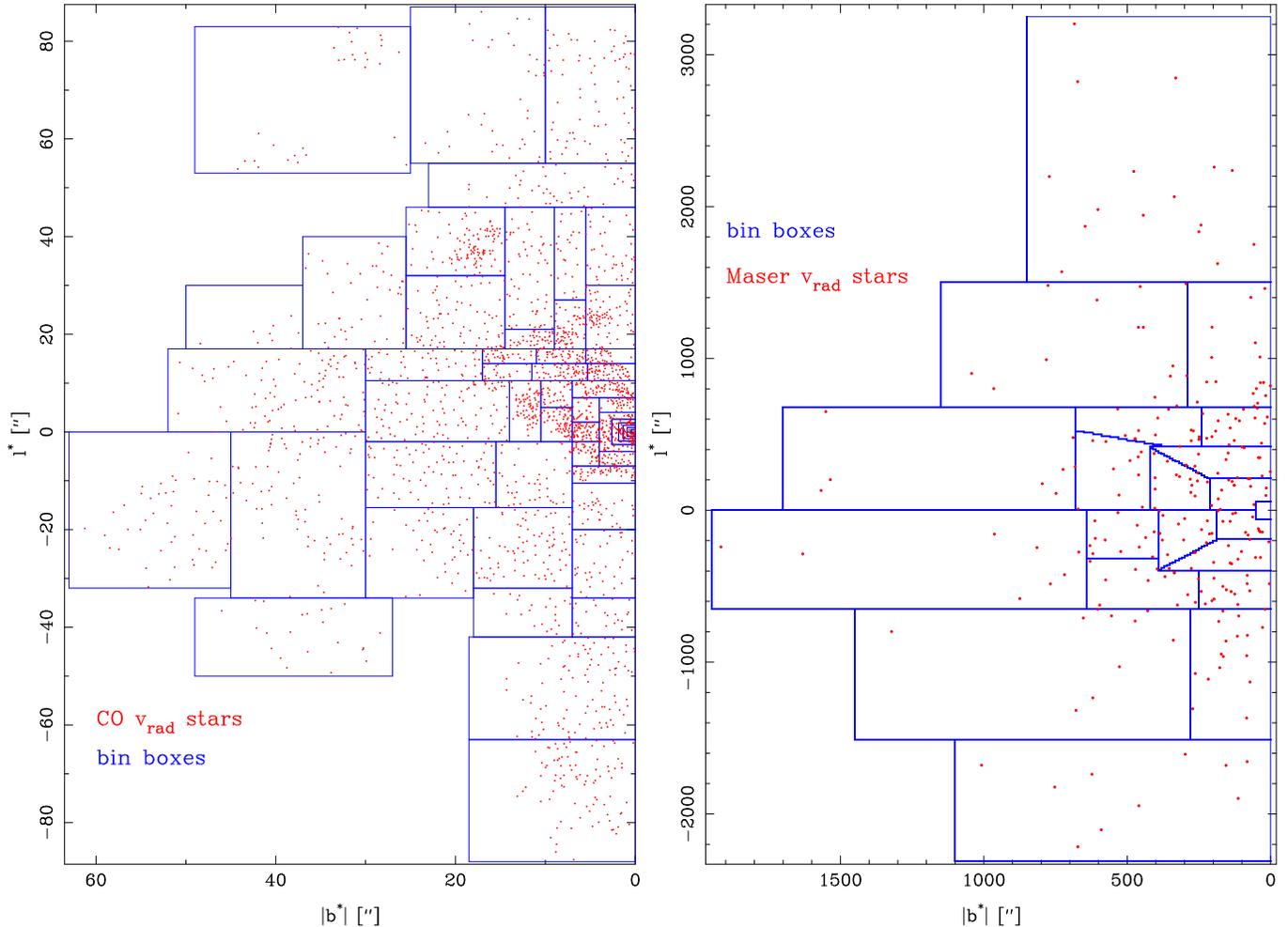

\begin{center}
 \includegraphics[width=0.725 \columnwidth,angle=-90]{fe2a.eps}
\includegraphics[width=0.725 \columnwidth,angle=-90]{fe2b.eps} 
\caption{
Binning of the radial velocity stars.
 These stars  are binned in l$^*$ and $|\mathrm{b}^*|$. On the left are the inner stars plotted, on the right the outer,
the maser sample from \citet{Lindqvist_92a,Deguchi_04}
} 
\label{fig:_bin_radv}
\end{center}
\end{figure}

The rotation of the cluster causes a change of the mean radial velocity with l$^*$ \citep{McGinn_89,Lindqvist_92b,Genzel_00} from negative 
velocities at negative l$^*$
to positive values for positive l$^*$. Our test for the rotation axis (Section~\ref{sec:rotation}) shows the radial velocity does not depend much
on the sign of b$^*$ as expected in an edge-on rotating system. Hence, a natural choice for the binning 
coordinates is (l$^*$, $|\mathrm{b}^*|$). We again choose our bins such that the bins are relatively 
evenly populated with stars, but the overall number of stars is smaller than for the proper motions. Outside the central 4$\arcsec$ there are 
on average 51 stars per bin. 
In the central 4$\arcsec$ we do not use the sign of l$^*$ since the rotation is there of minor importance  compared to the large radial velocity dispersion gradient. We choose quadrangular rings as bins to transit gap free to the rectangular bins
further out. The inner most bin includes all stars with $|\mathrm{l}^*|<0.5\arcsec$ and  
$|\mathrm{b}^*|<0.5\arcsec$, see Figure~\ref{fig:_bin_radv}. Further out, in the regime of the maser stars we adopted from the literature, 
we use more complex bins to achieve comparable star numbers per bins (Figure~\ref{fig:_bin_radv}).

 \section{F: Tabulated Values of the Cumulative Mass Profiles}
\label{sec_val_cum_mass}

Here, we present our mass models from Figure~\ref{fig:_nc2}  in Table~\ref{tab:cum_mass}.

\begin{deluxetable}{lllll} 
\tabletypesize{\scriptsize}
\tablecolumns{5}
\tablewidth{0pc}
\tablecaption{Cumulative mass distribution \label{tab:cum_mass}} 
\tablehead{
r[$\arcsec$]~~~~ & r[pc]~~~~ & M$_{\mathrm{A}}$  [10$^6$M$_{\odot}$] & M$_{\mathrm{B}}$  [10$^6$M$_{\odot}$]&  M$_{\mathrm{C}}$  [10$^6$M$_{\odot}$]
}
\startdata
     5 &  0.20   &      0.021   &     0.013       &    0.152\\ 
     7 &  0.28   &      0.044   &     0.029       &    0.230\\  
    10  &   0.40   &       0.092  &       0.066      &       0.357\\
    15  &   0.60  &        0.211  &        0.167     &   0.588\\
    20 &    0.80   &            0.374 &    0.316    &   0.839\\
    25  &   0.99    &          0.577 &        0.507    &   1.104 \\
    40  &   1.59 &  1.382  &  1.302   &  1.969\\
    50  &   1.99 &   2.046  &   1.974  &  2.594\\
    75 &    2.98 &   3.965  &   3.934  &  4.272\\
    
\enddata
\tablecomments{Cumulative mass of three different extended mass models. All assume R$_0=8.2\,$kpc.
 The overall preferred model M$_{\mathrm{A}}$ uses the preferred constant M/L model, with a inner slope of -0.81 (Row~5 in Table~\ref{tab:_mass_iso1}). 
 The model M$_{\mathrm{B}}$ is identical to the previous, however the inner slope is changed to the shallowest possible of -0.5.  
 To obtain M$_{\mathrm{C}}$ a power law model with $\delta_\mathrm{M}=1.232$ is used. All
are scaled to $6.09\times10^6\,$M$_{\odot}$ at 100$\arcsec$. At 40$\arcsec$ the mass difference between the power law case and the constant M/L case is maximal.
}
\end{deluxetable}

\section{G: Presentation of the Data Used}
\label{sec_used_data}

Here, we present our velocities used in the analysis in Table~\ref{tab:int_ext} in electronic form.

\begin{deluxetable}{llllllll} 
\tabletypesize{\scriptsize}
\tablecolumns{8}
\tablewidth{30pc}
\tablecaption{Radially sorted velocity data \label{tab:int_ext}}
\tablehead{R [$\arcsec$] & R.A. [$\arcsec$] &  Dec. [$\arcsec$]& v$_{\mathrm{R.A.}}$ [mas/yr] & v$_{\mathrm{Dec.}}$ [mas/yr] &  v$_{\mathrm{rad}}$ [km/s] & m$_\mathrm{H}$ & m$_\mathrm{Ks}$ }
\startdata
0.13 & 0.01 & -0.12 & 6.189 $\pm$ 0.1 & 23.017 $\pm$ 0.157 & 519.9 $\pm$ 43.9 & 17.8 & 15.87 \\
0.2 & -0.2 & 0.04 & -19.092 $\pm$ 0.71 & -12.852 $\pm$ 0.247 & 185 $\pm$ 70 & 19.1 & 17.15 \\
0.36 & -0.34 & -0.12 & -7.656 $\pm$ 0.208 & 5.091 $\pm$ 0.08 & 429.8 $\pm$ 12.7 & 18.79 & 16.76 \\
0.37 & 0.28 & 0.24 & -0.887 $\pm$ 0.148 & -1.138 $\pm$ 0.173 & 170 $\pm$ 22 & 18.9 & 16.97 \\
0.38 & 0.06 & -0.38 & -5.012 $\pm$ 0.012 & 2.981 $\pm$ 0.014 & 96.6 $\pm$ 1.6 & 16.13 & 14.04 \\
0.44 & -0.1 & -0.43 & -2.676 $\pm$ 0.042 & 0.984 $\pm$ 0.041 & -256 $\pm$ 2.5 & 17.51 & 15.06 \\
0.48 & -0.33 & -0.36 & -3.905 $\pm$ 0.027 & 0.497 $\pm$ 0.032 & -108.6 $\pm$ 5.8 & 18.63 & 16.51 \\
0.52 & 0.33 & 0.4 & -5.533 $\pm$ 0.115 & 9.866 $\pm$ 0.081 & 340 $\pm$ 5 & 18.6 & 16.62 \\
0.55 & 0.15 & 0.53 & 0.557 $\pm$ 0.025 & 2.944 $\pm$ 0.028 & -113.4 $\pm$ 2.5 & 17.54 & 15.53 \\
0.56 & 0.19 & -0.52 & -5.881 $\pm$ 0.057 & -4.647 $\pm$ 0.045 & 132.1 $\pm$ 1 & 17.51 & 15.44 \\
\enddata
\tablecomments{
The errors of the positions are irrelevant small for dynamics purposes. 
Thus, we do not give these errors. 
Parameters which are exactly zero and have no errors were not measured and not used.
(The full table is available in the electronic edition.)}
\end{deluxetable}

\end{document}